\documentclass[english,aps,prb,reprint]{revtex4-1}
\usepackage[T1]{fontenc}
\usepackage[latin9]{inputenc}
\usepackage{bm}
\usepackage{amsmath}
\usepackage{amssymb}
\usepackage{graphicx}

\makeatletter

\providecommand{\tabularnewline}{\\}

\usepackage[all]{xy}

\usepackage{variables}
\usepackage[bookmarks=false]{hyperref}

\makeatother

\usepackage{babel}
\begin{document}
\title{Qubit parametrization of the variational discrete action theory for
the multiorbital Hubbard model}
\author{Zhengqian Cheng and Chris A. Marianetti}
\affiliation{Department of Applied Physics and Applied Mathematics, Columbia University,
New York, NY 10027}
\date{\today}
\begin{abstract}
The variational discrete action theory (VDAT) at $\mathcal{N}=3$
is a potent tool for accurately capturing Mott and Hund physics at
zero temperature in $d=\infty$ at a cost comparable to the Gutzwiller
approximation, which is recovered by VDAT at $\mathcal{N}=2$. Here
we develop a qubit parametrization of the gauge constrained algorithm
of VDAT at $\mathcal{N}=3$ for the multiorbital Hubbard model with
general density-density interactions. The qubit parametrization yields
an explicit variational trial energy, and the variational parameters
consist of the momentum density distribution, the shape of a reference
fermi surface, and the pure state of a qubit system with dimension
of the local Hilbert space. To illustrate the power of the qubit parametrization,
we solve for the ground state properties of the multiorbital Hubbard
model with Hund coupling for local orbital number $N_{orb}=2-7$.
A Taylor series expansion of the partially optimized trial energy
is used to explain how the Hund's coupling changes the order of the
Mott transition. For the case of the $SU(2N_{orb})$ Hubbard model,
an explicit approach for computing the critical $U_{c}$ for the Mott
transition is provided, yielding an analytical expression for $U_{c}$
in the large $N_{orb}$ limit. Additionally, we provide an analytical
solution for the ground state properties of the single band Hubbard
model with a special density of states. Finally, we demonstrate that
the qubit parametrization can also be applied to $\mathcal{N}=2$,
for both G-type and B-type variants, where the G-type yields an identical
expression to the slave spin mean-field theory. The qubit parametrization
not only improves the efficiency and transparency of VDAT at $\mathcal{N}=3$,
but also provides the key advances for the construction of a one-body
reduced density matrix functional capable of capturing Mott and Hund
physics.
\end{abstract}
\maketitle

\section{Introduction\label{sec:Introduction}}

The multiorbital Hubbard model is a minimal model of interacting electrons
which accounts for key elements of typical strongly correlated electron
systems \cite{Imada19981039,Fernandes2017014503}, and the solution
in infinite dimensions provides a natural starting point for understanding
the solution in finite dimensions \cite{Georges199613,Kotliar2006865}.
The de facto standard for accurately solving the multiorbital Hubbard
model in $d=\infty$ is the dynamical mean-field theory (DMFT) \cite{Georges199613,Kotliar200453,Kotliar2006865,Vollhardt20121},
which requires self-consistently solving an effective Anderson impurity
model. Therefore, the proficiency of DMFT is determined by the state-of-the-art
for quantum impurity solvers \cite{Gull2011349}, which are still
extremely limited for the multiorbital case at zero temperature (see
introduction of Ref. \cite{Cheng2022205129}). Alternatively, the
standard for efficiently solving the multi-orbital Hubbard model is
the Gutzwiller approximation (GA), which semi-quantitatively describes
the Fermi liquid phase and the metal-insulator transition, but only
produces a crude description of the insulating phase \cite{Lu19945687,Okabe19972129,Bunemann19974011,Bunemann199829,Bunemann2007193104,Lanata2008155127}.
A long standing goal has been to improve upon GA while not incurring
the computational cost of DMFT. Fortunately, the variational discrete
action theory (VDAT) provides an alternate route to the exact ground
state properties in $d=\infty$ \cite{Cheng2021195138,Cheng2021206402,Cheng2022205129,Cheng2023035127}.
The variational ansatz of VDAT is the sequential product density matrix
(SPD), which is characterized by an integer $\mathcal{N}$ and as
either G-type or B-type, where G-type $\mathcal{N}=1$ recovers the
Hartree-Fock wave function, G-type $\mathcal{N}=2$ recovers the Gutzwiller
wave function, and $\mathcal{N}\rightarrow\infty$ recovers the exact
solution. A remarkable fact is that the G-type $\mathcal{N}=3$ SPD
accurately captures Mott and Hund physics in $d=\infty$ \cite{Cheng2022205129,Cheng2023035127},
while maintaining a computational cost comparable to the Gutzwiller
wave function.

An SPD with arbitrary $\mathcal{N}$ can be exactly evaluated in $d=\infty$
using the self-consistent canonical discrete action theory (SCDA)
\cite{Cheng2021195138,Cheng2021206402}. There are currently three
algorithms for executing VDAT within the SCDA, with the first two
being completely general and the third one having some restrictions.
The first approach is the most straightforward, requiring iterations
over two steps \cite{Cheng2021206402}. In the first step, given the
variational parameters, the SCDA equations can be self-consistently
solved, yielding the energy for the corresponding SPD. In the second
step, the variational parameters must be updated to minimize the energy.
This straightforward approach was successfully executed on the single
band Hubbard model in $d=\infty$. The second approach is the decoupled
minimization algorithm \cite{Cheng2022205129}, where one simultaneously
minimizes the variational parameters and updates the SCDA equations
towards self-consistency. This decoupled minimization algorithm was
executed for the G-type $\mathcal{N}=2-4$ SPD's in the two orbital
Hubbard model with the full rotationally invariant local interactions,
demonstrating that $\mathcal{N}=3$ accurately describes Mott and
Hund physics over all parameter space. The third approach is the gauge
constrained algorithm \cite{Cheng2023035127}, which was originally
applied to the case of a G-type $\mathcal{N}=2$ and $\mathcal{N}=3$
SPD with kinetic projectors that are diagonal in momentum space and
interacting projectors that do not introduce off-diagonal terms at
the single-particle level. The key feature of the gauge constrained
algorithm is that the SCDA self-consistency can be automatically satisfied,
greatly simplifying the task of minimizing over the variational parameters.
Furthermore, the momentum density distribution is used to reparametrize
the variational parameters in the kinetic projectors, and the trial
energy can be straightforwardly optimized over the momentum density
distribution. It is useful to note that the GA can only evaluate a
G-type $\mathcal{N}=2$ SPD under a Gutzwiller gauge \cite{Cheng2023035127}.
The SCDA has no such restrictions, which is formally appealing, but
imposing certain gauge constraints allows the SCDA self-consistency
to be automatically satisfied, while not reducing the variational
capacity. 

One shortcoming of the original gauge constrained algorithm for the
G-type $\mathcal{N}=3$ SPD is that some of the variational parameters
are unintuitive and there are two linear constraints on the momentum
density distribution per spin-orbital \cite{Cheng2023035127}. In
this paper, we introduce a new parametrization of the gauge constrained
algorithm, referred to as the qubit parametrization, which is mathematically
equivalent to the original gauge constrained algorithm and offers
several important advantages. First, there is only one linear constraint
per spin-orbital on the momentum density distribution, and therefore
the number of variational parameters is reduced by one per spin-orbital.
Second, the variational parameters are physically intuitive and facilitate
a deeper understanding of how the SPD captures Mott and Hund physics.
Therefore, the qubit parametrization achieves the long sought goal
of resolving the shortcomings of the Gutzwiller approximation while
maintaining the computational simplicity and physical appeal. The
convenience of the qubit parametrization has allowed for the construction
of a one-body reduced density matrix functional for the multi-orbital
Hubbard model which exactly encapsulates the VDAT result at $\mathcal{N}=3$,
and this is presented in a companion manuscript \cite{companion}. 

It is useful to compare to existing approaches in the literature which
seek to achieve our same goal of efficiently and accurately solving
the ground state properties of multiorbital Hubbard models. One approach
is the ghost Gutzwiller approximation (gGA) \cite{Lanata2017195126},
which introduces the Gutzwiller wave function in an extended Hilbert
space where the GA is then applied, though this comes with a substantial
increase in computational cost over the usual GA given that ghost
orbitals are introduced. In $d=\infty$, the gGA exactly evaluates
the Gutzwiller wave function in the extended Hilbert space, and therefore
provides an upper bound on the exact energy with increasing accuracy
as the extended Hilbert space grows. The gGA has not been proven to
converge to the exact solution of the multiorbital Hubbard model with
increasing number of ghost orbitals, though numerical evidence suggests
that this might be the case \cite{Lee2023L121104,Lee2023245147}.
Empirically, it has been demonstrated that at least two ghost orbitals
per physical orbital are needed for a global improvement over the
GA \cite{Lanata2017195126,Lee2023245147,Lee2023L121104}, which substantially
increases the computational cost of the gGA relative to GA. Therefore,
the important practical comparison between VDAT and gGA contrasts
VDAT at $\mathcal{N}=3$ to gGA with two ghost orbitals per physical
orbital. VDAT at $\mathcal{N}=3$ has been straightforwardly applied
up to the eight orbital Hubbard model \cite{Cheng2023035127}, while
the gGA has been applied up to the three orbital model \cite{MejutoZaera2023235150,Lee2023245147,Giuli2025020401}
using exact diagonalization and the five orbital model \cite{Lee2024115126}
using density matrix renormalization group (DMRG). The success of
the gGA in the multiorbital Hubbard model critically depends on the
ability of DMRG or related techniques to approximately obtain a ground
state in an expanded local Hilbert space. In terms of accuracy, both
approaches appear to be excellent over all parameter space, though
it seems possible that VDAT is more adept at capturing subtle differences
between competing phases (e.g. compare Fig. 4 in Ref. \cite{Cheng2022205129}
with Fig. 2 in Ref. \cite{MejutoZaera2023235150}). 

Another attempt to resolve the limitations of the GA is the slave
boson approach \cite{Kotliar19861362}, which formally recasts the
Hubbard model as a Hamiltonian of constrained fermions and auxiliary
bosons. The simplest saddle point approximation is typically referred
to as slave boson mean-field theory (SBMF), which exactly recovers
the GA in the multi-orbital Hubbard model \cite{Bunemann2007193104}
and has extensive applications \cite{Lechermann2007155102,Isidori2009115120,Bunemann2011203,medici2017167003,Piefke2018125154,Isidori2019186401,Chatzieleftheriou2020205127}.
Given that the slave boson approach is an exact formalism, there have
been various results including fluctuations beyond the mean-field
solution \cite{Lavagna1990142,Jolicoeur19912403,Li1991369,Raimondi199311453,Arrigoni19933178,Li199417837,Zimmermann199710097},
though none of these studies address corrections to the ground state
properties in the thermodynamic limit. A related idea is the time
dependent Gutzwiller approximation \cite{Seibold20012605,Lorenzana2003066404,Schiro2010076401,Oelsen2011076402,oelsen2011113031,Schiro2011165105,Bunemann2015550,Fabrizio2017075156},
which has mainly focussed on computing response functions. One study
did demonstrate that the ground state properties of the single orbital
Hubbard model can be improved by updating the double occupancy based
on the density-density response function \cite{Seibold20012605},
but this would be impractical in the multi-orbital Hubbard model,
and we are not aware of any such applications. Therefore, neither
slave bosons nor time dependent GA has provided a path beyond the
GA for ground state properties, though the idea of ghost orbitals
has recently been adopted into the slave boson methodology \cite{Lanata2022045111},
where the saddle point solution recovers the gGA. Another approach
related to SBMF is the slave spin approach \cite{De'medici2005205124},
which recasts the Hubbard model with density-density interactions
as a Hamiltonian of constrained fermions and auxiliary spins, and
the saddle point approximation also recovers the GA. As in the case
of slave bosons, the slave spin approach has produced many interesting
results on the multi-orbital Hubbard model \cite{Hassan2010035106,medici2017167003,Georgescu2015235117,Maurya2021425603,Maurya2022055602,Crispino2023155149,Gorni2023125166},
but has not offered a path beyond the mean-field solution. In the
present work, we demonstrate that the qubit parametrization for a
G-type $\mathcal{N}=2$ SPD is identical to the slave spin mean field
theory (see Appendix \ref{app:SSMF}). 

Finally, it is useful to discuss the off-shell effective energy theory
(OET) \cite{Cheng2020081105}, which can be reinterpreted through
the G-type and B-type SPD for $\mathcal{N}=2$. OET introduces an
approximation to evaluate the energy under the SPD, known as the central
point expansion (CPE), and this is applied to both the G-type and
B-type SPD for $\mathcal{N}=2$. Though the CPE is intrinsically different
than the SCDA, it also exactly evaluates the $\mathcal{N}=2$ SPD
in $d=\infty$, which is proven in the present work (see Appendix
\ref{app:CPE}). OET introduces corrections to the CPE for the G-type
and B-type SPD, guaranteeing the limiting behaviors for the weak and
strong coupling limits, respectively. Subsequently, one evaluates
the total energy in both cases and selects the solution with lower
energy. OET has yielded accurate results for the single band Hubbard
model in $d=1$, $d=2$, and $d=\infty$. The correction to the CPE
becomes far more challenging in the multiorbital Hubbard model, which
has not been pursued further. However, a similar idea of empirically
correcting the Gutzwiller approximation has been developed in the
context of the correlation matrix renormalization approximation \cite{Yao2014045131,Liu2021081113,Liu2021095902,Liu2022205124}.

The paper is organized in the following manner. Section \ref{subsec:Review-of-VDAT}
provides an overview of the SPD ansatz, the gauge symmetry of the
SPD, the SCDA, and the tensor product representation for evaluating
expectation values. Additionally, several methodological generalizations
are provided. Section \ref{sec:Overview-of-the} provides a high level
overview of the qubit energy form for the G-type and B-type SPD at
$\mathcal{N}=2$ and the G-type SPD at $\mathcal{N}=3$. Detailed
derivations of the results from Section \ref{sec:Overview-of-the}
are presented in Sections \ref{subsec:G-ansatz}, \ref{subsec:B-ansatz},
and \ref{subsec:GB-ansatz}. Finally, the qubit energy form is used
to examine the multi-orbital Hubbard model at half filling with $J/U\ge0$
in Section \ref{sec:Applications:-Multi-orbital-Hubb}. 

\section{\label{subsec:Review-of-VDAT}Review of the variational discrete
action theory }

Here we review several key ingredients of VDAT, including the sequential
product density matrix (SPD), the self-consistent canonical discrete
action theory (SCDA), and the tensor product representation for evaluating
local expectation values within the SCDA. In addition to reviewing
these concepts, we generalize the gauge symmetry of the SPD beyond
the case of a diagonal transformation, and generalize the tensor product
representation to arbitrary $\mathcal{N}$. These developments will
facilitate the subsequent analysis in this work. 

\subsection{The sequential product density matrix}

The SPD is the variational ansatz used within VDAT \cite{Cheng2021206402,Cheng2021195138},
and the SPD is a straightforward generalization of a variational wave
function ansatz first articulated in Ref. \cite{Dzierzawa19951993}.
In this work, we focus on the multiorbital Hubbard model, given as
\begin{equation}
\hat{H}=\hat{H}_{K}+\hat{H}_{loc}=\sum_{k\ell}\epsilon_{k\ell}\hat{n}_{k\ell}+\sum_{i,\ell<\ell'}U_{\ell\ell'}\hat{n}_{i\ell}\hat{n}_{i\ell'},\label{eq:multi-Hubbard}
\end{equation}
where $k$ and $i$ denote momentum and real-space site indices, respectively,
$\ell=1,\dots,2N_{orb}$ is the spin-orbital index within a local
site, and $U_{\ell\ell'}$ denotes the Coulomb interaction. In the
context of the multiorbital Hubbard model, the SPD with an integer
time step $\mathcal{N\geq}1$ is given as
\begin{equation}
\hat{\rho}=\hat{K}_{1}\hat{P}_{1}\dots\hat{K}_{\mathcal{N}}\hat{P}_{\mathcal{N}},\label{eq:spd}
\end{equation}
where $\hat{K}_{\tau}=\exp\left(\sum_{k\ell}\gamma_{k\ell\tau}\hat{n}_{k\ell}\right)$
is the kinetic projector, $\hat{P_{\tau}}=\exp(\sum_{i\Gamma}\upsilon_{\Gamma\tau}\hat{X}_{i\Gamma})$
is the interacting projector, $\tau=1,\dots,\mathcal{N}$ is the integer
time index, the diagonal Hubbard operator $\hat{X}_{i\Gamma}$ is
defined as 
\begin{align}
 & \hat{X}_{i\Gamma}=\prod_{\ell=1}^{2N_{orb}}\hat{X}_{i\Gamma;\ell},\label{eq:X decomposition}\\
 & \hat{X}_{i\Gamma;\ell}=\left(1-\hat{n}_{i\ell}\right)\delta_{0,\Gamma\left(\ell\right)}+\hat{n}_{i\ell}\delta_{1,\Gamma\left(\ell\right)},\label{eq:X decomposition component}
\end{align}
and $\Gamma\left(\ell\right)$ is $\ell$-th bit in the binary representation
of $\Gamma-1$, given as $\left(\Gamma\left(1\right)\dots\Gamma\left(2N_{orb}\right)\right)_{2}=\Gamma-1$,
and $\Gamma=1,\dots,2^{2N_{orb}}$. It should be noted that in general,
$\hat{P}_{\tau}$ is a direct product of local projectors from all
sites while $\hat{K}_{\tau}$ is a general non-interacting projector.
There are two schemes to ensure that the SPD is Hermitian and positive
semi-definite, denoted as G-type or B-type \cite{Cheng2021195138}.
In the following, we will examine both types for $\mathcal{N}\le3$.
For simplicity, we restrict $\gamma_{k\ell\tau}$ and $\upsilon_{\Gamma\tau}$
to be real numbers. It is also useful to reparametrize, up to a constant,
the $\hat{K}_{\tau}$ and $\hat{P_{\tau}}$ as
\begin{align}
 & \hat{K}_{\tau}=\prod_{k\ell}\big((1-\lambda_{k\ell\tau})(1-\hat{n}_{k\ell})+\lambda_{k\ell\tau}\hat{n}_{k\ell}\big),\\
 & \hat{P}_{\tau}=\prod_{i}\left(\sum_{\Gamma}u_{\Gamma\tau}\hat{X}_{i\Gamma}\right),
\end{align}
where the $\lambda_{k\ell\tau}$ and $u_{\Gamma\tau}$ are a reparametrization
of $\gamma_{k\ell\tau}$ and $\upsilon_{\Gamma\tau}$, and this form
is used to derive the qubit parametrization (see Section \ref{sec:Overview-of-the}). 

We first examine the $\mathcal{N}=1$ SPD. The G-type SPD for $\mathcal{N}=1$
is defined by $\hat{\rho}_{G1}=\hat{K}_{1}$, which is a non-interacting
many-body density matrix. If $\hat{K}_{1}=|\Psi_{0}\rangle\langle\Psi_{0}|$
where $|\Psi_{0}\rangle$ is a single Slater determinant, then $\hat{\rho}_{G1}$
corresponds with the Hartree-Fock wave function. Conversely, the B-type
SPD for $\mathcal{N}=1$ is defined as $\hat{\rho}_{B1}=\hat{P}_{1}$.
Normally $\hat{\rho}_{B1}$ corresponds to a mixed state, but if $\hat{P}_{1}=|\Psi_{at}\rangle\langle\Psi_{at}|$
with $|\Psi_{at}\rangle$ defined as a direct product of atomic states
from all sites, then $\hat{\rho}_{B1}$ corresponds to a pure state.

The G-type SPD with $\mathcal{N}=2$ is defined as
\begin{equation}
\hat{\rho}_{G2}=\hat{P}_{1}\hat{K}_{2}\hat{P}_{1}^{\dagger}.\label{eq:g2}
\end{equation}
If the center projector is taken as $\hat{K}_{2}=|\Psi_{0}\rangle\langle\Psi_{0}|$
with $|\Psi_{0}\rangle$ being a single Slater determinant, then $\hat{\rho}_{G2}=|\Psi_{G}\rangle\langle\Psi_{G}|$,
where $|\Psi_{G}\rangle=\hat{P}_{1}|\Psi_{0}\rangle$ is the Gutzwiller
wave function (GWF). The B-type SPD for $\mathcal{N}=2$ is defined
as 
\begin{equation}
\hat{\rho}_{B2}=\hat{K}_{1}\hat{P}_{1}\hat{K}_{1}^{\dagger}.\label{eq:b2}
\end{equation}
If the center projector is taken as $\hat{P}_{1}=|\Psi_{at}\rangle\langle\Psi_{at}|$
with $|\Psi_{at}\rangle$ defined as a direct product of atomic states
from all sites, then $\hat{\rho}_{B2}=|\Psi_{B}\rangle\langle\Psi_{B}|$
, where $|\Psi_{B}\rangle=\exp\left(\sum_{k\ell}\gamma_{k\ell}\hat{n}_{k\ell}\right)|\Psi_{at}\rangle$
and we have previously referred to $|\Psi_{B}\rangle$ as the Bearyswil
wave-function \cite{Cheng2021206402,Cheng2021195138}. However, it
should be noted that $|\Psi_{B}\rangle$ is distinct from the original
Bearyswil wave-function $|\Psi'_{B}\rangle$ \cite{Baeriswyl19870},
which is defined as $|\Psi'_{B}\rangle=\exp\left(\alpha\sum_{k\ell}\epsilon_{k\ell}\hat{n}_{k\ell}\right)|\Psi_{\infty}\rangle$,
where $\alpha$ is a variational parameter and $|\Psi_{\infty}\rangle$
is the fully projected GWF. Though $|\Psi'_{B}\rangle$ is technically
a special case of a G-type $\mathcal{N}=3$ SPD, which can also be
evaluated using the SCDA, it yields the same energy as $|\Psi_{B}\rangle$
for the insulating phase in $d=\infty$ for the multi-orbital Hubbard
model (see Section \ref{subsec:Recovering-the-N=00003D2-bytpe} and
Figure \ref{fig:compare_bethe_twopeak}).

The G-type SPD for $\mathcal{N}=3$ combines the variational power
of $\hat{\rho}_{G2}$ and $\hat{\rho}_{B2}$, and is defined as 
\begin{equation}
\hat{\rho}_{G3}=\hat{K}_{1}\hat{P}_{1}\hat{K}_{2}\hat{P}_{1}^{\dagger}\hat{K}_{1}^{\dagger}.\label{eq:g3}
\end{equation}
If $\hat{K}_{2}$ corresponds to a single Slater determinant $|\Psi_{0}\rangle$,
then $\hat{\rho}_{G3}=|\Psi_{GB}\rangle\langle\Psi_{GB}|$, where
$|\Psi_{GB}\rangle=\exp\left(\sum_{k\ell}\gamma_{k\ell}\hat{n}_{k\ell}\right)\hat{P}_{1}|\Psi_{0}\rangle$
is a mild generalization of the original Gutzwiller-Baeriswyl wave
function $|\Psi'_{GB}\rangle$ introduced by Otsuka (see Eq. (2.3)
in Ref. \cite{Otsuka19921645}), defined as $|\Psi'_{GB}\rangle=\exp\left(\alpha\sum_{k\ell}\epsilon_{k\ell}\hat{n}_{k\ell}\right)\hat{P_{1}}|\Psi_{0}\rangle$,
where $\alpha$ is a variational parameter. Finally, the $\mathcal{N}=3$
B-type SPD is defined as $\hat{\rho}_{B3}=\hat{P}_{1}\hat{K}_{2}\hat{P}_{2}\hat{K}_{2}^{\dagger}\hat{P}_{1}^{\dagger}$.
If $\hat{P}_{2}=|\Psi_{at}\rangle\langle\Psi_{at}|$, then $\hat{\rho}_{B3}=|\Psi_{BG}\rangle\langle\Psi_{BG}|$
where $|\Psi_{BG}\rangle=\hat{P}_{1}\exp\left(\sum_{k\ell}\gamma_{k\ell}\hat{n}_{k\ell}\right)|\Psi_{at}\rangle$
and we have previously referred to $|\Psi_{BG}\rangle$ as the Baeriswyl-Gutzwiller
wave-function. However, it should be noted that $|\Psi_{BG}\rangle$
is distinct from the original Baeriswyl-Gutzwiller wave-function $|\Psi_{BG}'\rangle$
\cite{Dzierzawa19951993}, which is defined as $|\Psi'_{BG}\rangle=\hat{P}_{1}\exp\left(\alpha\sum_{k\ell}\epsilon_{k\ell}\hat{n}_{k\ell}\right)|\Psi_{\infty}\rangle$,
where $\alpha$ is a variational parameter and $|\Psi_{\infty}\rangle$
is the fully projected GWF. The $|\Psi'_{BG}\rangle$ is technically
a special case of a G-type $\mathcal{N}=4$ SPD, which can also be
evaluated using the SCDA. 

It is useful to understand how the SPD recovers the exact solution
with increasing $\mathcal{N}$ \cite{Cheng2021195138}. The Trotter-Suzuki
decomposition \cite{Suzuki1976183}, given as 
\begin{equation}
e^{-\beta H}\approx\left(e^{-\frac{\beta}{\mathcal{N}}\left(\sum_{k\ell}\epsilon_{k\ell}\hat{n}_{k\ell}\right)}e^{-\frac{\beta}{\mathcal{N}}\left(\sum_{i,\ell<\ell'}U_{\ell\ell'}\hat{n}_{i\ell}\hat{n}_{i\ell'}\right)}\right)^{\mathcal{N}},\label{eq:ts}
\end{equation}
can be considered as a special case of the SPD where the variational
parameters are set by the parameters of the Hamiltonian and the temperature.
Given that the large $\mathcal{N}$ Trotter-Suzuki decomposition recovers
the exact finite temperature density matrix, the SPD in the large
$\mathcal{N}$ limit possesses sufficient variational power to solve
the Hamiltonian exactly. However, for a finite $\mathcal{N}$, the
SPD leverages the full variational freedom while maintaining the integer
time structure, leading to an observed exponential decrease of the
energy error with increasing $\mathcal{N}$ \cite{Cheng2021206402,Cheng2022205129}.
In contrast, the energy error for the Trotter-Suzuki decomposition
diminishes polynomially with increasing $\mathcal{N}$ \cite{Suzuki1976183}.
This key difference is the basis for the accuracy of the SPD for small
$\mathcal{N}$. 

Our initial efforts using VDAT have been exactly evaluating the SPD
in $d=\infty$ via the SCDA, and this same approach can be used as
a local approximation in finite dimensions. However, there is an existing
body of literature which used variational quantum Monte-Carlo (VMC)
to evaluate a simplified SPD in the Hubbard model. In the case of
$\hat{\rho}_{G2}$, it is most convenient to execute VMC in real space,
which has been used to evaluate the Gutzwiller wave function in the
Hubbard model \cite{Yokoyama19871490,Yokoyama19873582,Yokoyama19882482}.
For $\mathcal{N}>2$ in the one and two dimensional Hubbard model,
a restricted form of the SPD has been evaluated using quantum Monte-Carlo
to solve the single-orbital Hubbard model in two dimensions \cite{Otsuka19921645,Yamaji1998225,Yanagisawa19983867,Yanagisawa19993608,Koike200365,Yanagisawa2016114707,Yanagisawa2019054702,Yanagisawa20202040046,Yanagisawa202112,Yanagisawa2021127382,Sorella2023115133,Levy2024013237},
the $p$-$d$ model \cite{Yanagisawa202127004,Yanagisawa2001184509},
and selected molecules \cite{Chen20234484}. 

\subsection{Gauge Symmetry of the SPD\label{subsec:Gauge-Symmetry-of-SPD}}

A important aspect of the SPD is the corresponding gauge symmetry
\cite{Cheng2022205129,Cheng2023035127}, which will prove to be an
important tool for reducing the redundancy of the parametrization
of the SPD. Consider the general SPD $\hat{\rho}=\hat{K}_{1}\hat{P}_{1}\dots\hat{K}_{\mathcal{N}}\hat{P}_{\mathcal{N}}$,
where $\hat{K}_{\tau}=\exp\left(\bm{\gamma}_{\tau}\cdot\hat{\bm{n}}\right)$
is a kinetic projector and $\boldsymbol{\gamma}\cdot\hat{\boldsymbol{n}}\equiv\sum_{\ell\ell'}[\boldsymbol{\gamma}]_{\ell\ell'}[\hat{\boldsymbol{n}}]_{\ell\ell'}$
and $[\hat{\boldsymbol{n}}]_{\ell\ell'}=\hat{a}_{\ell}^{\dagger}\hat{a}_{\ell'}$,
and $\hat{P}_{\tau}$ is a general interacting projector. The essence
of the gauge symmetry lies in the fact that only the total product
$\hat{\rho}$ determines physical observables, while there is considerable
flexibility in the decomposition of $\hat{\rho}$ into different time
steps and the separation between kinetic and interacting projectors.
In order to analyze the gauge symmetry, we utilize the discrete action
theory \cite{Cheng2021195138}, which is a formalism for studying
integer time correlation functions of the SPD. The discrete action
theory can be seen as a generalization of the many-body Green's function
formalism to the case of integer time, including an integer time Dyson
equation given as
\begin{equation}
\boldsymbol{g}^{-1}-\boldsymbol{1}=\left(\bm{g}_{0}^{-1}-\boldsymbol{1}\right)\boldsymbol{S}
\end{equation}
where $\text{\ensuremath{\bm{g}}}_{0}$ is the non-interacting integer
time Green's functions, $\text{\ensuremath{\bm{g}}}$ is the interacting
integer time Green's functions, and $\boldsymbol{S}=\exp(-\boldsymbol{\Sigma}^{T})$
is the exponential form of the integer time self-energy, where $\boldsymbol{\Sigma}$
is the integer time self-energy. The integer time Dyson equation recovers
the usual Dyson when the SPD is chosen as the Trotter-Suzuki decomposition
in the large $\mathcal{N}$ limit \cite{Cheng2021195138}. The $\text{\ensuremath{\bm{g}}}_{0}$
and $\text{\ensuremath{\bm{g}}}$ are naturally defined in the compound
space \cite{Cheng2021195138} as $\text{\ensuremath{\bm{g}}}_{0}=\left\langle \barhat[\bm{n}]\right\rangle _{\spdcs_{0}}$
and $\text{\ensuremath{\bm{g}}}=\left\langle \barhat[\bm{n}]\right\rangle _{\spdcs}$,
where $[\barhat[\bm{n}]]_{\ell\tau,\ell'\tau'}=\barhat[a]_{\ell}^{\dagger(\tau)}\barhat[a]_{\ell'}^{(\tau')}$,
the underbar indicates an operator in the compound space, $\ell=1,\dots,L$
enumerates the spin-orbitals for the entire system, and $\tau=1,\dots,\mathcal{N}$
labels the integer time step. The discrete Dyson equation is a matrix
equation of dimension of $L\mathcal{N}\times L\mathcal{N}$, and it
exactly relates the interacting and noninteracting integer time Green's
function via $\boldsymbol{S}$. Given that $\boldsymbol{S}$ is more
convenient to work with, we refer to $\boldsymbol{S}$ as the integer
time self-energy for brevity. For the convenience of discussing the
gauge transformation of the SPD, it is useful to define several additional
quantities
\begin{align}
 & \bm{S}_{0}=\bm{S}_{Q}\bm{S}_{K},\\
 & \bm{S}_{F}=\bm{S}_{Q}\bm{S}_{K}\bm{S},
\end{align}
where $[\boldsymbol{S}_{Q}]_{\tau\tau'}=(-\delta_{\tau+1,\tau'}+\delta_{\tau-\discn+1,\tau'})\boldsymbol{1}$
and $\left[\bm{S}_{K}\right]_{\tau\tau'}=\delta_{\tau\tau'}\exp\left(-\bm{\gamma}_{\tau}^{T}\right)$
are $L\times L$ matrices. Using these definitions, $\bm{g}_{0}$
and $\bm{g}$ can be defined succinctly as
\begin{align}
 & \bm{g}_{0}=\left(\boldsymbol{1}+\bm{S}_{0}\right)^{-1},\\
 & \bm{g}=\left(\boldsymbol{1}+\bm{S}_{F}\right)^{-1}.
\end{align}

Generally, there are two types of gauge transformations of the SPD:
intra-time-step and inter-time-step. The intra-time-step transformation
occurs at the boundary of the kinetic and interacting projector for
time step $\tau$ in the following way: $\hat{K}_{\tau}\rightarrow\hat{K}_{\tau}\exp\left(\bm{\mu}_{\tau}\cdot\hat{\bm{n}}\right)$
and $\hat{P}_{\tau}\rightarrow\exp\left(-\bm{\mu}_{\tau}\cdot\hat{\bm{n}}\right)\hat{P}_{\tau}$.
Since the intra-time-step transformation does not alter the measurement
at integer times, $\bm{g}$ is invariant to this transformation. However,
given that $\hat{K}_{\tau}$ changes after the intra-time-step transformation,
we have $\bm{S}'_{K}=\bm{S}_{K}\bm{N}_{a}$, where $\left[\bm{N}_{a}\right]_{\tau\tau'}=\delta_{\tau\tau'}\exp\left(-\bm{\mu}_{\tau}^{T}\right)$,
while $\bm{S}'_{F}=\bm{S}_{F}$. Conversely, the inter-time-step transformation
is defined at the boundary of the interacting projector at $\tau$
and the kinectic projector at $\tau+1$ in the following way: $\hat{P}_{\tau}\rightarrow\hat{P}_{\tau}\exp\left(-\bm{\mu}_{\tau}\cdot\hat{\bm{n}}\right)$
and $\hat{K}_{\tau+1}\rightarrow\exp\left(\bm{\mu}_{\tau}\cdot\hat{\bm{n}}\right)\hat{K}_{\tau+1}$,
where $\bm{\mu}_{\mathcal{N}}\equiv0$. We first discuss how $\bm{g}$
changes under this inter-time-step transformation, and it is useful
to note the following identities (for details, see Eqns. (A22) and
(A23) in Ref. \cite{Cheng2021195138} and Eqns. (A1)-(A4) in Ref.
\cite{Cheng2022205129})
\begin{align}
 & e^{-\bm{\mu}_{\tau}\cdot\hat{\bm{n}}}\hat{a}_{\ell}^{\dagger}e^{\bm{\mu}_{\tau}\cdot\hat{\bm{n}}}=\sum_{\ell'}\left[e^{-\bm{\mu}_{\tau}^{T}}\right]_{\ell\ell'}\hat{a}_{\ell'}^{\dagger},\label{eq:adagger}\\
 & e^{-\bm{\mu}_{\tau}\cdot\hat{\bm{n}}}\hat{a}_{\ell}e^{\bm{\mu}_{\tau}\cdot\hat{\bm{n}}}=\sum_{\ell'}\hat{a}_{\ell'}\left[e^{\bm{\mu}_{\tau}^{T}}\right]_{\ell'\ell}.\label{eq:a}
\end{align}
Using the above two equations, we can determine the transformation
$\bm{g}'=\bm{N}_{b}\bm{g}\bm{N}_{b}^{-1}$, where $\left[\bm{N}_{b}\right]_{\tau\tau'}=\delta_{\tau\tau'}\exp\left(-\bm{\mu}_{\tau}^{T}\right)$,
and correspondingly $\bm{S}'_{F}=\bm{N}_{b}\bm{S}_{F}\bm{N}_{b}^{-1}$.
Using the transformation for the kinetic projector, we have $\bm{S}'_{K}=\tilde{\bm{N}}_{b}\bm{S}_{K}$,
where $\left[\tilde{\bm{N}}_{b}\right]_{\tau\tau'}=\delta_{\tau\tau'}\exp\left(-\bm{\mu}_{\tau-1}^{T}\right)$
and $\bm{\bm{\mu}}_{0}\equiv0$, and $\tilde{\bm{N}}_{b}$ can be
rewritten as $\tilde{\bm{N}}_{b}=\bm{S}_{Q}^{-1}\bm{N}_{b}\bm{S}_{Q}$.
It should be noted that a general gauge transformation includes both
intra-time-step and inter-time-step transformations, which can be
parametrized by $\bm{N}_{a}$ and $\bm{N}_{b}$, and the above results
can be synthesized as 
\begin{align}
 & \bm{g}'=\bm{N}_{b}\bm{g}\bm{N}_{b}^{-1},\label{eq:change g}\\
 & \bm{S}'_{F}=\left(\bm{g}'\right)^{-1}-1=\bm{N}_{b}\bm{S}_{F}\bm{N}_{b}^{-1},\label{eq:change Sf}\\
 & \bm{S}'_{K}=\tilde{\bm{N}}_{b}\bm{S}_{K}\bm{N}_{a}=\bm{S}_{Q}^{-1}\bm{N}_{b}\bm{S}_{Q}\bm{S}_{K}\bm{N}_{a}.\label{eq:change Sk}
\end{align}
Moreover, the form of the integer time Dyson equation is invariant
after the transformation, which requires 
\begin{align}
 & \bm{S}_{0}\bm{S}=\bm{S}_{Q}\bm{S}_{K}\bm{S}=\bm{S}_{F},\label{eq:dyson1}\\
 & \bm{S'}_{0}\bm{S}'=\bm{S}_{Q}\bm{S}'_{K}\bm{S}'=\bm{S}'_{F}.\label{eq:dyson2}
\end{align}
Using Eq. (\ref{eq:change Sk}), we have

\begin{align}
 & \bm{S'}_{0}=\bm{S}_{Q}\bm{S}'_{K}=\bm{N}_{b}\bm{S}_{Q}\bm{S}_{K}\bm{N}_{a}=\bm{N}_{b}\bm{S}_{0}\bm{N}_{a},\label{eq:change S0}
\end{align}
while using Eqns. (\ref{eq:change S0}), (\ref{eq:change Sf}), (\ref{eq:dyson1}),
and (\ref{eq:dyson2}), we have 
\begin{align}
\bm{S}' & =\left(\bm{S'}_{0}\right)^{-1}\bm{S}'_{F}=\left(\bm{N}_{b}\bm{S}_{0}\bm{N}_{a}\right)^{-1}\bm{N}_{b}\bm{S}_{F}\bm{N}_{b}^{-1}\nonumber \\
 & =\bm{N}_{a}^{-1}\bm{S}\bm{N}_{b}^{-1},\label{eq:change S}
\end{align}
which can also be directly verified by assuming $\hat{P}_{\tau}$
is non-interacting. Correspondingly, the non-interacting integer time
Green's function changes as 
\begin{equation}
\bm{g}'_{0}=\left(\boldsymbol{1}+\bm{S}'_{0}\right)^{-1}=\left(\boldsymbol{1}+\bm{N}_{b}\left(\bm{g}_{0}^{-1}-\boldsymbol{1}\right)\bm{N}_{a}\right)^{-1}.\label{eq:change g0}
\end{equation}
Notice that Eqns. (\ref{eq:change g})-(\ref{eq:change g0}) are exact
relations. Within the SCDA \cite{Cheng2021195138}, $\bm{\mathcal{G}}$
will transform similarly to $\bm{g}_{0}$, where $\bm{N}_{a}$ and
$\bm{N}_{b}$ are local. These gauge transformations provide a rigorous
theoretical foundation for simplifying the task of achieving self-consistency
within the SCDA. Moreover, choosing the appropriate gauge condition
will also help to improve the numerical stability when minimizing
over the variational parameters. 

\subsection{Review of the SCDA \label{subsec:Review-of-SCDA}}

The discrete action theory is a formalism designed to study integer
time correlation functions of the SPD, and the self-consistent canonical
discrete action theory (SCDA) is the integer time analogue of the
dynamical mean-field theory \cite{Cheng2021206402,Cheng2021195138}.
The SCDA exactly evaluates the SPD in infinite dimensions, and can
be used as a local approximation for the SPD in finite dimensions.
To appreciate the idea of the SCDA, we define the discrete action
which encodes the integer time correlation of the SPD as 
\begin{equation}
\spdcs=\spdcs_{0}\barhat[P],\label{eq:discrete action perturb}
\end{equation}
where $\spdcs_{0}=\barhat[Q]\prod_{\tau=1}^{\mathcal{N}}\barhat[K]_{\tau}^{\left(\tau\right)}$
and $\barhat[P]=\prod_{\tau=1}^{\mathcal{N}}\barhat[P]_{\tau}^{\left(\tau\right)}$
are the non-interacting and interacting parts of the discrete action
$\spdcs$, and the underbar denotes operators in the compound space.

The SCDA can be formulated as a scheme to evaluate the kinetic energy
and local interacting energy through two self-consistent approximations
of $\spdcs$ \cite{Cheng2022205129}, denoted as $\barhat[\rho]_{K}$
and $\barhat[\rho]_{loc}$, respectively. Within the SCDA, the total
energy is given as 
\begin{equation}
\langle\barhat[H]^{(\mathcal{N})}\rangle_{\spdcs}=\langle\barhat[H]_{K}^{\left(\mathcal{N}\right)}\rangle_{\barhat[\rho]_{K}}+\langle\barhat[H]_{loc}^{\left(\mathcal{N}\right)}\rangle_{\barhat[\rho]_{loc}},
\end{equation}
where $\barhat[\rho]_{K}$ and $\barhat[\rho]_{loc}$ are defined
as 

\begin{align}
\barhat[\rho]_{K} & =\spdcs_{0}\exp(-\ln\boldsymbol{S}^{T}\cdot\barhat[\bm{n}]),\label{eq:rhok}\\
\barhat[\rho]_{loc} & =\exp\left(-\ln\left(\bm{\mathcal{G}}^{-1}-1\right)^{T}\cdot\barhat[\bm{n}]\right)\barhat[P],\label{eq:rholoc-1}
\end{align}
where $\bm{\mathcal{G}}$ is the non-interacting integer time Green's
function for the local canonical discrete action $\barhat[\rho]_{loc}$.
Both $\bm{S}$ and $\bm{\mathcal{G}}$ are $L\mathcal{N}\times L\mathcal{N}$
matrices, where $L$ is the number of spin-orbitals of the entire
system. In the SCDA, the key assumption is that $\bm{S}$ and $\bm{\mathcal{G}}$
are block diagonal in real space clusters, given by $\left[\bm{S}\right]_{ij}=\bm{S}_{i}\delta_{ij}$
and $\left[\bm{\mathcal{G}}\right]_{ij}=\bm{\mathcal{G}}_{i}\delta_{ij}$
where $i$ and $j$ are real space cluster indices. To determine $\bm{S}_{i}$
and $\bm{\mathcal{G}}_{i}$, we have the following two sets of conditions
for each cluster index $i$, 
\begin{enumerate}
\item The integer time Dyson equation, given as
\begin{equation}
\left(\boldsymbol{g}_{i}^{-1}-\boldsymbol{1}\right)=\left(\bm{\mathcal{G}}_{i}^{-1}-\boldsymbol{1}\right)\boldsymbol{S}_{i},\label{eq: integer time dyson}
\end{equation}
provides a way to compute $\bm{S}_{i}$ by evaluating $\barhat[\rho]_{loc;i}=\text{Tr}_{/i}\barhat[\rho]_{loc}$,
which can be expressed as 
\begin{equation}
\barhat[\rho]_{loc;i}=\exp\left(-\ln\left(\bm{\mathcal{G}}_{i}^{-1}-1\right)^{T}\cdot\barhat[\bm{n}]_{i}\right)\barhat[P]_{i}.\label{eq:rholoci}
\end{equation}
\item The self-consistency condition for the local integer time Green's
function, given as
\begin{equation}
\bm{g}_{i}=\bm{g}'_{i},\label{eq:self-consistency g g'}
\end{equation}
where $\boldsymbol{g}_{i}=\left\langle \barhat[\boldsymbol{n}]_{i}\right\rangle _{\barhat[\rho]_{loc}}$
and $\boldsymbol{g}'_{i}=\left\langle \barhat[\boldsymbol{n}]_{i}\right\rangle _{\barhat[\rho]_{K}}$. 
\end{enumerate}
It is interesting to examine the sufficiency of Eq. (\ref{eq: integer time dyson})
and (\ref{eq:self-consistency g g'}) to determine both $\bm{\mathcal{G}}$
and $\bm{S}$ by counting the number of unknown entries and equations.
We denote the number of sites in the lattice as $N_{site}$ and the
number of spin orbitals on site $i$ as $N_{i}$. The number of unknowns
in $\bm{\mathcal{G}}$ and $\bm{S}$ is $2\sum_{i=1}^{N_{site}}\left(N_{i}\mathcal{N}\right)^{2}$.
For a given $i$, the two matrix equations given by Eq. (\ref{eq: integer time dyson})
and (\ref{eq:self-consistency g g'}) provide $2\left(N_{i}\mathcal{N}\right)^{2}$
entries. Therefore, we have a sufficient number of equations to determine
$\bm{\mathcal{G}}$ and $\bm{S}$. 

\subsection{The tensor product representation for expectation values under $\barhat[\rho]_{loc;i}$ }

The key computational cost of implementing the SCDA is to compute
local observables under $\barhat[\rho]_{loc;i}$, and a convenient
approach was devised for $\mathcal{N}=3$ using a tensor product representation
\cite{Cheng2023035127}. Here we generalize the tensor product representation
to arbitrary $\mathcal{N}$. We assume that the interacting projector
has the form $\hat{P}_{\tau;i}=\sum_{\Gamma}u_{\tau i\Gamma}\hat{X}_{i\Gamma}$
and $\boldsymbol{\mathcal{G}}_{i}$ is diagonal in the spin orbital
index $\ell$, given as $\left[\bm{\mathcal{G}}_{i}\right]_{\ell\ell}=\delta_{\ell\ell'}\mathcal{G}_{i,\ell}$,
where $\mathcal{G}_{i,\ell}$ is an $\mathcal{N\times\mathcal{N}}$
matrix. For a local operator $\barhat[O]$ at site $i$, the expectation
value can be written as 
\begin{equation}
\langle\barhat[O]\rangle_{\barhat[\rho]_{loc;i}}=\frac{\langle\barhat[P]_{i}\barhat[O]\rangle_{\barhat[\rho]_{loc;i,0}}}{\langle\barhat[P]_{i}\rangle_{\barhat[\rho]_{loc;i,0}}}=\frac{\left(\barhat[O]\right)_{u}\cdot\left(\otimes_{\tau}u_{\tau}\right)}{\left(\barhat[1]\right)_{u}\cdot\left(\otimes_{\tau}u_{\tau}\right)},\label{eq:local_expect_values}
\end{equation}
where the non-interacting discrete action for $\barhat[\rho]_{loc;i,0}$
is given by 
\begin{equation}
\barhat[\rho]_{loc;i,0}=\exp(-\ln\left(\bm{\mathcal{G}}_{i}^{-1}-\boldsymbol{1}\right)^{T}\cdot\barhat[\bm{n}]_{i}),
\end{equation}
where the vector $u_{\tau}=\left(u_{\tau i1},\dots,u_{\tau iN_{\Gamma}}\right)$
enumerates over the index $\Gamma$ with $N_{\Gamma}$ as the number
of local diagonal Hubbard operators at site $i$, the $\mathcal{N}$
dimensional tensor $\left(\barhat[O]\right)_{u}$ is defined as 
\begin{equation}
\left[\left(\barhat[O]\right)_{u}\right]_{\Gamma_{1},\dots,\Gamma_{\mathcal{N}}}=\left\langle \left(\prod_{\tau=1}^{\mathcal{N}}\barhat[X]_{i\Gamma_{\tau}}^{\left(\tau\right)}\right)\barhat[O]\right\rangle _{\barhat[\rho]_{loc;i,0}},\label{eq:ou}
\end{equation}
the direct product is defined as $\left[\otimes_{\tau=1}^{\mathcal{N}}u_{\tau}\right]_{\Gamma_{1},\dots,\Gamma_{\mathcal{N}}}=\prod_{\tau=1}^{\mathcal{N}}u_{\tau i\Gamma_{\tau}}$,
and $\bm{A}\cdot\bm{B}$ is the contraction between two $\mathcal{N}$-dimensional
tensors as 
\begin{equation}
\bm{A}\cdot\bm{B}=\sum_{\Gamma_{1},\dots,\Gamma_{\mathcal{N}}}\left[\bm{A}\right]_{\Gamma_{1},\dots,\Gamma_{\mathcal{N}}}\left[\bm{B}\right]_{\Gamma_{1},\dots,\Gamma_{\mathcal{N}}}.\label{eq:contraction A B}
\end{equation}
We now consider operators that can be written as a product over the
spin orbitals, given as $\barhat[O]=\prod_{\ell}\barhat[O]_{\ell}$.
Assuming that $\mathcal{\bm{\mathcal{G}}}_{i}$ is diagonal in the
spin orbital index $\ell$, we have 

\begin{equation}
\left[\left(\barhat[O]\right)_{u}\right]_{\Gamma_{1},\dots,\Gamma_{\mathcal{N}}}=\prod_{\ell=1}^{2N_{orb}}\left\langle \left(\prod_{\tau=1}^{\mathcal{N}}\barhat[X]_{i\Gamma_{\tau};\ell}^{\left(\tau\right)}\right)\barhat[O]_{\ell}\right\rangle _{\barhat[\rho]_{loc;i}},
\end{equation}
where $\hat{X}{}_{i\Gamma;\ell}$ is defined in Eq. (\ref{eq:X decomposition component}).
Therefore, we can write $\left(\barhat[O]\right)_{u}$ as a direct
product 
\begin{equation}
\left(\barhat[O]\right)_{u}=\otimes_{\ell=1}^{2N_{orb}}\left(\barhat[O]_{\ell}\right)_{u;\ell},\label{eq:direct_product_structure}
\end{equation}
where the direct product is defined as 
\begin{equation}
\left[\otimes_{\ell=1}^{2N_{orb}}\boldsymbol{M}_{\ell}\right]_{\Gamma_{1},\dots,\Gamma_{\mathcal{N}}}=\prod_{\ell=1}^{2N_{orb}}\left[\boldsymbol{M}_{\ell}\right]_{\Gamma_{1}(\ell),\dots,\Gamma_{\mathcal{N}}(\ell)},
\end{equation}
with $\left(\barhat[O]_{\ell}\right)_{u;\ell}$ defined as 
\begin{equation}
\left[\left(\barhat[O]_{\ell}\right)_{u;\ell}\right]_{\Gamma_{1}(\ell),\dots,\Gamma_{\mathcal{N}}(\ell)}\equiv\left\langle \left(\prod_{\tau=1}^{\mathcal{N}}\barhat[X]_{i\Gamma_{\tau};\ell}^{\left(\tau\right)}\right)\barhat[O]_{\ell}\right\rangle _{\barhat[\rho]_{loc;i,0}},
\end{equation}
which only depends $\left[\bm{\mathcal{G}}_{i}\right]_{\ell\ell}$.
For some cases, there is no interacting projector on a given time
step, and the corresponding dimension of $\left(\barhat[O]_{\ell}\right)_{u;\ell}$
can be traced out. Moreover, Eq. (\ref{eq:local_expect_values}) can
be conveniently rewritten into a matrix product for all cases we study,
which will be further discussed in Sections \ref{subsec:G-ansatz},
\ref{subsec:B-ansatz}, and \ref{subsec:GB-ansatz}.

\section{Overview of the qubit energy form\label{sec:Overview-of-the}}

\global\long\def\rhoeff{\bm{\rho}}%

The SCDA can exactly evaluate an SPD in $d=\infty$ with arbitrary
$\mathcal{N}$ via a self-consistent solution, as outlined in Section
\ref{subsec:Review-of-VDAT}, and therefore the total energy can only
be numerically evaluated for a given set of variational parameters
in general. However, the recently developed gauge constrained SCDA
algorithm \cite{Cheng2023035127} allows for an explicit evaluation
of a G-type SPD with $\mathcal{N}\le3$, circumventing the need for
self-consistent solution (see Section \ref{subsec:Review-of-the-gauge-const-SCDA}
for a review). In the present work, we offer a further refinement
of the gauge constrained SCDA by transforming all variational parameters
into physically intuitive variables, referred to as the qubit parametrization.
Moreover, the qubit parametrization analytically resolves one constraint,
reducing the number of variational parameters by one per spin orbital.
Given the complexity of deriving the qubit parametrization, we collect
all key results in this section, providing a self-contained presentation
of all details needed to implement a practical calculation. A G-type
and B-type SPD at $\mathcal{N}$ will be denoted as $\hat{\rho}_{G\mathcal{N}}$
and $\hat{\rho}_{B\mathcal{N}}$, respectively. For pedagogical purposes,
we first present results for $\hat{\rho}_{G2}$ and $\hat{\rho}_{B2}$
before finally considering $\hat{\rho}_{G3}$, illustrating that $\hat{\rho}_{G3}$
can be seen as a unification of $\hat{\rho}_{G2}$ and $\hat{\rho}_{B2}$.
Detailed derivations of all results for $\hat{\rho}_{G2}$, $\hat{\rho}_{B2}$,
and $\hat{\rho}_{G3}$ are provided in Sections \ref{subsec:G-ansatz},
\ref{subsec:B-ansatz}, and \ref{subsec:GB-ansatz}, respectively,
while explicit applications are presented in Section \ref{sec:Applications:-Multi-orbital-Hubb}.
It should be emphasized that all qubit parametrization results are
mathematically identical to corresponding VDAT results.

For the sake of generality, we consider the multiorbital Hubbard Hamiltonian
given as
\begin{equation}
\hat{H}=\sum_{k\ell}\epsilon_{k\ell}\hat{n}_{k\ell}+\sum_{i}H_{loc}\left(\left\{ \hat{n}_{i\ell}\right\} \right),
\end{equation}
where $H_{loc}$ is a polynomial function of the local density operators,
given as
\begin{align}
H_{loc}\left(\left\{ \hat{n}_{i\ell}\right\} \right)= & \sum_{\ell}c_{\ell}\hat{n}_{i\ell}+\sum_{\ell_{1}\ell_{2}}c_{\ell_{1}\ell_{2}}\hat{n}_{i\ell_{1}}\hat{n}_{i\ell_{2}}\nonumber \\
 & +\sum_{\ell_{1}\ell_{2}\ell_{3}}c_{\ell_{1}\ell_{2}\ell_{3}}\hat{n}_{i\ell_{1}}\hat{n}_{i\ell_{2}}\hat{n}_{i\ell_{3}}+\dots,\label{eq:Hloc polynomial}
\end{align}
where $c_{\ell_{1}\dots\ell_{n}}$ are parameters that define the
n-body interactions. For a typical Hubbard model, only two-body interactions
will be needed. 

We begin by defining the local Hilbert space and the corresponding
fermionic operators, and the corresponding map to qubit operators
via the Jordan-Wigner transformation. For a given site $i$, the atomic
configurations $\{|\Gamma\rangle\}$ form a complete basis for the
local Hilbert space with dimension $2^{2N_{orb}}$, and we use $\hat{a}_{\ell}^{\dagger}$
and $\hat{a}_{\ell}$ to represent the creation and annihilation operators
for spin-orbital $\ell$. Then we express the fermionic operators
in terms of spin operators via the Jordan-Wigner transformation defined
as 

\begin{align}
 & \hat{a}_{\ell}^{\dagger}=\left(\prod_{\ell'<\ell}\hat{\sigma}_{\ell'}^{z}\right)\hat{\sigma}_{\ell}^{-},\label{eq:Jordan Wigner1}\\
 & \hat{a}_{\ell}=\left(\prod_{\ell'<\ell}\hat{\sigma}_{\ell'}^{z}\right)\hat{\sigma}_{\ell}^{+},\label{eq:Jordan Wigner2}\\
 & \hat{a}_{\ell}^{\dagger}\hat{a}_{\ell}=\hat{n}_{\ell}=\frac{1}{2}\left(1-\hat{\sigma}_{\ell}^{z}\right),\label{eq:nell_sigmaz}
\end{align}
where $\hat{\sigma}_{\ell}^{\pm}=\left(\hat{\sigma}_{\ell}^{x}\pm i\hat{\sigma}_{\ell}^{y}\right)/2$
and $\hat{\sigma}_{\ell}^{\mu}$ is defined as 
\begin{equation}
\hat{\sigma}_{\ell}^{\mu}=\hat{I}_{2^{\ell-1}}\otimes\hat{\sigma}^{\mu}\otimes\hat{I}_{2^{2N_{orb}-\ell}},\label{eq:sigma_dp}
\end{equation}
where $\mu\in\{x,y,z\}$ and $\hat{\sigma}^{\mu}$ is a standard $2\times2$
Pauli matrix. It should be noted that we choose a convention for the
Jordan-Wigner transformation such that $|0\rangle$ is associated
with $|\uparrow\rangle$ and $|1\rangle$ is associated with $|\downarrow\rangle$.
This local Jordan-Wigner transformation can be used to transform the
usual Gutzwiller energy form into the qubit energy form, though spin
operators will naturally emerge when using the tensor representation
of local observables within the SCDA.

We proceed by evaluating the ground state energy under $\hat{\rho}_{G2}$,
where $\hat{\rho}_{G2}$ is parametrized as
\begin{align}
 & \hat{\rho}_{G2}=\hat{P}\left(\{u_{\Gamma}\}\right)\hat{K}\left(\{n_{k\ell,0}\}\right)\hat{P}\left(\{u_{\Gamma}\}\right),\label{eq:g2-1}\\
 & \hat{P}\left(\{u_{\Gamma}\}\right)=\prod_{i}\left(\sum_{\Gamma}u_{\Gamma}\hat{X}_{i\Gamma}\right),\\
 & \hat{K}\left(\{n_{k\ell,0}\}\right)=\prod_{k\ell}\big((1-n_{k\ell,0})(1-\hat{n}_{k\ell})+n_{k\ell,0}\hat{n}_{k\ell}\big),
\end{align}
where the variational parameters $\{u_{\Gamma}\}$ and $\{n_{k\ell,0}\}$
are real and $n_{k\ell,0}\in[0,1]$. The variational parameters $\{u_{\Gamma}\}$
can be reparametrized as a pure state of an effective $2N_{orb}$
qubit system (i.e. a $2N_{orb}$ spin $1/2$ system) characterized
by a many-body density matrix $\rhoeff$. The total trial energy $E(\rhoeff,\{n_{k\ell,0}\})$
for $\hat{\rho}_{G2}$ can then be written as
\begin{align}
 & E(\rhoeff,\{n_{k\ell,0}\})=\sum_{\ell}\int dk\epsilon_{k\ell}n_{k\ell}+\left\langle H_{loc}\left(\left\{ \hat{n}_{\ell}\right\} \right)\right\rangle _{\rhoeff},\label{eq:energygansatz}\\
 & n_{k\ell}=\left\langle \hat{n}_{\ell}\right\rangle _{\rhoeff}+\frac{\xi_{\ell}^{2}}{\xi_{\ell,0}^{2}}\left(n_{k\ell,0}-\left\langle \hat{n}_{\ell}\right\rangle _{\rhoeff}\right),\label{eq:nkl rhog2}\\
 & \xi_{\ell}=\frac{1}{2}\langle\hat{\sigma}_{\ell}^{x}\rangle_{\rhoeff},\\
 & \xi_{\ell,0}=\sqrt{\left\langle \hat{n}_{\ell}\right\rangle _{\rhoeff}\left\langle 1-\hat{n}_{\ell}\right\rangle _{\rhoeff}},\label{eq:energygansatz_final}
\end{align}
with a constraint for each $\ell$
\begin{equation}
\int dkn_{k\ell;0}=\left\langle \hat{n}_{\ell}\right\rangle _{\rhoeff}.\label{eq:ga constraint}
\end{equation}
Here we have taken the continuum limit of the discretized $n_{k\ell}$
and choose the convention $\int dk=1$. Equations (\ref{eq:energygansatz})-(\ref{eq:ga constraint})
give an \textit{explicit} functional form for the trial energy as
a function of the variational parameters. The qubit energy form in
Eqns. (\ref{eq:energygansatz})-(\ref{eq:ga constraint}) is equivalent
to a previous result obtained using the slave spin mean-field method
\cite{De'medici2005205124} (see Appendix \ref{app:SSMF} for a detailed
comparison). It should be emphasized that Eqns. (\ref{eq:energygansatz})-(\ref{eq:ga constraint})
are simply a transformation of the SCDA algorithm, and therefore are
completely equivalent to the usual Gutzwiller approximation. 

We now proceed to study $\hat{\rho}_{B2}$, which is the dual of $\hat{\rho}_{G2}$,
given as
\begin{equation}
\hat{\rho}_{B2}=\hat{K}\left(\{\lambda_{k\ell}\}\right)\hat{P}\left(\{u_{\Gamma}\}\right)\hat{K}\left(\{\lambda_{k\ell}\}\right),\label{eq:b2-1}
\end{equation}
where the variational parameters $\{u_{\Gamma}\}$ and $\{\lambda_{k\ell}\}$
are real and $u_{\Gamma}\ge0$. The qubit parametrization reparametrizes
the $\{u_{\Gamma}\}$ and $\{\lambda_{k\ell}\}$ into $\rhoeff$ and
$n_{k\ell}\in[0,1]$, where $\rhoeff$ is a diagonal, positive semi-definite
$2^{2N_{orb}}\times2^{2N_{orb}}$ matrix and $n_{k\ell}$ is the physical
single particle density matrix, yielding the following total energy:
\begin{align}
 & E(\rhoeff,\{n_{k\ell}\})=\sum_{\ell}\int dk\epsilon_{k\ell}n_{k\ell}+\left\langle H_{loc}\left(\left\{ \hat{n}_{eff,\ell}\right\} \right)\right\rangle _{\rhoeff},\label{eq:energybanstz}\\
 & \hat{n}_{eff,\ell}=\left\langle \hat{n}_{\ell}\right\rangle _{\rhoeff}+\frac{A_{\ell}^{2}}{\xi_{\ell,0}^{2}}\left(\hat{n}_{\ell}-\left\langle \hat{n}_{\ell}\right\rangle _{\rhoeff}\right),\label{eq:nhat_eff}\\
 & A_{\ell}=\int dk\sqrt{n_{k\ell}\left(1-n_{k\ell}\right)},\label{eq:Aell rhob2}\\
 & \xi_{\ell,0}=\sqrt{\left\langle \hat{n}_{\ell}\right\rangle _{\rhoeff}\left\langle 1-\hat{n}_{\ell}\right\rangle _{\rhoeff}},
\end{align}
with a constraint for each $\ell$
\begin{equation}
\int dkn_{k\ell}=\left\langle \hat{n}_{\ell}\right\rangle _{\rhoeff}.
\end{equation}
It should be noted that the operators $\hat{n}_{eff,\ell}$ and $\hat{n}_{\ell}$
have the same expectation values under $\rhoeff$, which is evident
from Eq. (\ref{eq:nhat_eff}). This energy form provides a minimal
description for the Mott insulating state in the multiorbital Hubbard
model in $d=\infty$, much like the result of the Gutzwiller approximation
for the metallic phase. 

We now proceed to study $\hat{\rho}_{G3}$, which combines the variational
capacity of both $\hat{\rho}_{G2}$ and $\hat{\rho}_{B2}$. The key
result of this paper is to recast the previously obtained explicit
evaluation of the energy under $\hat{\rho}_{G3}$ into a physically
intuitive form which can be viewed as a combination of the variational
parameters of the qubit energy form for $\hat{\rho}_{G2}$ and $\hat{\rho}_{B2}$.
The $\hat{\rho}_{G3}$ is given as

\begin{align}
\hat{\rho}_{G3}= & \hat{K}\left(\{\lambda_{k\ell}\}\right)\hat{P}\left(\{u_{\Gamma}\}\right)\hat{K}\left(\{n_{k\ell,0}\}\right)\nonumber \\
 & \times\hat{P}\left(\{u_{\Gamma}\}\right)\hat{K}\left(\{\lambda_{k\ell}\}\right),
\end{align}
where the variational parameters $\{\lambda_{k\ell}\}$, $\{u_{\Gamma}\}$,
and $\{n_{k\ell,0}\}$ are real and $n_{k\ell;0}\in[0,1]$. As demonstrated
in Ref. \cite{Cheng2023035127}, when $\hat{K}\left(\{n_{k\ell,0}\}\right)$
corresponds to a Slater determinant (i.e. $n_{k\ell;0}=0$ or $n_{k\ell;0}=1$),
the total energy can be explicitly evaluated, and therefore we follow
this condition. Under this restricted form of $\{n_{k\ell,0}\}$,
it is useful to introduce the concept of a reference Fermi surface,
which delineates the boundary between $n_{k\ell;0}=0$ and $n_{k\ell;0}=1$
throughout the Brillouin zone. We can reparametrize $\{\lambda_{k\ell}\}$
and $\{u_{\Gamma}\}$ in terms of $\{n_{k\ell}\}$ and $\rhoeff$,
where $\rhoeff$ is $2^{2N_{orb}}\times2^{2N_{orb}}$ matrix corresponding
to a pure state of a $2N_{orb}$ qubit system and $n_{k\ell}\in[0,1]$
is the physical momentum density distribution. The total trial energy
is given as

\begin{align}
 & E(\rhoeff,\{n_{k\ell}\},\{n_{k\ell,0}\})=\sum_{\ell}\int dk\epsilon_{k\ell}n_{k\ell}\nonumber \\
 & +\left\langle H_{loc}\left(\left\{ \hat{n}_{eff,\ell}\right\} \right)\right\rangle _{\rhoeff},\label{eq:energy_n3_qubit}\\
 & \hat{n}_{eff,\ell}=f_{\ell,0}+f_{\ell,x}\hat{\sigma}_{\ell}^{x}+f_{\ell,z}\hat{\sigma}_{\ell}^{z},
\end{align}
and there are two constraints for each $\ell$ given as
\begin{equation}
\int_{<}dk=\int dkn_{k\ell}=\left\langle \hat{n}_{\ell}\right\rangle _{\rho},\label{eq:restriction1}
\end{equation}
where the symbol $<$ indicates that the integration is over the region
where $n_{k\ell,0}=1$ for a given $\ell$, and the symbol $>$ indicates
the region where $n_{k\ell,0}=0$. For a given $\ell$, the quantities
$f_{\ell,0}$, $f_{\ell,x}$ and $f_{\ell,z}$ are nontrivial analytical
functions of the following five variables 
\begin{align}
 & n_{\ell}=\left\langle \hat{n}_{\ell}\right\rangle _{\rhoeff},\\
 & \xi_{\ell}=\frac{1}{2}\langle\hat{\sigma}_{\ell}^{x}\rangle_{\rhoeff},\\
 & \Delta_{\ell}=\int_{>}dkn_{k\ell}=\int_{<}dk\left(1-n_{k\ell}\right),\label{eq:delta definition}\\
 & \mathcal{A}_{<\ell}=\int_{<}dk\sqrt{n_{k\ell}\left(1-n_{k\ell}\right)},\label{eq:A<ell}\\
 & \mathcal{A}_{>\ell}=\int_{>}dk\sqrt{n_{k\ell}\left(1-n_{k\ell}\right)},\label{eq:A>ell}
\end{align}
where $\rhoeff$ and $n_{k\ell}$ are constrained such that
\begin{equation}
\left|\xi_{\ell}\right|\leq\sqrt{(1-n_{\ell})n_{\ell}-(1-\Delta_{\ell})\Delta_{\ell}},\label{eq:general_constraint}
\end{equation}
which ensures that $f_{\ell,0}$, $f_{\ell,x}$ and $f_{\ell,z}$
are real numbers (see Eq. (\ref{eq:thetaell})). The explicit functional
dependence is given by the following list of equations 

\allowdisplaybreaks

\begin{align}
 & \xi_{\ell,0}=\sqrt{\left(1-n_{\ell}\right)n_{\ell}},\label{eq:algolist1}\\
 & \delta n_{\ell}=n_{\ell}-\frac{1}{2},\\
 & \theta_{\ell}=\cos^{-1}\left(\frac{\xi_{\ell}}{\sqrt{\xi_{\ell,0}^{2}+\left(\Delta_{\ell}-1\right)\Delta_{\ell}}}\right),\label{eq:thetaell}\\
 & g_{\ell,12}=\Delta_{\ell}\cot\left(\theta_{\ell}\right),\label{eq:g12}\\
 & c_{\ell}=\sqrt{1-\frac{2\text{\ensuremath{\delta}n}_{\ell}}{g_{\ell,12}^{2}+n_{\ell}^{2}}},\label{eq:cell}\\
 & \phi_{\ell}=\tan^{-1}\left(\frac{g_{\ell,12}}{g_{\ell,12}^{2}-\xi_{\ell,0}^{2}}\right),\label{eq:phiell}\\
 & \phi_{\ell,0}=\phi_{\ell}-\theta_{\ell},\label{eq:phiell0}\\
 & \mathcal{G}_{\ell,12}=-\frac{1}{2}\tan\left(\frac{\phi_{\ell,0}}{2}\right),\label{eq:g012}\\
 & \mathcal{I}_{\ell}=g_{\ell,12}\mathcal{G}_{\ell,12}+\frac{n_{\ell}}{2},\label{eq:Iell}\\
 & \mathcal{J}_{\ell}=\frac{g_{\ell,12}}{2}-n_{\ell}\mathcal{G}_{\ell,12},\\
 & \mathcal{A}'_{<,\ell}=\mathcal{A}_{<,\ell}+\sin\left(\theta_{\ell}\right)\mathcal{A}_{>,\ell},\\
 & \mathcal{A}'_{>,\ell}=\cos\left(\theta_{\ell}\right)\mathcal{A}_{>,\ell},\label{eq:Aprime>ell}\\
 & \mathcal{G}_{\ell,13}=-\mathcal{G}_{\ell,32}=\frac{\mathcal{I}_{\ell}\mathcal{A}'_{>,\ell}-\mathcal{J}_{\ell}\mathcal{A}'_{<,\ell}}{\left(g_{\ell,12}^{2}+n_{\ell}^{2}\right)\sqrt{c_{\ell}\sin\left(\theta_{\ell}\right)}},\label{eq:g013}\\
 & \mathcal{G}_{\ell,23}=-\mathcal{G}_{\ell,31}=\frac{\mathcal{I}_{\ell}\mathcal{A}'_{<,\ell}+\mathcal{J}_{\ell}\mathcal{A}'_{>,\ell}}{\left(g_{\ell,12}^{2}+n_{\ell}^{2}\right)\sqrt{c_{\ell}\sin\left(\theta_{\ell}\right)}},\label{eq:g023}\\
 & i_{\ell}=\frac{8\mathcal{G}_{\ell,13}\mathcal{G}_{\ell,23}}{4\mathcal{G}_{\ell,12}^{2}+1},\label{eq:iell}\\
 & j_{\ell}=\frac{4\left(\mathcal{G}_{\ell,13}^{2}-\mathcal{G}_{\ell,23}^{2}\right)}{4\mathcal{G}_{\ell,12}^{2}+1},\label{eq:jell}\\
 & \mathcal{G}_{\ell,33}=n_{\ell}+j_{\ell}\left(g_{\ell,12}-\mathcal{G}_{\ell,12}\right)-i_{\ell}\text{\ensuremath{\delta}n}_{\ell},\label{eq:Gell33}\\
 & p_{\ell}=\frac{\xi_{\ell}\left(\cot\left(\phi_{\ell,0}\right)-2g_{\ell,12}\right)-2\text{\ensuremath{\delta}n}_{\ell}^{2}}{2\left(\xi_{\ell}^{2}+\text{\ensuremath{\delta}n}_{\ell}^{2}\right)},\label{eq:pl}\\
 & q_{\ell}=\frac{\text{\ensuremath{\delta}n}_{\ell}\left(\cot\left(\phi_{\ell,0}\right)-2g_{\ell,12}+2\xi_{\ell}\right)}{2\left(\xi_{\ell}^{2}+\text{\ensuremath{\delta}n}_{\ell}^{2}\right)},\label{eq:ql}\\
 & f_{\ell,0}=\mathcal{G}_{\ell,33}-\frac{1}{2}j_{\ell}\csc\left(\phi_{\ell,0}\right),\label{eq:fell0}\\
 & f_{\ell,z}=\frac{1}{2}\left(i_{\ell}p_{\ell}-j_{\ell}q_{\ell}\right),\label{eq:fellz}\\
 & f_{\ell,x}=\frac{1}{2}\left(i_{\ell}q_{\ell}+j_{\ell}p_{\ell}\right).\label{eq:fellx}
\end{align}
Several important points should be noted. First, the $\{n_{k\ell}\}$
only enter the local interaction energy through $\Delta_{\ell}$,
$\mathcal{A}_{<\ell}$, $\mathcal{A}_{>\ell}$, and the constraint
$\int dkn_{k\ell}=n_{\ell}$. Second, the $\{n_{k\ell,0}\}$ only
enter the local interaction energy through the regions of integration
(i.e. $>$ and $<$) and the constraint $\int dkn_{k\ell,0}=n_{\ell}$.
Third, the operators $\hat{n}_{eff,\ell}$ and $\hat{n}_{\ell}$ have
the same expectation values under $\rhoeff$, as in the case of the
$\hat{\rho}_{B2}$. Finally, it should be clear that evaluating the
$\hat{\rho}_{G3}$ has a similar computational cost as compared to
$\hat{\rho}_{G2}$ and $\hat{\rho}_{B2}$, where the largest computational
cost is associated with evaluating the local interaction energy. 

In the following, we briefly discuss how to numerically minimize the
qubit energy form for $\hat{\rho}_{G3}$. For a multiorbital Hubbard
model with $2N_{orb}$ spin-orbitals, the formal variational parameters
are $\left\{ n_{k\ell}\right\} $, $\left\{ n_{k\ell;0}\right\} $,
and $\rhoeff$ with $\ell=1,\dots,2N_{orb}$, and there are two local
density constraints per spin orbital (see Eqn. (\ref{eq:restriction1})).
We choose $\left\{ n_{k\ell;0}\right\} $ corresponding to the momentum
density distribution for the non-interacting Hamiltonian with local
density $\left\{ n_{\ell}\right\} $, and therefore $\left\{ n_{k\ell;0}\right\} $
is determined from $\left\{ n_{\ell}\right\} $. It is important to
realize that the local interaction energy $\left\langle H_{loc}\left(\left\{ \hat{n}_{eff,\ell}\right\} \right)\right\rangle _{\rhoeff}$
does not depend on the full details of $\{n_{k\ell}\}$, but instead
only on $\left\{ n_{\ell}\right\} $, $\left\{ \Delta_{\ell}\right\} ,$
$\left\{ \mathcal{A}_{<\ell}\right\} $, and $\left\{ \mathcal{A}_{>\ell}\right\} $.
Therefore, for a given spin orbital $\ell$, four Lagrange multipliers,
$a_{<\ell}$, $b_{<\ell}$, $a_{>\ell}$, and $b_{>\ell}$, can be
used to obtain the partially optimized $n_{k\ell}$ as \cite{Cheng2023035127}
\begin{equation}
n_{k\ensuremath{\ell}}\big|_{k\in X}=\frac{1}{2}\left(\frac{a_{\text{X\ensuremath{\ell}}}-\epsilon_{k\ensuremath{\ell}}}{\sqrt{\left(a_{\text{X\ensuremath{\ell}}}-\epsilon_{k\ensuremath{\ell}}\right){}^{2}+b_{\text{X\ensuremath{\ell}}}^{2}}}+1\right),\label{eq:nkl_optimized}
\end{equation}
where $X$ is either $<$ or $>$. Interestingly, the optimized $n_{k\ell}$
yields a non-trivial dependence on $\epsilon_{k\ell}$, in contrast
to the Gutzwiller approximation (see Figure \ref{fig:momentum-density-distribution}).
There is a local density constraint between $\{n_{k\ell}\}$ and $\rhoeff$
given by Eqn. (\ref{eq:restriction1}), and there are two strategies
to enforce this. The first strategy is to start from $\{n_{k\ell}\}$,
which are parametrized by $5\times2N_{orb}$ variational parameters
$\{n_{\ell}\}$, $\{a_{<\ell}\}$, $\{a_{>\ell}\}$, $\{b_{<\ell}\}$
and $\{b_{>\ell}\}$, though only $4\times2N_{orb}$ are independent
given Eqn. (\ref{eq:restriction1}). Subsequently, the $\rhoeff$
can be parametrized by $2^{2N_{orb}}-2N_{orb}-1$ independent variational
parameters \cite{Cheng2023035127}. Therefore, there are a total of
$2^{2N_{orb}}+3\times2N_{orb}-1$ independent variational parameters.
This strategy motivates the construction of a one-body reduced density
matrix functional, as presented in our companion paper \cite{companion}.
The second strategy is to start from $\rhoeff$, which can be parametrized
by $2^{2N_{orb}}-1$ independent variational parameters, which determines
$\{n_{\ell}\}$. Subsequently, the $\{\Delta_{\ell}\}$ , $\{b_{<\ell}\}$,
and $\{b_{>\ell}\}$ are chosen as independent variational parameters,
which determines $\{a_{X\ell}\}$, yielding $2^{2N_{orb}}+3\times2N_{orb}-1$
independent variational parameters. The second strategy allows for
a trivial implementation of the inequality constraint among $\Delta_{\ell}$,
$n_{\ell}$, and $\xi_{\ell}$ (see Eq. (\ref{eq:xiellmax})), for
a given $\ell$, which is very important in practice. In either case,
one is left with minimizing over $2^{2N_{orb}}+3\times2N_{orb}-1$
variational parameters, which may be achieved using a variety of standard
approaches. 

Finally, we summarize the applications that are studied in Section
\ref{subsec:Understanding-the-Mott-single-band-Delta}. Consider the
multiorbital Hubbard model with density-density interactions given
as

\begin{equation}
\hat{H}_{loc}=U\hat{O}_{1}+\left(U-2J\right)\hat{O}_{2}+\left(U-3J\right)\hat{O}_{3},\label{eq:Hubbard_multi_orbital}
\end{equation}
where $\hat{O}_{1}=\sum_{\alpha}\delta\hat{n}_{\alpha\uparrow}\delta\hat{n}_{\alpha\downarrow}$,
$\hat{O}_{2}=\sum_{\alpha<\beta,\sigma}\delta\hat{n}_{\alpha\sigma}\delta\hat{n}_{\beta\bar{\sigma}}$,
$\hat{O}_{3}=\sum_{\alpha<\beta,\sigma}\delta\hat{n}_{\alpha\sigma}\delta\hat{n}_{\beta\sigma}$,
and $\delta\hat{n}_{\alpha\sigma}=\hat{n}_{\alpha\sigma}-\frac{1}{2}$,
with the orbital indices $\alpha,\beta$ taking values of $1,\dots,N_{orb}$
and $\sigma\in\{\uparrow,\downarrow\}$. We consider the special case
where $\epsilon_{k\ell}$ is independent of $\ell$ and has particle-hole
symmetry. In this case, we show that Eq. (\ref{eq:energy_n3_qubit})
can be optimized over $\rhoeff$ and $\{n_{k\ell;0}\}$, with the
assumption that the optimized $n_{k\ell;0}$ is given as $\theta(n_{k\ell}-n_{\ell}^{\star})$
where $\theta$ is the Heaviside function and $n_{\ell}^{\star}$
is chosen such that $\int dkn_{k\ell;0}=n_{\ell}$, yielding a trial
energy purely as a functional of $\{n_{k\ell}\}$ as 
\begin{equation}
E(\{n_{k\ell}\})=\sum_{\ell}\int dk\epsilon_{k\ell}n_{k\ell}+UA^{4}\widetilde{O}\left(\Delta\right),
\end{equation}
where $\widetilde{O}\left(\Delta\right)$ depends on $N_{orb}$ and
$J/U$, and $n_{k\ell}$ is independent of $\ell$ and has particle-hole
symmetry. Here we have dropped the spin-orbital index $\ell$ (e.
g. $\Delta_{\ell}\rightarrow\Delta$) and define $A=\mathcal{A}_{<}+\mathcal{A}_{>}$.
An important result is that $\widetilde{O}\left(\Delta\right)$ is
non-analytic at $\Delta=\Delta_{c}$, which can be used to infer the
nature of the Mott transition. For $N_{orb}>1$, the $\tilde{O}\left(\Delta\right)$
can be numerically represented by a one dimensional spline function
\cite{companion}, while $N_{orb}=1$ has an explicit expression which
can be derived as
\begin{align}
 & \tilde{O}\left(\Delta\right)=\begin{cases}
-\frac{1-4d_{1}\left(\Delta\right)}{4(2-4\Delta)^{2}\Delta^{2}}, & \Delta<\Delta_{c}\\
-4, & \Delta\ge\Delta_{c}
\end{cases},\label{eq:otilde-one-band}\\
 & d_{1}(\Delta)=\frac{1}{4}-\frac{2\sqrt{\frac{2}{3}}\Delta^{2}\left(-4\Delta+h_{2}+2\right){}^{2}}{3\left(4(\Delta-1)\Delta+(2\Delta-1)h_{1}+3\right){}^{3/2}},\label{eq:d1_twopeak}\\
 & h_{1}=\sqrt{4(\Delta-1)\Delta+9},\label{eq:h1-1}\\
 & h_{2}=\sqrt{4\Delta\left(10\Delta+3h_{1}-10\right)-6h_{1}+18},\\
 & \Delta_{c}=\frac{1}{6}\left(3-\sqrt{3}\right).
\end{align}
The above equations finally deliver a closed form expression for a
trial energy which qualitatively and quantitatively captures the Mott
transition in the single band Hubbard model at half-filling in $d=\infty$,
exactly reproducing the VDAT results for a G-type SPD at $\mathcal{N}=3$.

\section{\label{subsec:G-ansatz}Derivation of the qubit energy form for $\hat{\rho}_{G2}$}

In this section, we derive Eq. (\ref{eq:energygansatz}) using several
different approaches. In Section \ref{subsec:GA-heuristic-qubit},
we review the GA using both a heuristic derivation and the central
point expansion (CPE), and convert the standard energy form of the
GA to the qubit energy form using the Jordan-Wigner transformation.
In Section \ref{subsec:SCDA qubit}, we use the SCDA within the Gutzwiller
gauge to evaluate $\hat{\rho}_{G2}$, and the tensor product representation
is used to obtain the qubit energy form. 

\subsection{The derivation of the qubit energy form from the GA\label{subsec:GA-heuristic-qubit}}

In this section, we provide an elementary derivation of the qubit
energy form, which consists of two steps. In the first step, we derive
the standard form of the GA, using both a heuristic argument and the
CPE \cite{Cheng2020081105}, where the energy is parametrized by $\left\{ n_{k\ell;0}\right\} $
and $\rho_{loc}$, where $\rho_{loc}$ is the local reduced density
matrix of the ansatz, which is a diagonal matrix in the basis of $|\Gamma\rangle$.
The quasi-particle weight is constructed from $\rho_{loc}$ and the
fermionic creation and annihilation operators. In the second step,
we convert the standard form of the GA to the qubit energy form which
is parametrized by $\left\{ n_{k\ell;0}\right\} $ and $\rhoeff$,
where $\rhoeff$ is a pure state in a $2N_{orb}$-qubit space, and
the quasi-particle weight is constructed from $\rhoeff$ and the Pauli
spin operators. 

\subsubsection{Derivation of the GA: a heuristic argument and the CPE\label{subsec:heuristic-argument-GA}}

While the Gutzwiller approximation is well known, it is normally applied
to the special case where $\hat{\rho}_{G2}$ is a pure state. Here
we provide a heuristic derivation for a general $\hat{\rho}_{G2}$.
Additionally, we use the CPE to derive the same result from a different
perspective. We begin by presenting the heuristic derivation of the
GA. First consider the expectation values for the diagonal Hubbard
operator $\hat{X}_{i\Gamma}$ measured under $\hat{K}_{2}=\hat{K}\left(\left\{ n_{k\ell;0}\right\} \right)$,
given as 
\begin{equation}
\langle\hat{X}_{i\Gamma}\rangle_{\hat{K}_{2}}=\prod_{\ell=1}^{N}\left(\delta_{\Gamma_{\ell},0}\left(1-n_{\ell}\right)+\delta_{\Gamma_{\ell},1}n_{\ell}\right),\label{eq:xig k2}
\end{equation}
where $n_{\ell}=\left\langle \hat{n}_{i\ell}\right\rangle _{\hat{K}_{2}}$
denotes the local density at a given site $i$ for the spin orbital
$\ell$ under $\hat{K}_{2}$. The first assumption of the GA is that
the atomic configuration distribution under $\hat{\rho}_{G2}$ can
be approximated as
\begin{equation}
\left\langle \hat{X}_{i\Gamma}\right\rangle _{\hat{\rho}_{G2}}=u_{\Gamma}^{2}\left\langle \hat{X}_{i\Gamma}\right\rangle _{\hat{K}_{2}}\bigg/\sum_{\Gamma}u_{\Gamma}^{2}\left\langle \hat{X}_{i\Gamma}\right\rangle _{\hat{K}_{2}},\label{eq:pgamma}
\end{equation}
which ignores off-site contributions in $\hat{K}_{2}$. Additionally,
a constraint is applied to $\hat{P}_{1}$ such that the local density
is invariant, given as
\begin{equation}
\left\langle \hat{n}_{i\ell}\right\rangle _{\hat{\rho}_{G2}}=\left\langle \hat{n}_{i\ell}\right\rangle _{\hat{K}_{2}},\label{eq:Gutzwiller constraint}
\end{equation}
which we refer to as the Gutzwiller constraint. The denominator in
Eq. (\ref{eq:pgamma}) ensures that the sum of all expectation values
for the diagonal Hubbard operators are normalized. Given that the
off-diagonal Hubbard operators have zero expectation value due to
the restriction of the SPD used in this study, the expectation values
for all diagonal Hubbard operators allow the evaluation of any local
observable. The second assumption of the GA lies within the evaluation
for the hopping term between two distinct sites. The idea is to break
a hopping term like $\hat{a}_{i\ell}^{\dagger}\hat{a}_{j\ell}$ into
two processes: first annihilating an electron at site $j$ and then
creating an electron at site $i$, and the GA assumes the probabilities
for the two steps are independent and only depend on the atomic distributions
on site $i$ and $j$. The probability of creating or destroying an
electron is renormalized by the interacting projector, and the GA
assumes this is obtained by counting all relevant one-particle excitation
processes as 
\begin{equation}
\mathcal{R}_{\ell}=\frac{\sum_{\Gamma\Gamma'}\sqrt{\left\langle \hat{X}_{i\Gamma}\right\rangle _{\hat{\rho}_{G2}}\left\langle \hat{X}_{i\Gamma'}\right\rangle _{\hat{\rho}_{G2}}}\left|\left\langle \Gamma'|\hat{a}_{i\ell}^{\dagger}|\Gamma\right\rangle \right|^{2}}{\sum_{\Gamma\Gamma'}\sqrt{\left\langle \hat{X}_{i\Gamma}\right\rangle _{\hat{K}_{2}}\left\langle \hat{X}_{i\Gamma'}\right\rangle _{\hat{K}_{2}}}\left|\left\langle \Gamma'|\hat{a}_{i\ell}^{\dagger}|\Gamma\right\rangle \right|^{2}}.\label{eq:r component}
\end{equation}
To simply this expression, we introduce $\rho_{loc}$ and $\rho_{loc;0}$,
which are defined within the local Hilbert space as 
\begin{alignat}{1}
 & \left[\rho_{loc}\right]_{\Gamma\Gamma'}=\delta_{\Gamma\Gamma'}\langle\hat{X}_{i\Gamma}\rangle_{\hat{\rho}_{G2}},\label{eq:rholoc}\\
 & \left[\rho_{loc;0}\right]_{\Gamma\Gamma'}=\delta_{\Gamma\Gamma'}\langle\hat{X}_{i\Gamma}\rangle_{\hat{K}_{2}},\label{eq:rholoc0}
\end{alignat}
which are matrices of dimension $2^{2N_{orb}}\times2^{2N_{orb}}$,
yielding
\begin{equation}
\mathcal{R}_{\ell}=\frac{\text{Tr}\left(\sqrt{\rho_{loc}}\hat{a}_{\ell}^{\dagger}\sqrt{\rho_{loc}}\hat{a}_{\ell}\right)}{\text{Tr}\left(\sqrt{\rho_{loc;0}}\hat{a}_{\ell}^{\dagger}\sqrt{\rho_{loc;0}}\hat{a}_{\ell}\right)}.\label{eq:r matrix}
\end{equation}
The Gutzwiller constraint, Eq. (\ref{eq:Gutzwiller constraint}),
can be rewritten as $\langle\hat{n}_{\ell}\rangle_{\rho_{loc}}=\langle\hat{n}_{\ell}\rangle_{\rho_{loc;0}}$.
Given the second assumption of the GA, the single particle density
matrix between two different sites $j$ and $j'$ is renormalized
as 
\begin{equation}
\mathcal{Z}_{\ell}\equiv\frac{\langle\hat{a}_{j\ell}^{\dagger}\hat{a}_{j'\ell}\rangle_{\hat{\rho}_{G2}}}{\langle\hat{a}_{j\ell}^{\dagger}\hat{a}_{j'\ell}\rangle_{\hat{K}_{2}}}=\mathcal{R}_{\ell}^{2}.\label{eq:z}
\end{equation}
Combining Eq. (\ref{eq:z}) with the Gutzwiller constraint, the single
particle density matrix for arbitrary $j$ and $j'$ is given as 
\begin{equation}
\left\langle \hat{a}_{j\ell}^{\dagger}\hat{a}_{j'\ell}\right\rangle _{\hat{\rho}_{G2}}=\mathcal{Z}_{\ell}\left\langle \hat{a}_{j\ell}^{\dagger}\hat{a}_{j'\ell}\right\rangle _{\hat{K}_{2}}+\delta_{jj'}\left(1-\mathcal{Z}_{\ell}\right)n_{\ell}.
\end{equation}
The momentum density distribution $n_{k\ell}=\langle\hat{a}_{k\ell}^{\dagger}\hat{a}_{k\ell}\rangle_{\hat{\rho}_{G2}}$
can then be computed as 
\begin{alignat}{1}
n_{k\ell} & =\frac{1}{N_{site}}\sum_{jj'}e^{ik\left(j-j'\right)}\left\langle \hat{a}_{j\ell}^{\dagger}\hat{a}_{j'\ell}\right\rangle _{\hat{\rho}_{G2}}\\
 & =\mathcal{Z}_{\ell}n_{k\ell;0}+\left(1-\mathcal{Z}_{\ell}\right)n_{\ell},\label{eq:nk}
\end{alignat}
where $N_{site}$ is the number of $k$-points, $\hat{a}_{k}^{\dagger}=\left(1/\sqrt{N_{site}}\right)\sum_{j}e^{ik\cdot j}\hat{a}_{j}^{\dagger}$,
and $n_{k\ell;0}=\langle a_{k\ell}^{\dagger}a_{k\ell}\rangle_{\hat{K}_{2}}$.
Therefore, the standard form for the total energy per site of the
GA is given by 
\begin{align}
 & E(\{n_{k\ell,0}\},\rho_{loc})=\sum_{\ell}\int dk\epsilon_{k\ell}n_{k\ell}+\left\langle H_{loc}\left(\left\{ \hat{n}_{\ell}\right\} \right)\right\rangle _{\rho_{loc}},\label{eq:ga energy}\\
 & n_{k\ell}=\left\langle \hat{n}_{\ell}\right\rangle _{\rho_{loc}}+\mathcal{R}_{\ell}^{2}\left(n_{k\ell,0}-\left\langle \hat{n}_{\ell}\right\rangle _{\rho_{loc}}\right),\\
 & \mathcal{R}_{\ell}=\frac{\text{Tr}\left(\sqrt{\rho_{loc}}\hat{a}_{\ell}^{\dagger}\sqrt{\rho_{loc}}\hat{a}_{\ell}\right)}{\sqrt{\left\langle \hat{n}_{\ell}\right\rangle _{\rho_{loc}}\left\langle 1-\hat{n}_{\ell}\right\rangle _{\rho_{loc}}}},\label{eq:GA Rell}
\end{align}
with the density constraint 
\begin{equation}
\int dkn_{k\ell;0}=\left\langle \hat{n}_{\ell}\right\rangle _{\rho_{loc}}.
\end{equation}
The variational parameters satisfy $n_{kl;0}\in\left[0,1\right]$
and $\rho_{loc}$ is a diagonal positive semi-definite matrix. 

We now provide an overview of how to derive Eq. (\ref{eq:ga energy})
using the CPE (see Appendix \ref{appendix:The-CPE-for-G}). We begin
by making several observations about the GA. First, when $n_{k\ell;0}=n_{\ell}$,
the $\hat{\rho}_{G2}$ describes a collection of atoms, and in this
case Eq. (\ref{eq:ga energy}) yields an exact evaluation in any dimension.
Second, there is a linear relation between $n_{k\ell}$ and $n_{k\ell;0}$
given by 
\begin{equation}
n_{k\ell}-n_{\ell}=\mathcal{Z}_{\ell}\left(n_{k\ell;0}-n_{\ell}\right).\label{eq:ga delta nkl}
\end{equation}
Third, $\rho_{loc}$ is directly determined by $n_{\ell}$ and $\{u_{\Gamma}\}$,
and is independent of $n_{k\ell;0}$. Recall that the $\hat{\rho}_{G2}$
is defined as $\hat{\rho}_{G2}=\hat{P}_{1}\hat{K}_{2}\hat{P}_{1}$,
where $\hat{K}_{2}=\hat{K}\left(\left\{ n_{k\ell;0}\right\} \right)$
and $\hat{P}_{1}=\hat{P}\left(\left\{ u_{\Gamma}\right\} \right)$.
We now introduce $\hat{K}_{2}^{\star}=\hat{K}\left(\left\{ n_{k\ell;0}=n_{\ell}\right\} \right)$,
which has the same local density matrix as $\hat{K}_{2}$, and $\rho_{G2}^{\star}=\hat{P}_{1}\hat{K}_{2}^{\star}\hat{P}_{1}$,
which describes a collection of atoms. The $\{u_{\Gamma}\}$ can be
reparametrized by the local reduced density matrix $\rho_{loc}^{\star}$
of $\rho_{G2}^{\star}$, where $\left[\rho_{loc}^{\star}\right]_{\Gamma\Gamma'}=\delta_{\Gamma\Gamma'}\langle\hat{X}_{i\Gamma}\rangle_{\rho_{G2}^{\star}}$
and is constrained by $n_{\ell}=\langle\hat{a}_{\ell}^{\dagger}\hat{a}_{\ell}\rangle_{\rho_{loc}^{\star}}.$
Therefore, $n_{k\ell}$ and $\rho_{loc}$ are functionals of $\left\{ n_{k\ell;0}\right\} $
and $\rho_{loc}^{\star}$. The CPE amounts to the expansion of observables
in terms of $\{n_{k\ell;0}\}$ about $\left\{ n_{k\ell;0}=n_{\ell}\right\} $,
and up to the first order, one recovers Eq. (\ref{eq:ga delta nkl})
and $\rho_{loc}=\rho_{loc}^{\star}$, proving that the first-order
CPE recovers the GA. 

\subsubsection{Converting the GA energy into the qubit form using the Jordan-Wigner
transformation}

Here we discuss how to convert the standard form of the GA energy
into the qubit form, which is mathematically equivalent. The qubit
form provides a unified view of the energy evaluated using $\hat{\rho}_{G2}$,
$\hat{\rho}_{B2}$, and $\hat{\rho}_{G3}$. The qubit parametrization
consists of two steps. First, we introduce a purified many-body density
matrix $\rhoeff$ from $\rho_{loc}$. Second, we perform the Jordan-Wigner
transformation, which converts $\mathcal{R}_{\ell}$ (Eq. (\ref{eq:GA Rell}))
from an expression involving fermionic operators $\hat{a}_{\ell}^{\dagger}$
and $\hat{a}_{\ell}$ into an expression involving spin operators
$\hat{\sigma}_{\ell}^{x}$. We begin by defining $\rhoeff=|\Psi\rangle\langle\Psi|$,
where $|\Psi\rangle=\sum_{\Gamma}\sqrt{\left[\rho_{loc}\right]_{\Gamma\Gamma}}|\Gamma\rangle$,
which yields 
\begin{equation}
\left[\rhoeff\right]_{\Gamma\Gamma'}=\sqrt{\left[\rho_{loc}\right]_{\Gamma\Gamma}\left[\rho_{loc}\right]_{\Gamma'\Gamma'}}.
\end{equation}
 Using the Jordan-Wigner transformation (see Eqns. (\ref{eq:Jordan Wigner1})
and (\ref{eq:Jordan Wigner2})), we obtain
\begin{align}
 & \left|\left\langle \Gamma'|\hat{a}_{\ell}^{\dagger}|\Gamma\right\rangle \right|^{2}=\langle\Gamma'|\hat{a}_{\ell}^{\dagger}|\Gamma\rangle\langle\Gamma|\hat{a}_{\ell}|\Gamma'\rangle\\
 & =\langle\Gamma'|\left(\prod_{\ell'<\ell}\hat{\sigma}_{\ell'}^{z}\right)\hat{\sigma}_{\ell}^{+}|\Gamma\rangle\langle\Gamma|\left(\prod_{\ell'<\ell}\hat{\sigma}_{\ell'}^{z}\right)\hat{\sigma}_{\ell}^{-}|\Gamma'\rangle\\
 & =\langle\Gamma'|\hat{\sigma}_{\ell}^{+}|\Gamma\rangle\langle\Gamma|\hat{\sigma}_{\ell}^{-}|\Gamma'\rangle=\left|\langle\Gamma'|\hat{\sigma}_{\ell}^{+}|\Gamma\rangle\right|^{2}.\label{eq:a to sigma}
\end{align}
Therefore, we can rewrite the numerator of $\mathcal{R}_{\ell}$ (see
Eq. (\ref{eq:r matrix})) as
\begin{align}
 & \sum_{\Gamma\Gamma'}\sqrt{\left[\rho_{loc}\right]_{\Gamma\Gamma}\left[\rho_{loc}\right]_{\Gamma'\Gamma'}}\left|\left\langle \Gamma'|\hat{a}_{\ell}^{\dagger}|\Gamma\right\rangle \right|^{2}\\
 & =\sum_{\Gamma\Gamma'}\left[\rhoeff\right]_{\Gamma\Gamma'}\left|\langle\Gamma'|\sigma_{\ell}^{+}|\Gamma\rangle\right|^{2}\\
 & =\frac{1}{2}\sum_{\Gamma\Gamma'}\left[\rhoeff\right]_{\Gamma\Gamma'}\langle\Gamma'|\hat{\sigma}_{\ell}^{x}|\Gamma\rangle=\frac{1}{2}\left\langle \hat{\sigma}_{\ell}^{x}\right\rangle _{\rhoeff}=\xi_{\ell},
\end{align}
where we used the fact that $\rhoeff$ is a real symmetric matrix
and the following relation
\begin{equation}
\langle\Gamma'|\hat{\sigma}_{\ell}^{x}|\Gamma\rangle=\left|\langle\Gamma'|\hat{\sigma}_{\ell}^{+}|\Gamma\rangle\right|^{2}+\left|\langle\Gamma|\hat{\sigma}_{\ell}^{+}|\Gamma'\rangle\right|^{2}.
\end{equation}
Moreover, the denominator of $\mathcal{R}_{\ell}$ (see Eq. (\ref{eq:r matrix}))
is given as
\begin{equation}
\text{Tr}\left(\sqrt{\rho_{loc;0}}\hat{a}_{\ell}^{\dagger}\sqrt{\rho_{loc;0}}\hat{a}_{\ell}\right)=\sqrt{n_{\ell}\left(1-n_{\ell}\right)}=\xi_{\ell;0}.
\end{equation}
Therefore, we have demonstrated that $\mathcal{R}_{\ell}=\xi_{\ell}/\xi_{\ell;0}$,
and the local interaction can be written as 
\begin{equation}
\left\langle H_{loc}\left(\left\{ \hat{n}_{\ell}\right\} \right)\right\rangle _{\rho_{loc}}=\left\langle H_{loc}\left(\left\{ \hat{n}_{\ell}\right\} \right)\right\rangle _{\rhoeff}
\end{equation}
Using the expression for $\mathcal{R}_{\ell}$ and the local energy,
we arrive at the energy expressions given in Eqns. (\ref{eq:energygansatz})-(\ref{eq:energygansatz_final}).
This proof demonstrates that the usual Gutzwiller approximation can
be straightforwardly transformed into the qubit energy form, which
is equivalent to the result of the slave spin mean-field theory \cite{De'medici2005205124}
(see Appendix \ref{app:SSMF} for additional details). 

\subsection{The derivation of the qubit energy form using the SCDA \label{subsec:SCDA qubit}}

In this section, we use the gauge constrained SCDA to derive the qubit
energy form. The derivation consists of two steps. First, in Section
\ref{subsec:The-SCDA-gutz gauge}, the Gutzwiller gauge is used to
rederive the standard form of the GA energy. Second, in Section \ref{subsec:Derivation-of-qubit-SCDA},
we derive the qubit energy form using the tensor representation, which
does not rely on the Jordan-Wigner transformation. Additionally, in
Section \ref{subsec:SCDA-under-gauge}, we discuss how the quantities
$\boldsymbol{\mathcal{G}}$ , $\boldsymbol{S}$, and $\boldsymbol{g}$
change under a general gauge transformation.

\subsubsection{\label{subsec:The-SCDA-gutz gauge}The SCDA within the Gutzwiller
gauge}

In this section, we demonstrate how the gauge of the SPD can be used
to automatically satisfy the SCDA self-consistency condition \cite{Cheng2021195138},
and we utilize some notation from the CPE. Interestingly, the Gutzwiller
constraint of the GA can be used to define an appropriate gauge for
the SPD, which we refer to as the Gutzwiller gauge. Starting from
$\hat{\rho}_{G2}=\hat{P}_{1}\hat{K}_{2}\hat{P}_{1}$ with an arbitrary
gauge, a gauge transformation $\hat{P}_{1}\rightarrow\hat{P}_{1}\hat{N}^{-1}$
and $\hat{K}_{2}\rightarrow\hat{N}\hat{K}_{2}\hat{N}$ can always
be performed to ensure that $\left\langle \hat{n}_{i\ell}\right\rangle _{\hat{\rho}_{G2}}=\left\langle \hat{n}_{i\ell}\right\rangle _{\hat{K}_{2}}=n_{\ell}$,
where $\hat{N}=\exp\left(\sum_{i\ell}\mu_{\ell}\hat{n}_{i\ell}\right)$
is chosen to satisfy the Gutzwiller constraint. Within this Gutzwiller
gauge, we can choose $\left[\bm{\mathcal{G}}_{i}\right]_{\ell\ell'}=\delta_{\ell\ell'}\mathcal{G}_{\ell}$,
where the component for the spin-orbital $\ell$ is given by 
\begin{equation}
\mathcal{G}_{\ell}=\left(\begin{array}{cc}
n_{\ell} & 1-n_{\ell}\\
-n_{\ell} & n_{\ell}
\end{array}\right).\label{eq:mathcalg choice}
\end{equation}
We now prove that this $\mathcal{G}_{\ell}$ ensures that the SCDA
self-consistency condition is automatically satisfied.  The first
step of the proof relies on the fact that under the Gutzwiller gauge
$\barhat[\rho]_{loc}$ within the SCDA is the discrete action of the
central point of the SPD $\rho_{G2}^{\star}=\hat{P}_{1}\hat{K}_{2}^{\star}\hat{P}_{1}$,
which can be shown as follows. The local reduced density matrix of
$\rho_{G2}^{\star}$ at site $i$ is $\hat{\rho}_{G2;i}^{\star}=\hat{P}_{1;i}\hat{K}_{2;i}^{\star}\hat{P}_{1;i}$,
which yields a discrete action $\barhat[\rho]_{G2;i}^{\star}=\barhat[\rho]_{G2;i;0}^{\star}\barhat[P]_{1;i}^{\left(1\right)}\barhat[P]_{1;i}^{\left(2\right)}$,
where $\barhat[\rho]_{G2;i;0}^{\star}=\barhat[Q]\barhat[K]_{2;i}^{\star}$.
Given that $\left\langle \hat{n}_{i\ell}\right\rangle _{\hat{K}_{2;i}^{\star}}=n_{\ell}$,
the integer time Green's function for $\barhat[\rho]_{G2;i;0}^{\star}$
is given by Eq. (\ref{eq:mathcalg choice}), and correspondingly $\barhat[\rho]_{G2;i}^{\star}$
is equivalent to the local discrete action $\barhat[\rho]_{loc;i}$
for the SCDA, proving $\barhat[\rho]_{G2}^{\star}=\barhat[\rho]_{loc}$.
Local observables within the SCDA can then be evaluated under the
central point of the SPD as
\begin{equation}
\langle\hat{X}_{i\Gamma}\rangle_{\hat{\rho}_{G2;i}}=\langle\barhat[X]_{i\Gamma}^{\left(2\right)}\rangle_{\barhat[\rho]_{loc;i}}=\langle\hat{X}_{i\Gamma}\rangle_{\hat{\rho}_{G2;i}^{\star}},\label{eq:X=00003DX gutz}
\end{equation}
consistent with the relation derived from the CPE (see Eq. (\ref{eq:x=00003Dx})).
Similarly, the local interacting integer time Green's function $\left[\bm{g}_{i}\right]_{\ell\ell'}=\delta_{\ell\ell'}g_{\ell}$
can be evaluated as 
\begin{equation}
g_{\ell}=\left(\begin{array}{cc}
n_{\ell} & \left(1-n_{\ell}\right)\mathcal{R}_{\ell,12}\\
-n_{\ell}\mathcal{R}_{\ell,21} & n_{\ell}
\end{array}\right),\label{eq:gloc n2}
\end{equation}
where 
\begin{equation}
\mathcal{R}_{\ell,12}\equiv\frac{\text{Tr}\left(\hat{P}_{1;i}\hat{a}_{i\ell}^{\dagger}\hat{K}_{2;i}^{\star}\hat{P}_{1;i}\hat{a}_{i\ell}\right)/\text{Tr}\left(\hat{\rho}_{G2;i}^{\star}\right)}{\text{Tr}\left(\hat{a}_{i\ell}^{\dagger}\hat{K}_{2;i}^{\star}\hat{a}_{i\ell}\right)/\text{Tr}\left(\hat{K}_{2;i}^{\star}\right)},\label{eq:r12}
\end{equation}
and
\begin{equation}
\mathcal{R}_{\ell,21}\equiv\frac{\text{Tr}\left(\hat{P}_{1;i}\hat{a}_{i\ell}\hat{K}_{2;i}^{\star}\hat{P}_{1;i}\hat{a}_{i\ell}^{\dagger}\right)/\text{Tr}\left(\hat{\rho}_{G2;i}^{\star}\right)}{\text{Tr}\left(\hat{a}_{i\ell}\hat{K}_{2;i}^{\star}\hat{a}_{i\ell}^{\dagger}\right)/\text{Tr}\left(\hat{K}_{2;}^{\star}\right)},\label{eq:r21}
\end{equation}
and the Gutzwiller gauge ensures that the diagonal part of $g_{\ell}$
is $n_{\ell}$. To connect $\mathcal{R}_{\ell,12}$ and $\mathcal{R}_{\ell,21}$
with $\mathcal{R}_{\ell}$, we use the following relations 
\begin{align}
 & \hat{a}_{i\ell}^{\dagger}\sqrt{\hat{K}_{2;i}^{\star}}=\sqrt{\left(1-n_{\ell}\right)/n_{\ell}}\sqrt{\hat{K}_{2;i}^{\star}}\hat{a}_{i\ell}^{\dagger},\\
 & \hat{a}_{i\ell}\sqrt{\hat{K}_{2;i}^{\star}}=\sqrt{n_{\ell}/\left(1-n_{\ell}\right)}\sqrt{\hat{K}_{2;i}^{\star}}\hat{a}_{i\ell},
\end{align}
which yields $\mathcal{R}_{\ell,12}=\mathcal{R}_{\ell,21}=\mathcal{R}_{\ell}$.
The integer time self-energy can then be computed as $\left[\bm{S}_{i}\right]_{\ell\ell'}=\delta_{\ell\ell'}S_{\ell}$,
using $S_{\ell}=\left(\mathcal{G}_{\ell}^{-1}-1\right)^{-1}\left(g_{\ell}^{-1}-1\right)$,
which yields

\begin{align}
S_{\ell} & =\frac{1}{n_{\ell}-\left(n_{\ell}-1\right)\mathcal{R}_{\ell}^{2}}\nonumber \\
 & \times\left(\begin{array}{cc}
\mathcal{R}_{\ell} & \left(n_{\ell}-1\right)\left(\mathcal{R}_{\ell}^{2}-1\right)\\
n_{\ell}\left(\mathcal{R}_{\ell}^{2}-1\right) & \mathcal{R}_{\ell}
\end{array}\right).\label{eq:sloc n2}
\end{align}
Therefore, the interacting integer time Green's function for a given
$k$-point is given by $\left[\bm{g}_{k}\right]_{\ell\ell'}=\delta_{\ell\ell'}g_{k\ell}$,
where $g_{k\ell}=\left(1+\left(g_{k\ell;0}^{-1}-1\right)S_{\ell}\right)^{-1}$,
which yields 
\begin{equation}
g_{k\ell}=\left(\begin{array}{cc}
n_{k\ensuremath{\ell},0} & \mathcal{R}_{\ell}\left(1-n_{k\ensuremath{\ell},0}\right)\\
-\mathcal{R}_{\ell}n_{k\ensuremath{\ell},0} & \mathcal{R}_{\ell}^{2}\left(n_{k\ensuremath{\ell},0}-n_{\ell}\right)+n_{\ell}
\end{array}\right),\label{eq:gkl}
\end{equation}
and the momentum density distribution $n_{k\ell}=\left[g_{k\ell}\right]_{22}$
is in agreement with Eq. (\ref{eq:nk}). Finally, the SCDA self-consistency
can be verified as $\frac{1}{N_{site}}\sum_{k}g_{k\ell}=g_{\ell}.$

\subsubsection{Derivation of the qubit energy form\label{subsec:Derivation-of-qubit-SCDA}}

Here we show how to derive the qubit energy form, given in Eqns. (\ref{eq:energygansatz})-(\ref{eq:ga constraint}),
using the tensor product representation. We begin by evaluating the
relevant observables under $\barhat[\rho]_{loc}$, where the tensor
product representation in Eq. (\ref{eq:local_expect_values}) simplifies
to 
\begin{equation}
(\barhat[O])_{u}\cdot\left(u_{1}\otimes u_{2}\right)=u^{T}(\barhat[O])_{u}u,
\end{equation}
given that $u^{T}=u_{1}=u_{2}$. The relevant components needed to
evaluate the local integer time Green's function in the Gutzwiller
gauge are given by
\begin{align}
\left(\barhat[1]\right)_{u;\ell} & =\left(\begin{array}{cc}
1-n_{\ell} & 0\\
0 & n_{\ell}
\end{array}\right),\\
\left(\barhat[a]_{\ell}^{\dagger\left(1\right)}\barhat[a]_{\ell}^{\left(1\right)}\right)_{u;\ell} & =\left(\begin{array}{cc}
0 & 0\\
0 & n_{\ell}
\end{array}\right),\\
\left(\barhat[a]_{\ell}^{\dagger\left(1\right)}\barhat[a]_{\ell}^{\left(2\right)}\right)_{u;\ell} & =\left(\begin{array}{cc}
0 & 0\\
1-n_{\ell} & 0
\end{array}\right),\\
\left(\barhat[a]_{\ell}^{\dagger\left(2\right)}\barhat[a]_{\ell}^{\left(1\right)}\right)_{u;\ell} & =\left(\begin{array}{cc}
0 & -n_{\ell}\\
0 & 0
\end{array}\right),\\
\left(\barhat[a]_{\ell}^{\dagger\left(2\right)}\barhat[a]_{\ell}^{\left(2\right)}\right)_{u;\ell} & =\left(\begin{array}{cc}
0 & 0\\
0 & n_{\ell}
\end{array}\right).
\end{align}
A linear transformation $u=Vw$ can be introduced such that 

\begin{equation}
w^{T}(\barhat[O])_{w}w=u^{T}(\barhat[O])_{u}u,
\end{equation}
where 

\begin{align}
 & (\barhat[O])_{w}=(\barhat[O]_{1})_{w;1}\otimes\dots\otimes(\barhat[O]_{M})_{w;2N_{orb}},\\
 & (\barhat[O]_{\ell})_{w;\ell}\equiv V_{\ell}^{T}(\barhat[O]_{\ell})_{u;\ell}V_{\ell},\\
 & V=V_{1}\otimes\dots\otimes V_{2N_{orb}}.
\end{align}
We choose $V_{\ell}$ such that the identity operator in the $w$-representation
is an identity matrix, which is given by 
\begin{equation}
V_{\ell}=\left(\begin{array}{cc}
\frac{1}{\sqrt{1-n_{\ell}}} & 0\\
0 & \frac{1}{\sqrt{n_{\ell}}}
\end{array}\right),
\end{equation}
and correspondingly, the components in the $w$-representation are
given by
\begin{align}
\left(\barhat[1]\right)_{w;\ell} & =\left(\begin{array}{cc}
1 & 0\\
0 & 1
\end{array}\right),\label{eq:identity_w2}\\
\left(\barhat[a]_{\ell}^{\dagger\left(1\right)}\barhat[a]_{\ell}^{\left(1\right)}\right)_{w;\ell} & =\left(\begin{array}{cc}
0 & 0\\
0 & 1
\end{array}\right),\\
\left(\barhat[a]_{\ell}^{\dagger\left(1\right)}\barhat[a]_{\ell}^{\left(2\right)}\right)_{w;\ell} & =\left(\begin{array}{cc}
0 & 0\\
\frac{\sqrt{1-n_{\ell}}}{\sqrt{n_{\ell}}} & 0
\end{array}\right),\label{eq:g12_w2}\\
\left(\barhat[a]_{\ell}^{\dagger\left(2\right)}\barhat[a]_{\ell}^{\left(1\right)}\right)_{w;\ell} & =\left(\begin{array}{cc}
0 & -\frac{\sqrt{n_{\ell}}}{\sqrt{1-n_{\ell}}}\\
0 & 0
\end{array}\right),\label{eq:g21_w2}\\
\left(\barhat[a]_{\ell}^{\dagger\left(2\right)}\barhat[a]_{\ell}^{\left(2\right)}\right)_{w;\ell} & =\left(\begin{array}{cc}
0 & 0\\
0 & 1
\end{array}\right).\label{eq:g22_w}
\end{align}
To connect with the qubit energy form, we define the many-body density
matrix corresponding to a pure state of the qubit system as 
\begin{equation}
\rhoeff=ww^{T},
\end{equation}
where the renormalization factor $\mathcal{R}_{\ell}$ can now be
rewritten as 

\begin{align}
\mathcal{R}_{\ell} & =\frac{1}{2}\left\langle \frac{\left(\barhat[a]_{\ell}^{\dagger\left(1\right)}\barhat[a]_{\ell}^{\left(2\right)}\right)_{w}}{1-n_{\ell}}-\frac{\left(\barhat[a]_{\ell}^{\dagger\left(2\right)}\barhat[a]_{\ell}^{\left(1\right)}\right)_{w}}{n_{\ell}}\right\rangle _{\rhoeff}\\
 & =\frac{\frac{1}{2}\left\langle \hat{\sigma}_{\ell}^{x}\right\rangle _{\rhoeff}}{\sqrt{\left(1-n_{\ell}\right)n_{\ell}}}=\frac{\xi_{\ell}}{\xi_{\ell;0}},
\end{align}
using Eqns. (\ref{eq:g12_w2}), (\ref{eq:g21_w2}), and (\ref{eq:gloc n2}).
It should be emphasized that $\hat{\sigma}_{\ell}^{x}$ is a $2^{2N_{orb}}\times2^{2N_{orb}}$
matrix, defined in Eq. (\ref{eq:sigma_dp}). Additionally, the local
expectation value of the interaction energy can be rewritten as 
\begin{align}
\left\langle H_{loc}\left(\left\{ \barhat[n]_{\ell}^{(2)}\right\} \right)\right\rangle _{\barhat[\rho]_{loc}} & =\frac{\textrm{Tr}\left(\rhoeff H_{loc}\left(\left\{ \left(\barhat[n]_{\ell}^{(2)}\right)_{w}\right\} \right)\right)}{\textrm{Tr}(\rhoeff\left(\barhat[1]\right)_{w})}\label{eq:Hloc_equality-1}\\
 & =\left\langle H_{loc}\left(\left\{ \hat{n}_{\ell}\right\} \right)\right\rangle _{\rhoeff},
\end{align}
using Eq. (\ref{eq:g22_w}) and (\ref{eq:identity_w2}), where $\hat{n}_{\ell}$
is defined in Eq. (\ref{eq:nell_sigmaz}).

\subsubsection{SCDA under a general gauge transformation\label{subsec:SCDA-under-gauge}}

Here we discuss the SCDA within an arbitrary gauge, where $\hat{\rho}_{G2}=\hat{P}_{1}\hat{K}_{2}\hat{P}_{1}^{\dagger}$,
as this will be important to understanding the gauge constrained algorithm
for $\mathcal{N}=3$. A general gauge transformation which maintains
a G-type form is defined as $\hat{P}_{1}\rightarrow\hat{P}_{1}\hat{N}^{-1}$,
$\hat{K}_{2}\rightarrow\hat{N}\hat{K}_{2}\hat{N}^{\dagger}$ , $\hat{P}_{1}^{\dagger}\rightarrow\left(\hat{N}^{\dagger}\right)^{-1}\hat{P}_{1}^{\dagger}$,
where $\hat{N}=\exp\left(\bm{\mu}\cdot\hat{\bm{n}}\right)$. Using
the results of Section \ref{subsec:Gauge-Symmetry-of-SPD}, this gauge
transformation can be decomposed in terms of an intra-time-step transformation
given by $\bm{N}_{a}=\text{diag}\left(\boldsymbol{1},\exp\left(-\bm{\mu}^{*}\right)\right)$,
and an inter-time-step transformation given by $\bm{N}_{b}=\text{diag}\left(\exp\left(-\bm{\mu}^{T}\right),\boldsymbol{1}\right)$.
The corresponding transformations for the integer time Green's functions
can be found in Section \ref{subsec:Gauge-Symmetry-of-SPD}. In the
following, we focus on the transformation with the form $\left[\bm{\mu}\right]_{i\ell,i'\ell}=\delta_{ii'}\delta_{\ell\ell'}\mu_{\ell}$
where $\mu_{\ell}$ is a real number. Using Eq. (\ref{eq:change g}),
the transformation of $\left[\bm{g}_{i}\right]_{\ell\ell'}=\delta_{\ell\ell'}g_{\ell}$
with the component $g_{\ell}$ is given as 
\begin{equation}
g'_{\ell}=\left(\begin{array}{cc}
n_{\ell} & e^{-\mu_{\ell}}\left(1-n_{\ell}\right)\mathcal{R}_{\ell}\\
-e^{\mu_{\ell}}n_{\ell}\mathcal{R}_{\ell} & n_{\ell}
\end{array}\right).\label{eq:glp}
\end{equation}
Using Eq. (\ref{eq:change S}), the transformation for $\bm{S}_{i}$
is given by

\begin{align}
S'_{\ell} & =\frac{1}{n_{\ell}+\left(1-n_{\ell}\right)\mathcal{R}_{\ell}^{2}}\nonumber \\
 & \times\left(\begin{array}{cc}
e^{\mu_{\ell}}\mathcal{R}_{\ell} & \left(n_{\ell}-1\right)\left(\mathcal{R}_{\ell}^{2}-1\right)\\
e^{2\mu_{\ell}}n_{\ell}\left(\mathcal{R}_{\ell}^{2}-1\right) & e^{\mu_{\ell}}\mathcal{R}_{\ell}
\end{array}\right).
\end{align}
Finally, the transformation for $\bm{\mathcal{G}}_{i}$ is given by
\begin{equation}
\mathcal{G}'_{\ell}=\frac{1}{n_{\ell}+e^{-2\mu_{\ell}}\left(1-n_{\ell}\right)}\left(\begin{array}{cc}
n_{\ell} & e^{-2\mu_{\ell}}\left(1-n_{\ell}\right)\\
-n_{\ell} & n_{\ell}
\end{array}\right).
\end{equation}

We proceed by exploring a particular choice of gauge for the SPD,
referred to as the anti-symmetric gauge, which will motivate the gauge
choice for the case of $\mathcal{N}=3$. The essence of the anti-symmetric
gauge is to choose $\mu_{\ell}$ such that $\left[g'_{\ell}\right]_{12}=-\left[g'_{\ell}\right]_{21}$,
which can be accomplished as 
\begin{equation}
\mu_{\ell}=\frac{1}{2}\ln\frac{1-n_{\ell}}{n_{\ell}}.\label{eq:mul_antisymmetric}
\end{equation}
 Under the anti-symmetric gauge, we have 

\begin{align}
 & \mathcal{G}'_{\ell}=\left(\begin{array}{cc}
\frac{1}{2} & \frac{1}{2}\\
-\frac{1}{2} & \frac{1}{2}
\end{array}\right),\label{eq:mathcalg antisymetric n2}\\
 & S_{\ell}'=\frac{1}{n_{\ell}+\left(1-n_{\ell}\right)\mathcal{R}_{\ell}^{2}}\\
 & \times\left(\begin{array}{cc}
\sqrt{\frac{1}{n_{\ell}}-1}\mathcal{R}_{\ell} & \left(n_{\ell}-1\right)\left(\mathcal{R}_{\ell}^{2}-1\right)\\
-\left(n_{\ell}-1\right)\left(\mathcal{R}_{\ell}^{2}-1\right) & \sqrt{\frac{1}{n_{\ell}}-1}\mathcal{R}_{\ell}
\end{array}\right),\label{eq:sl antisymmetric}\\
 & g'_{\ell}=\left(\begin{array}{cc}
n_{\ell} & \sqrt{\left(1-n_{\ell}\right)n_{\ell}}\mathcal{R}_{\ell}\\
-\sqrt{\left(1-n_{\ell}\right)n_{\ell}}\mathcal{R}_{\ell} & n_{\ell}
\end{array}\right).\label{eq:gl antisymmetric}
\end{align}
The anti-symmetric gauge can also automatically satisfy the SCDA self-consistency
condition, and be used to derive the qubit energy form.

\section{\label{subsec:B-ansatz}Derivation of the qubit energy form for $\hat{\rho}_{B2}$}

In this section, we derive Eq. (\ref{eq:energybanstz}) using several
different approaches. In Section \ref{subsec:n2b cpe}, we present
two derivations: a heuristic approach and the central point expansion
(CPE). In Section \ref{subsec:The-derivation-of-SCDA-rhoB2}, we use
the SCDA to evaluate $\hat{\rho}_{B2}$, and the tensor product representation
is used to obtain the qubit energy form.

\subsection{\label{subsec:n2b cpe}Derivation of the qubit energy form: a heuristic
approach and the CPE}

To begin, it is useful to rewrite the local interaction Hamiltonian
given in Eq. (\ref{eq:Hloc polynomial}) as
\begin{equation}
H_{loc}\left(\left\{ \hat{n}_{i\ell}\right\} \right)=\sum_{I}E_{I}\delta\hat{D}_{iI},\label{eq:fluctuation_form_H_int}
\end{equation}
where $\delta\hat{D}_{iI}=\prod_{\ell\in I}\delta\hat{n}_{i\ell}$,
the density fluctuation is defined as $\delta\hat{n}_{i\ell}=\hat{n}_{i\ell}-n_{\ell}$,
the index $I$ enumerates all possible subsets of the local spin orbitals,
with $\delta\hat{D}_{iI}=1$ when $I=\left\{ \right\} $, and the
parameters $E_{I}$ reparametrize the coefficients in Eq. (\ref{eq:Hloc polynomial}).
Using the alternative form of the local interaction Hamiltonian given
in Eq. (\ref{eq:fluctuation_form_H_int}), the corresponding qubit
form of the trial energy is given as 

\begin{align}
 & E\left(\left\{ n_{k\ell}\right\} ,\rhoeff\right)=\sum_{\ell}\int dk\epsilon_{k\ell}n_{k\ell}+\sum_{I}E_{I}\delta D_{I},\label{eq:alternative_qubit_form_rhoB2}\\
 & \delta D_{I}=\left(\prod_{\ell\in I}\mathcal{F}_{\ell}\right)\langle\delta\hat{D}_{I}\rangle_{\rhoeff},\label{eq:renormalization n2b}\\
 & \mathcal{F}_{\ell}=\frac{A_{\ell}^{2}}{\xi_{\ell,0}^{2}}=\left(\frac{\int dk\sqrt{n_{k\ell}\left(1-n_{k\ell}\right)}}{\sqrt{n_{\ell}\left(1-n_{\ell}\right)}}\right)^{2},
\end{align}
where the variational parameters $n_{k\ell}\in\left[0,1\right]$,
the $\rhoeff$ is a many-body density matrix for a $2N_{orb}$ qubit
system which is diagonal in the Pauli-Z basis, $\delta\hat{D}_{I}=\prod_{\ell\in I}(\hat{n}_{\ell}-n_{\ell})$,
and $n_{\ell}=\int dkn_{k\ell}$. For a given $\ell$, there is constraint
given by 
\begin{equation}
\int dkn_{k\ell}=\left\langle \hat{n}_{\ell}\right\rangle _{\rhoeff}.
\end{equation}
Below, we present two approaches for deriving Eq. (\ref{eq:alternative_qubit_form_rhoB2}).

A heuristic approach for deriving Eq. (\ref{eq:alternative_qubit_form_rhoB2})
is via the formal duality between $\hat{\rho}_{G2}$ and $\hat{\rho}_{B2}$.
First, in $\hat{\rho}_{G2}$, the center projector $\hat{K}_{2}$
is constrained to have the same local density as $\hat{\rho}_{G2}$,
i.e., $\left\langle \hat{n}_{i\ell}\right\rangle _{\hat{K}_{2}}=\left\langle \hat{n}_{i\ell}\right\rangle _{\hat{\rho}_{G2}}$.
Similarly, the center projector $\hat{P}_{1}$ in $\hat{\rho}_{B2}$
is constrained to have the same local density as $\hat{\rho}_{B2}$,
i.e., $\left\langle \hat{n}_{i\ell}\right\rangle _{\hat{P}_{1}}=\left\langle \hat{n}_{i\ell}\right\rangle _{\hat{\rho}_{B2}}$.
Second, in $\hat{\rho}_{G2}$ (see Eq. (\ref{eq:energygansatz})),
the momentum density fluctuation $\delta n_{k\ell}=n_{k\ell}-n_{\ell}$
is renormalized from the bare momentum density fluctuation $\delta n_{k\ell;0}=n_{k\ell;0}-n_{\ell}$
with a factor $\mathcal{Z}=\xi_{\ell}^{2}/\xi_{\ell;0}^{2}$, which
depends on $\rhoeff$. To the contrary, $\langle\delta\hat{D}_{iI}\rangle_{\hat{\rho}_{B2}}$
is renormalized from the reference value $\langle\delta\hat{D}_{iI}\rangle_{\hat{P}_{1}}$
with a factor $\prod_{\ell\in I}\mathcal{F_{\ell}}$, where $\mathcal{F}_{\ell}$
depends on $n_{k\ell}$. The expression for $\mathcal{F}_{\ell}$
can be determined using a counting scheme similar to the one used
within the GA. We begin by transforming $\delta\hat{D}_{iI}$ into
momentum space, which requires a summation over terms consisting of
$N_{I}$ creation and $N_{I}$ annihilation operators, where $N_{I}$
is the number of spin orbitals in set $I$. Each creation or annihilation
process for $k\ell$ will be scaled by 
\begin{align}
 & \frac{\sum_{\Gamma\Gamma'}\sqrt{\left\langle \hat{X}_{k\ell\Gamma}\right\rangle _{\hat{\rho}_{B2}}\left\langle \hat{X}_{k\ell\Gamma'}\right\rangle _{\hat{\rho}_{B2}}}\left|\left\langle \Gamma'|\hat{a}_{k\ell}^{\dagger}|\Gamma\right\rangle \right|^{2}}{\sum_{\Gamma\Gamma'}\sqrt{\left\langle \hat{X}_{k\ell\Gamma}\right\rangle _{\hat{P}_{1}}\left\langle \hat{X}_{k\ell\Gamma'}\right\rangle _{\hat{P}_{1}}}\left|\left\langle \Gamma'|\hat{a}_{k\ell}^{\dagger}|\Gamma\right\rangle \right|^{2}}\nonumber \\
 & =\frac{\sqrt{n_{k\ell}\left(1-n_{k\ell}\right)}}{\sqrt{n_{\ell}\left(1-n_{\ell}\right)}},
\end{align}
where $\Gamma\in\{1,2\}$ enumerates the empty and occupied states
for a given $k\ell$ and 
\begin{align}
 & \hat{X}_{k\ell\Gamma}=\delta_{\Gamma-1,0}(1-\hat{n}_{k\ell})+\delta_{\Gamma-1,1}\hat{n}_{k\ell}.
\end{align}
Using the infinite dimensional approximation where momentum conservation
can be neglected, we obtain $\langle\delta\hat{D}_{iI}\rangle_{\hat{\rho}_{B2}}=\left(\prod_{\ell\in I}\mathcal{F}_{\ell}\right)\langle\delta\hat{D}_{iI}\rangle_{\hat{P}_{1}}$.
By identifying $\rhoeff$ as the local reduced density matrix of $\hat{P}_{1}$,
we obtain the form given in Eq. (\ref{eq:renormalization n2b}).

A more rigorous way to derive the factor in Eq. (\ref{eq:renormalization n2b})
is to use the CPE. We outline the derivation here, and the details
can be found in Appendix \ref{subsec:Appendix-the-CPE-for-B}. The
CPE for $\hat{\rho}_{B2}$ is a dual version to the CPE of $\hat{\rho}_{G2}$,
where the latter is described in Section \ref{subsec:heuristic-argument-GA}.
We start by defining the central point as $\hat{\rho}_{B2}^{\star}=\hat{K}_{1}\hat{P}_{1}^{\star}\hat{K}_{1}$,
where $\hat{P}_{1}^{\star}$ is a non-interacting projector chosen
such that $n_{\ell}\equiv\left\langle \hat{n}_{i\ell}\right\rangle _{\hat{P}_{1}}=\left\langle \hat{n}_{i\ell}\right\rangle _{\hat{P}_{1}^{\star}}$.
The kinetic projector $\hat{K}_{1}$ can be parametrized using $n_{k\ell}^{\star}=\langle\hat{n}_{k\ell}\rangle_{\hat{\rho}_{B2}^{\star}}$
with the constraint$\int dkn_{k\ell}^{\star}=n_{\ell},$ while $\hat{P}_{1}$
can be parametrized using $\delta D_{iI;0}=\langle\delta\hat{D}_{iI}\rangle_{\hat{P}_{1}}$
and $\{n_{\ell}\}$. Therefore, the observables under $\hat{\rho}_{B2}$
are functionals of $n_{k\ell}^{\star}$ and $\left\{ \delta D_{iI;0}\right\} $.
Performing a first order expansion in terms of $\left\{ \delta D_{iI;0}\right\} $
about $\left\{ \delta D_{iI;0}=0\right\} $, we obtain $n_{k\ell}=n_{k\ell}^{\star}$
and 
\begin{equation}
\delta D_{iI}=\left(\prod_{\ell\in I}\mathcal{F}_{\ell}\right)\delta D_{iI;0}.\label{eq: fluctation form rhoB2}
\end{equation}

Finally, we explain how to obtain Eqns. (\ref{eq:energybanstz})-(\ref{eq:Aell rhob2})
from Eq. (\ref{eq: fluctation form rhoB2}). We begin by defining
the effective density operator $\hat{n}_{eff,\ell}=n_{\ell}+\mathcal{F}_{\ell}\delta\hat{n}_{\ell}$
where $\delta\hat{n}_{\ell}=\hat{n}_{\ell}-n_{\ell}$, the corresponding
fluctuation form $\delta\hat{n}_{eff,\ell}\equiv\hat{n}_{eff,\ell}-\hat{n}_{\ell}=\mathcal{F}_{\ell}\delta\hat{n}_{\ell}$,
and a diagonal many-body density matrix $\rhoeff$ for a $2N_{orb}$
qubit system $\left[\rhoeff\right]_{\Gamma\Gamma'}=\delta_{\Gamma\Gamma'}\langle\hat{X}_{i\Gamma}\rangle_{\hat{P}_{1}}$.
Using $\delta D_{iI,0}=\left\langle \prod_{i\in I}\delta\hat{n}_{\ell}\right\rangle _{\rhoeff}$
and Eq. (\ref{eq: fluctation form rhoB2}), we have $\delta D_{iI}=\left\langle \prod_{i\in I}\delta\hat{n}_{eff,\ell}\right\rangle _{\rhoeff}$.
Furthermore, $n_{\ell}=\left\langle \hat{n}_{\ell}\right\rangle _{\rhoeff}$
implies that $\left\langle \delta\hat{n}_{\ell}\right\rangle _{\rhoeff}=\left\langle \delta\hat{n}_{eff,\ell}\right\rangle _{\rhoeff}=0$.
Therefore, the expectation value of $H_{loc}\left(\left\{ \hat{n}_{i\ell}\right\} \right)$
is given by $\left\langle H_{loc}\left(\left\{ \hat{n}_{eff,\ell}\right\} \right)\right\rangle _{\rhoeff}$,
providing the connection to Eqns. (\ref{eq:energybanstz})-(\ref{eq:Aell rhob2}).

\subsection{The derivation of the qubit energy form via the SCDA\label{subsec:The-derivation-of-SCDA-rhoB2}}

In this section, we use the gauge constrained SCDA to derive the qubit
energy form. The derivation consists of two steps. First, in Section
\ref{subsec:SCDA-for-B-ansatz}, we use the Gutzwiller gauge to automatically
satisfy the SCDA self-consistency condition. Second, in Section \ref{subsec:Deriving-the-density-fluctuation},
we derive the qubit energy form using the tensor representation. Additionally,
in Section \ref{subsec:Gauge-symmetry-of}, we discuss how the quantities
$\boldsymbol{\mathcal{G}}$ , $\boldsymbol{S}$, and $\boldsymbol{g}$
change under a general gauge transformation. 

\subsubsection{\label{subsec:SCDA-for-B-ansatz}SCDA within the Gutzwiller gauge}

In this section, we use the gauge constrained SCDA to evaluate $\hat{\rho}_{B2}=\hat{K}_{1}\hat{P}_{1}\hat{K}_{1}$,
where $\hat{P}_{1}=\exp\left(\sum_{i\Gamma}\upsilon_{\Gamma}\hat{X}_{i\Gamma}\right)$
and $\hat{K}_{1}=\exp\left(\sum_{k\ell}\gamma_{k\ell}\hat{n}_{k\ell}\right)$,
recovering the results from Section \ref{subsec:n2b cpe}. The key
observation is that for $\hat{\rho}_{B2}$, the interacting projectors
only act on the first integer time step. Consider a gauge transformation
for $\hat{\rho}_{B2}$ given as $\hat{K}_{1}\rightarrow\hat{K}_{1}\hat{N}$
and $\hat{P_{1}}\rightarrow\hat{N}^{-1}\hat{P}_{1}\hat{N}^{-1}$ where
$\hat{N}=\exp\left(\sum_{i\ell}\mu_{\ell}\hat{n}_{i\ell}\right)$.
We choose the gauge transformation such that $\bm{\mathcal{G}}_{i}$
is given as $\left[\bm{\mathcal{G}}_{i}\right]_{\ell\ell'}=\delta_{\ell\ell'}\mathcal{G}_{\ell}$,
where 
\begin{equation}
\mathcal{G}_{\ell}=\left(\begin{array}{cc}
n_{\ell} & \mathcal{G}_{\ell,1,2}\\
\mathcal{G}_{\ell,2,1} & \mathcal{G}_{\ell,2,2}
\end{array}\right),\label{eq:g0 b2}
\end{equation}
and $n_{\ell}=\left\langle \hat{n}_{i\ell}\right\rangle _{\hat{\rho}_{B2;0}}$
is the local density for the non-interacting SPD $\hat{\rho}_{B2;0}=\hat{K}_{1}\hat{K_{1}}$.
Under this gauge choice, $\hat{P}_{1}$ will ensure that $\left[\bm{g}_{i}\right]_{\ell\ell'}=\delta_{\ell\ell'}g_{\ell}$
with $\left[g_{\ell}\right]_{11}=\left[\mathcal{G}_{\ell}\right]_{11}=n_{\ell}$.
The integer time self-energy $\left[\bm{S}_{i}\right]$$_{\ell\ell'}=\delta_{\ell\ell'}S_{\ell}$
can then be determined as 
\begin{equation}
S_{\ell}=\left(\begin{array}{cc}
1 & 0\\
0 & 1
\end{array}\right).\label{eq:Sl b2}
\end{equation}
The remaining entries of Eq. (\ref{eq:g0 b2}) should be determined
from the self-consistency condition in Eq. (\ref{eq:self-consistency g g'}).
Given that $S_{\ell}$ is the identity matrix, the non-interacting
and interacting integer time Green's functions are given by $\left[\bm{g}_{k;0}\right]_{\ell\ell'}=\delta_{\ell\ell'}g_{k\ell;0}$
and $\left[\bm{g}_{k}\right]_{\ell\ell'}=\delta_{\ell\ell'}g_{k\ell}$
where 

\begin{equation}
g_{k\ell;0}=g_{k\ell}=\left(\begin{array}{cc}
n_{k\ensuremath{\ell}} & \sqrt{\left(1-n_{k\ensuremath{\ell}}\right)n_{k\ensuremath{\ell}}}\\
-\sqrt{\left(1-n_{k\ensuremath{\ell}}\right)n_{k\ensuremath{\ell}}} & n_{k\ensuremath{\ell}}
\end{array}\right),\label{eq:gkl in nkl b2}
\end{equation}
and $n_{k\ell}$ is the momentum density distribution. We can then
verify that the self-consistency condition $\bm{g}_{i}=\left(1/N_{\textrm{site}}\right)\sum_{k}\bm{g}_{k}$
is fulfilled, given that 
\begin{equation}
g_{\ell}=\mathcal{G}_{\ell}=\left(\begin{array}{cc}
n_{\ell} & A_{\ell}\\
-A_{\ell} & n_{\ell}
\end{array}\right),\label{eq:gl b2}
\end{equation}
where $n_{\ell}=\left(1/N_{site}\right)\sum_{k}n_{k\ell}$ and the
$A_{\ell}$ is defined as 
\begin{equation}
A_{\ell}=\frac{1}{N_{site}}\sum_{k}\sqrt{\left(1-n_{k\ensuremath{\ell}}\right)n_{k\ensuremath{\ell}}}.\label{eq:Al b2}
\end{equation}
We analogously refer to this gauge as the Gutzwiller gauge given that
$g_{\ell}=\mathcal{G}_{\ell}$, and therefore the diagonal elements
of the two matrices are the same.

\subsubsection{\label{subsec:Deriving-the-density-fluctuation}Derivation of the
qubit energy form }

In this section, we use the tensor product representation to derive
the qubit energy form, paralleling the procedure in Section \ref{subsec:Derivation-of-qubit-SCDA}.
Given that we only have interacting projectors at the first integer
time step, operators in the $u$-representation $\left(\barhat[O]\right)_{u}$
can be reduced from a matrix to a vector, given as 
\begin{equation}
\left[\left(\barhat[O]\right)_{u}\right]_{\Gamma}=\langle\barhat[X]_{i\Gamma}^{\left(1\right)}\barhat[O]\rangle_{\barhat[\rho]_{loc,i;0}},
\end{equation}
and the expectation value of $\barhat[O]$ at site $i$ is given as
\begin{equation}
\langle\barhat[O]\rangle_{\barhat[\rho]_{loc;i}}=\frac{\left(\barhat[O]\right)_{u}\cdot u}{\left(\barhat[1]\right)_{u}\cdot u},
\end{equation}
where $u=u_{1}$ and the dot denotes the normal dot product between
two vectors. Given the direct product structure of $\left(\barhat[O]\right)_{u}$,
we only need to compute the component for a given spin-orbital. Here,
we only list the relevant matrix elements needed to derive the $w$-representation,
given as 
\begin{align}
 & \left(\barhat[1]\right)_{u;\ell}=\left(\begin{array}{cc}
1-n_{\ell} & n_{\ell}\end{array}\right),\\
 & \left(\barhat[a]_{\ell}^{\dagger\left(1\right)}\barhat[a]_{\ell}^{\left(1\right)}\right)_{u;\ell}=\left(\begin{array}{cc}
0 & n_{\ell}\end{array}\right),\\
 & \left(\barhat[a]_{\ell}^{\dagger\left(2\right)}\barhat[a]_{\ell}^{\left(2\right)}\right)_{u;\ell}=\left(\begin{array}{cc}
-A_{\ell}^{2}-n_{\ell}^{2}+n_{\ell} & A_{\ell}^{2}+n_{\ell}^{2}\end{array}\right).
\end{align}
The $u$-representation and $w$-representation are related by $u=Vw$
such that 
\begin{align}
 & \left(\barhat[O]\right)_{w}w=\left(\barhat[O]\right)_{u}u,
\end{align}
and therefore
\begin{align}
 & \left(\barhat[O]\right)_{w}=\left(\barhat[O]\right)_{u}V.
\end{align}
When $V$ has a direct product form $V=V_{1}\otimes\dots\otimes V_{2N_{orb}},$
then 
\begin{equation}
\left(\barhat[O]\right)_{w;\ell}=\left(\barhat[O]\right)_{u;\ell}V_{\ell},
\end{equation}
and $V_{\ell}$ is chosen to obtain $\left(\barhat[1]\right)_{w;\ell}$
as a vector of ones. The matrix elements are given as

\begin{align}
 & V_{\ell}=\left(\begin{array}{cc}
\frac{1}{1-n_{\ell}} & 0\\
0 & \frac{1}{n_{\ell}}
\end{array}\right),\\
 & \left(\barhat[1]\right)_{w;\ell}=\left(\begin{array}{cc}
1 & 1\end{array}\right),\\
 & \left(\barhat[a]_{\ell}^{\dagger\left(1\right)}\barhat[a]_{\ell}^{\left(1\right)}\right)_{w;\ell}=\left(\begin{array}{cc}
0 & 1\end{array}\right),\\
 & \left(\barhat[a]_{\ell}^{\dagger\left(2\right)}\barhat[a]_{\ell}^{\left(2\right)}\right)_{w;\ell}=\left(\begin{array}{cc}
\frac{A_{\ell}^{2}}{n_{\ell}-1}+n_{\ell} & \frac{A_{\ell}^{2}}{n_{\ell}}+n_{\ell}\end{array}\right).
\end{align}
From the preceding equations, we have 

\begin{align}
 & \left(\barhat[a]_{\ell}^{\dagger\left(2\right)}\barhat[a]_{\ell}^{\left(2\right)}-n_{\ell}\barhat[1]\right)_{w;\ell}=\left(\barhat[a]_{\ell}^{\dagger\left(1\right)}\barhat[a]_{\ell}^{\left(1\right)}-n_{\ell}\barhat[1]\right)_{w;\ell}\mathcal{F}_{\ell},\label{eq:wrep_Fell}
\end{align}
where 
\begin{equation}
\mathcal{F}_{\ell}=\frac{A_{\ell}^{2}}{\left(1-n_{\ell}\right)n_{\ell}},
\end{equation}
and therefore we have 
\begin{equation}
\left\langle \delta\barhat[D]_{iI}^{\left(2\right)}\right\rangle _{\barhat[\rho]_{loc;i}}=\left(\prod_{\ell\in I}\mathcal{F}_{\ell}\right)\left\langle \delta\barhat[D]_{iI}^{\left(1\right)}\right\rangle _{\barhat[\rho]_{loc;i}}.
\end{equation}
We now proceed to reinterpret the local energy in terms of the qubit
representation, identifying $\left(\barhat[a]_{\ell}^{\dagger\left(1\right)}\barhat[a]_{\ell}^{\left(1\right)}\right)_{w}=\textrm{diag}(\hat{n}_{\ell})$
and $\left(\barhat[a]_{\ell}^{\dagger\left(2\right)}\barhat[a]_{\ell}^{\left(2\right)}\right)_{w}=\textrm{diag}(\hat{n}_{eff,\ell})$,
Eq. (\ref{eq:wrep_Fell}) becomes
\begin{align}
 & \hat{n}_{eff,\ell}-n_{\ell}=\left(\hat{n}_{\ell}-n_{\ell}\right)\mathcal{F}_{\ell}.
\end{align}
Furthermore, we define a diagonal many-body density matrix $\rhoeff$
with $\textrm{diag}(\rhoeff)=w$, resulting in
\begin{align}
\langle H_{loc}(\{\barhat[n]_{\ell}^{(2)}\})\rangle_{\barhat[\rho]_{loc;i}} & =\frac{\left(H_{loc}(\{\barhat[n]_{\ell}^{(2)}\})\right)_{w}\cdot w}{\left(\barhat[1]\right)_{w}\cdot w}\\
 & =\langle H_{loc}(\{n{}_{eff,\ell}\})\rangle_{\rhoeff}.
\end{align}
Therefore, the qubit energy form has been recovered.

\subsubsection{\label{subsec:Gauge-symmetry-of}SCDA under a general gauge transformation}

Here we discuss the SCDA within an arbitrary gauge, analogous to Section
\ref{subsec:SCDA-under-gauge}. Given $\hat{\rho}_{B2}=\hat{K}_{1}\hat{P}_{1}\hat{K}_{1}^{\dagger}$,
a general gauge transformation is given as: $\hat{K}_{1}\rightarrow\hat{K}_{1}\hat{N}$,
$\hat{P}_{1}\rightarrow\hat{N}^{-1}\hat{P}_{1}\left(\hat{N}^{\dagger}\right)^{-1}$,
$\hat{K}_{1}^{\dagger}\rightarrow\hat{N}^{\dagger}\hat{K}_{1}^{\dagger}$,
where $\hat{N}=\exp\left(\bm{\mu}\cdot\hat{\bm{n}}\right)$. Therefore,
we have $\bm{N}{}_{a}=\text{diag}\left(\exp\left(-\bm{\mu}^{T}\right),\boldsymbol{1}\right)$
and $\bm{N}_{b}=\text{diag}\left(\exp\left(-\bm{\mu}^{*}\right),\boldsymbol{1}\right)$.
Assuming $\left[\bm{\mu}\right]_{\ell\ell'}=\delta_{\ell\ell'}\mu_{\ell}$,
where $\mu_{\ell}$ is real, we can obtain the transformation for
$g_{\ell}$, $S_{\ell}$, and $\mathcal{G}{}_{\ell}$ as 

\begin{align}
g'_{\ell} & =\left(\begin{array}{cc}
n_{\ell} & A_{\ell}e^{-\mu_{\ell}}\\
-A_{\ell}e^{\mu_{\ell}} & n_{\ell}
\end{array}\right),\\
S'_{\ell} & =\left(\begin{array}{cc}
e^{2\mu_{\ell}} & 0\\
0 & 1
\end{array}\right),\\
\mathcal{G}'_{\ell} & =\frac{1}{\left(e^{2\mu_{\ell}}-1\right)n_{\ell}+1}\label{eq:gp in n2}\\
 & \times\left(\begin{array}{cc}
e^{2\mu_{\ell}}n_{\ell} & A_{\ell}e^{\mu_{\ell}}\\
A_{\ell}\left(-e^{\mu_{\ell}}\right) & \left(e^{2\mu_{\ell}}-1\right)\left(A_{\ell}^{2}+n_{\ell}^{2}\right)+n_{\ell}
\end{array}\right).
\end{align}
We can define an anti-symmetric gauge, similar to the $\hat{\rho}_{G2}$
case, where we choose $\left[\mathcal{G}_{\ell}'\right]_{11}=1/2$,
leading to $\mu_{\ell}=\frac{1}{2}\ln\left(\frac{1-n_{\ell}}{n_{\ell}}\right)$,
yielding 

\begin{align}
g_{\ell}'= & \left(\begin{array}{cc}
n_{\ell} & A_{\ell}/\sqrt{\frac{1}{n_{\ell}}-1}\\
-A_{\ell}\sqrt{\frac{1}{n_{\ell}}-1} & n_{\ell}
\end{array}\right),\\
S_{\ell}'= & \left(\begin{array}{cc}
\frac{1-n_{\ell}}{n_{\ell}} & 0\\
0 & 1
\end{array}\right),\\
\mathcal{G}'_{\ell}= & \left(\begin{array}{cc}
\frac{1}{2} & \frac{A_{\ell}}{2\sqrt{-\left(n_{\ell}-1\right)n_{\ell}}}\\
-\frac{A_{\ell}}{2\sqrt{-\left(n_{\ell}-1\right)n_{\ell}}} & \frac{A_{\ell}^{2}\left(2n_{\ell}-1\right)}{2\left(n_{\ell}-1\right)n_{\ell}}+n_{\ell}
\end{array}\right).
\end{align}
The anti-symmetric gauge can also automatically satisfy the SCDA self-consistency
condition, and be used to derive the qubit energy form.

\section{\label{subsec:GB-ansatz}Derivation of the qubit energy form for
$\hat{\rho}_{G3}$}

In this section, we derive Eq. (\ref{eq:energy_n3_qubit}) using the
gauge constrained SCDA. In section \ref{subsec:Comparing-original-gauge},
we provide a high level comparison of the original gauge constrained
algorithm and the qubit parametrization. In section \ref{subsec:Review-of-the-gauge-const-SCDA},
we provide a review of the original gauge constrained algorithm, which
is necessary to understand the qubit parametrization in this work.
In section \ref{subsec:Derivation-of-qubit-N3}, we propose the qubit
parametrization which yields the qubit energy form. In section \ref{subsec:Examining-special-cases},
we examine the case of half-filling, and explore how the qubit energy
form for $\hat{\rho}_{G3}$ can recover the cases of $\hat{\rho}_{G2}$
and $\hat{\rho}_{B2}$.

\subsection{\label{subsec:Comparing-original-gauge}Comparing the original gauge
constrained trial energy to the qubit energy form}

In this section, we outline how the qubit parametrization improves
the original gauge constrained algorithm \cite{Cheng2023035127}.
In the original gauge constrained algorithm, the trial energy under
$\hat{\rho}_{G3}$ is given as 
\begin{equation}
\sum_{\ell}\int dk\epsilon_{k\ell}n_{k\ell}+E_{loc}\left(\left\{ \mathcal{G}_{12,\ell}\right\} ,\{u_{\Gamma}\},\left\{ \mathcal{A}_{<\ell}\right\} ,\left\{ \mathcal{A}_{>\ell}\right\} \right),\label{eq:energy_unminimized}
\end{equation}
which is a function of $\{n_{k\ell}\}$, $\left\{ \mathcal{G}_{12,\ell}\right\} $,
$\{u_{\Gamma}\}$, and $\{n_{k\ell;0}\}$, given that $\left\{ \mathcal{A}_{<\ell}\right\} $
and $\left\{ \mathcal{A}_{>\ell}\right\} $ are functions of $\{n_{k\ell}\}$
and $\{n_{k\ell;0}\}$. It should be noted that $n_{k\ell;0}\in\{0,1\}$
and $n_{k\ell}\in[0,1]$, and there are three constraints for a given
$\ell$

\begin{align}
 & \int_{<}dk=n_{\ell}\left(\left\{ \mathcal{G}_{12;\ell}\right\} ,\{u_{\Gamma}\}\right),\label{eq:linearc1}\\
 & \int n_{k\ell}dk=n_{\ell}\left(\left\{ \mathcal{G}_{12;\ell}\right\} ,\{u_{\Gamma}\}\right),\label{eq:linearc2}\\
 & \Delta_{\ell}\equiv\int_{>}n_{k\ell}dk=\Delta_{\ell}\left(\left\{ \mathcal{G}_{12;\ell}\right\} ,\{u_{\Gamma}\}\right),\label{eq:linearc3}
\end{align}
where the functions $n_{\ell}\left(\left\{ \mathcal{G}_{12;\ell}\right\} ,\{u_{\Gamma}\}\right)$
and $\Delta_{\ell}\left(\left\{ \mathcal{G}_{12;\ell}\right\} ,\{u_{\Gamma}\}\right)$
are explicitly defined in Ref. \cite{Cheng2023035127}. While this
parametrization allows for an explicit evaluation of the total energy,
there are several shortcomings of this parametrization. First, the
function $n_{\ell}\left(\left\{ \mathcal{G}_{12;\ell}\right\} ,\{u_{\Gamma}\}\right)$
is highly non-trivial and thus the minimization under a fixed density
is cumbersome. Previously, this problem was addressed by introducing
a linear transformation over $\{u_{\Gamma}\}$, known as the $w$
representation. Second, the function $\Delta_{\ell}\left(\left\{ \mathcal{G}_{12;\ell}\right\} ,\{u_{\Gamma}\}\right)$
may yield a value outside the allowed bounds for $\Delta_{\ell}$.
This problem can be addressed by imposing appropriate restrictions
on $\left\{ \mathcal{G}_{12;\ell}\right\} $. Finally, $\left\{ \mathcal{G}_{12;\ell}\right\} $
makes the physical interpretation of the total energy expression somewhat
obscure. 

In this paper, the aforementioned shortcomings are resolved using
the qubit parametrization (see Section \ref{sec:Overview-of-the}).
The qubit parametrization has two important differences. First, the
qubit parametrization employs an effective many-body density matrix
$\rhoeff,$ having dimension $2^{2N_{orb}}\times2^{2N_{orb}}$, corresponding
to a pure state of a $2N_{orb}$ qubit system. The $\rhoeff$ is constructed
such that the density of $\rhoeff$ is the same as the physical local
density, and it can be viewed as a function of $\{u_{\Gamma}\}$ and
$\left\{ \mathcal{G}_{12,\ell}\right\} $. Second, the qubit parametrization
uses $\Delta_{\ell}\left(\left\{ \mathcal{G}_{12;\ell}\right\} ,\{u_{\Gamma}\}\right)$
to solve $\left\{ \mathcal{G}_{12,\ell}\right\} $ as a function of
$\rhoeff$ and $\left\{ \Delta_{\ell}\right\} $, reducing the total
number of constraints per spin orbital from three to two. 

\subsection{Review of the gauge constrained SCDA algorithm\label{subsec:Review-of-the-gauge-const-SCDA}}

In this section, we use original gauge constrained SCDA to evaluate
$\hat{\rho}_{G3}$. It should be noted that there are several restrictions
on the variational freedom of the SPD when using the gauge constrained
SCDA. First, the kinetic projector must be diagonal in momentum space.
Second, the interacting projector may not introduce off-diagonal terms
at the single-particle level. These two restrictions guarantee that
the local integer time self-energy and Green's function are diagonal
in the original basis. For Hamiltonians with density-density interactions
and hopping parameters that are diagonal in the orbital index, such
as the ones treated in this paper, the aforementioned variational
restrictions do not limit the variational power of the SPD. There
are two critical insights in the gauge constrained algorithm. First,
the integer time self-energy only has non-trivial values within the
time steps containing the interacting projector, and therefore $\bm{\mathcal{G}}$
only needs to be specified in the corresponding regions. Second, the
gauge symmetry can be used to restrict the form of $\bm{\mathcal{G}}$. 

We start by examining the gauge symmetry of $\hat{\rho}_{G3}=\hat{K}_{1}\hat{P}_{1}\hat{K}_{2}\hat{P}_{1}^{\dagger}\hat{K}_{1}^{\dagger}$,
where the gauge transformation is given by $\hat{K}_{1}\rightarrow\hat{K}_{1}\hat{N}_{1}$,
$\hat{P}_{1}\rightarrow\hat{N}_{1}^{-1}\hat{P}_{1}\hat{N}_{2}^{-1}$,
and $\hat{K}_{2}\rightarrow\hat{N}_{2}\hat{K}_{2}\hat{N}_{2}^{\dagger}$,
where $\hat{N}_{1}=\exp\left(\bm{\mu}_{1}\cdot\hat{\bm{n}}\right)$
and $\hat{N}_{2}=\exp\left(\bm{\mu}_{2}\cdot\hat{\bm{n}}\right)$.
The gauge transformation can be parametrized by 
\begin{align}
\bm{N}_{a} & =\text{diag}\left(\exp\left(-\bm{\mu}_{1}^{T}\right),\exp\left(-\bm{\mu}_{2}^{*}\right),1\right),\\
\bm{N}_{b} & =\text{diag}\left(\exp\left(-\bm{\mu}_{2}^{T}\right),\exp\left(-\bm{\mu}_{1}^{\star}\right),1\right),
\end{align}
as explained in Section \ref{subsec:Gauge-Symmetry-of-SPD}. In the
following, we assume $\left[\bm{\mu}_{i}\right]_{\ell\ell'}=\delta_{\ell\ell'}\mu_{i,\ell}$
where $\mu_{i,\ell}$ is a real number, yielding 
\begin{align}
 & g'_{\ell}=\text{diag}\left(e^{-\mu_{2,\ell}},e^{-\mu_{1,\ell}},1\right)g_{\ell}\text{diag}\left(e^{\mu_{2,\ell}},e^{\mu_{1,\ell}},1\right),\label{eq:g l}\\
 & S'_{\ell}=\text{diag}\left(e^{\mu_{1,\ell}},e^{\mu_{2,\ell}},1\right)S_{\ell}\text{diag}\left(e^{\mu_{2,\ell}},e^{\mu_{1,\ell}},1\right),\label{eq: S l}\\
 & \left(\mathcal{G}'_{\ell}\right)^{-1}-1=\text{diag}\left(e^{-\mu_{2,\ell}},e^{-\mu_{1,\ell}},1\right)\nonumber \\
 & \hspace{3em}\times\left(\mathcal{G}{}_{\ell}^{-1}-1\right)\text{diag}\left(e^{-\mu_{1,\ell}},e^{-\mu_{2,\ell}},1\right).\label{eq:mathcalG l}
\end{align}
Notice that the interacting projectors only act on the first and second
time step, and therefore it is useful to split $\mathcal{G}_{\ell}$
into the following block structure:

\begin{align}
\mathcal{G}_{\ell} & =\left(\begin{array}{cc|c}
[\mathcal{G}_{\ell}]_{11} & [\mathcal{G}_{\ell}]_{12} & [\mathcal{G}_{\ell}]_{13}\\{}
[\mathcal{G}_{\ell}]_{21} & [\mathcal{G}_{\ell}]_{22} & [\mathcal{G}_{\ell}]_{23}\\
\hline \,[\mathcal{G}_{\ell}]_{31} & [\mathcal{G}_{\ell}]_{32} & [\mathcal{G}_{\ell}]_{33}
\end{array}\right)=\begin{pmatrix}\mathcal{G}_{\ell;A} & \mathcal{G}_{\ell;B}\\
\mathcal{G}_{\ell;C} & \mathcal{G}_{\ell;D}
\end{pmatrix}.\label{eq:Gl block}
\end{align}
A similar block structure is adopted for $g_{\ell}$ and $S_{\ell}$. 

The first step is to focus on the $A$ block, which is sufficient
to determine the integer time self-energy. Similar to the case of
$\hat{\rho}_{B2}$, we only need to specify $\mathcal{G}_{\ell;A}$,
which is sufficient to determine $g_{\ell;A}$ and $S_{\ell;A}$ and
therefore $S_{\ell}$. Moreover, similar to the derivation of Eq.
(\ref{eq:mathcalG l}), we have
\begin{align}
\left(\mathcal{G}'_{\ell;A}\right)^{-1}-1 & =\text{diag}\left(e^{-\mu_{2,\ell}},e^{-\mu_{1,\ell}}\right)\nonumber \\
 & \times\left(\mathcal{G}{}_{\ell;A}^{-1}-1\right)\text{diag}\left(e^{-\mu_{1,\ell}},e^{-\mu_{2,\ell}}\right).\label{eq:mathcalG l A}
\end{align}
The gauge transformation can be used to further restrict the form
of $\mathcal{G}_{\ell;A}$. Given that Eq. (\ref{eq:mathcalG l A})
involves the inverse of $\mathcal{G}_{\ell;A}$, it is more convenient
to first use Eq. (\ref{eq:g l}). Notice that $\left[g_{\ell}\right]_{11}=\left[g_{\ell}\right]_{22}$
given translation symmetry and the fact that the total particle number
of a given spin-orbital $\ell$ commutes with $\hat{P}_{1}$. Therefore,
Eq. (\ref{eq:g l}) indicates that the diagonal elements of $g_{\ell}$
are invariant, while $\mu_{1;\ell}-\mu_{2;\ell}$ can be chosen such
that $\left[g'_{\ell}\right]_{12}=-\left[g'_{\ell}\right]_{21}$.
Notice that the interacting projectors are the same for the first
and second time step, and therefore $\left[\mathcal{G}_{\ell}'\right]_{11}=\left[\mathcal{G}_{\ell}'\right]_{22}$
and $\left[\mathcal{G}_{\ell}'\right]_{12}=-\left[\mathcal{G}_{\ell}'\right]_{21}$.
Now consider a gauge transformation with $\mu_{1;\ell}=\mu_{2;\ell}=\mu_{\ell}$,
which still preserves $\left[\mathcal{G}''_{\ell}\right]_{11}=\left[\mathcal{G}''_{\ell}\right]_{22}$
and $\left[\mathcal{G}''_{\ell}\right]_{12}=-\left[\mathcal{G}''_{\ell}\right]_{21}$,
and we can choose $\mu_{\ell}$ such that $\left[\mathcal{G}''_{\ell}\right]_{11}=1/2$.
Therefore, after fully exploring the gauge symmetry of the SPD, it
is sufficient to use just one parameter $\mathcal{G}_{\ell,12}$ to
parametrize $\mathcal{G}_{\ell;A}$ as 
\begin{equation}
\mathcal{G}_{\ell;A}=\left(\begin{array}{cc}
\frac{1}{2} & \mathcal{G}_{\ell,12}\\
-\mathcal{G}_{\ell,12} & \frac{1}{2}
\end{array}\right),\label{eq:GlA choice}
\end{equation}
 and correspondingly
\begin{equation}
g_{\ell;A}=\left(\begin{array}{cc}
n_{\ell} & g_{\ell,12}\\
-g_{\ell,12} & n_{\ell}
\end{array}\right).\label{eq:glA}
\end{equation}
It is useful to define a new quantity 
\begin{equation}
S_{0,\ell}^{\left(A\right)}\equiv\mathcal{G}_{\ell;A}^{-1}-1=\left(\begin{array}{cc}
\frac{2}{4\mathcal{G}_{\ell,12}^{2}+1}-1 & -\frac{4\mathcal{G}_{\ell,12}}{4\mathcal{G}_{\ell,12}^{2}+1}\\
\frac{\mathcal{G}_{\ell,12}}{\mathcal{G}_{\ell,12}^{2}+\frac{1}{4}} & \frac{2}{4\mathcal{G}_{\ell,12}^{2}+1}-1
\end{array}\right),\label{eq:s0lA}
\end{equation}
which is different from the $A$ block of $S_{0,\ell}=\mathcal{G}_{\ell}^{-1}-1$,
denoted as $S_{0,\ell;A}$. A similar quantity $S_{F,\ell}^{\left(A\right)}$
can also be defined as 
\begin{equation}
S_{F,\ell}^{\left(A\right)}=g_{\ell;A}^{-1}-1=\left(\begin{array}{cc}
\frac{n_{\ell}}{g_{\ell,12}^{2}+n_{\ell}^{2}}-1 & -\frac{g_{\ell,12}}{g_{\ell,12}^{2}+n_{\ell}^{2}}\\
\frac{g_{\ell,12}}{g_{\ell,12}^{2}+n_{\ell}^{2}} & \frac{n_{\ell}}{g_{\ell,12}^{2}+n_{\ell}^{2}}-1
\end{array}\right),\label{eq:SlAF}
\end{equation}
which is different from the $A$ block of $S_{F,\ell}=g_{\ell}^{-1}-1$,
denoted as $S_{F,\ell;A}$. Using the results of Section \ref{subsec:Block-structure-of-integer-time-Dyson},
the integer time Dyson equation within the $A$ block may be written
as 
\begin{equation}
S_{F,\ell}^{\left(A\right)}=S_{0,\ell}^{\left(A\right)}S_{\ell;A},\label{eq:dyson A block}
\end{equation}
and $S_{\ell;A}$ can be determined as 
\[
S_{\ell;A}=\left(\begin{array}{cc}
S_{\ell,11} & S_{\ell,12}\\
-S_{\ell,12} & S_{\ell,11}
\end{array}\right),
\]
where Eqns. (\ref{eq:GlA choice}) and (\ref{eq:glA}) can be used
to obtain 
\begin{align}
 & S_{\ell,11}=\left(4\mathcal{G}_{\ell,12}^{2}+1\right)^{-1}\left(g_{\ell,12}^{2}+n_{\ell}^{2}\right)^{-1}\big(4g_{\ell,12}\mathcal{G}_{\ell,12}\nonumber \\
 & +4\mathcal{G}_{\ell,12}^{2}\left(g_{\ell,12}^{2}+\left(n_{\ell}-1\right)n_{\ell}\right)-g_{\ell,12}^{2}-n_{\ell}^{2}+n_{\ell}\big),\\
 & S_{\ell,12}=\left(4\mathcal{G}_{\ell,12}^{2}+1\right)^{-1}\left(g_{\ell,12}^{2}+n_{\ell}^{2}\right)^{-1}\big(4\left(1-n_{\ell}\right)n_{\ell}\mathcal{G}_{\ell,12}\nonumber \\
 & +g_{\ell,12}\left(4\mathcal{G}_{\ell,12}\left(\mathcal{G}_{\ell,12}-g_{\ell,12}\right)-1\right)\big).
\end{align}
 The $S_{\ell}$ is a $3\times3$ matrix obtained from $S_{F,\ell}=S_{0,\ell}S_{\ell}$,
yielding 
\begin{equation}
S_{\ell}=\begin{pmatrix}S_{\ell;A} & 0\\
0 & 1
\end{pmatrix}.
\end{equation}

The second step is to examine the lattice integer time Green's function
and use the self-consistency condition to resolve the $B$, $C$,
and $D$ blocks. The lattice integer time Green's function can be
parametrized by the physical momentum density distribution $n_{k\ell}\equiv\langle\barhat[n]_{k\ell}^{(3)}\rangle_{\barhat[\rho]_{K}}$
and the integer time self-energy $S_{\ell}$ by assuming $\hat{K}_{2}$
is a single Slater determinant with $\left\langle \hat{n}_{k\ell}\right\rangle _{\hat{K}_{2}}=1$
for $k\in<$ and $\left\langle \hat{n}_{k\ell}\right\rangle _{\hat{K}_{2}}=0$
for $k\in>$, where the symbols $<$ or $>$ denote the occupied and
unoccupied regions of $\hat{K}_{2}$. The local integer time Green's
function for the lattice is denoted as $\left[\bm{g}'_{i}\right]_{\ell\ell'}=\delta_{\ell\ell'}g'_{\ell}$,
where 
\begin{equation}
g'_{\ell}=\begin{pmatrix}g'_{\ell,A} & g'_{\ell,B}\\
g'_{\ell,C} & n_{\ell}
\end{pmatrix},\label{eq:gllattice}
\end{equation}
where $g'_{\ell,A}$ , $g'_{\ell,B}$, and $g'_{\ell,C}$ are $2\times2$,
$2\times1$, and $1\times2$ matrices, respectively, defined as 
\begin{align}
 & g'_{\ell,A}=\left(\begin{array}{cc}
n_{\ell} & \frac{S_{\ell,11}}{S_{\ell,12}}\Delta_{\ell}\\
-\frac{S_{\ell,11}}{S_{\ell,12}}\Delta_{\ell} & n_{\ell}
\end{array}\right),\label{eq:gA}\\
 & g'_{\ell,B}=\left(\begin{array}{c}
\frac{S_{\ell,11}}{\sqrt{S_{\ell,12}\left(S_{\ell,11}^{2}+S_{\ell,12}^{2}\right)}}\mathcal{A}_{>,\ell}\\
\frac{1}{\sqrt{S_{\ell,12}}}\mathcal{A}_{<,\ell}+\sqrt{\frac{S_{\ell,12}}{S_{\ell,11}^{2}+S_{\ell,12}^{2}}}\mathcal{A}_{>,\ell}
\end{array}\right),\label{eq:gB}\\
 & g'_{\ell,C}=\left(\begin{array}{c}
-\frac{S_{\ell,11}}{\sqrt{S_{\ell,12}}}\mathcal{A}_{<,\ell}\\
-\sqrt{S_{\ell,12}}\mathcal{A}_{<,\ell}-\sqrt{\frac{S_{\ell,11}^{2}+S_{\ell,12}^{2}}{S_{\ell,12}}}\mathcal{A}_{>,\ell}
\end{array}\right)^{T},\label{eq:gC}
\end{align}
where $\mathcal{A}_{<\ell}$, $\mathcal{A}_{>\ell}$, and $\Delta_{\ell}$
are defined in Eqns. (\ref{eq:A<ell}), (\ref{eq:A>ell}), and (\ref{eq:delta definition}).

Using Eqns. (\ref{eq:gllattice}), (\ref{eq:g0B}), (\ref{eq:g0C}),
and (\ref{eq:g0D}), the $B$, $C$, and $D$ blocks of Eq. (\ref{eq:Gl block})
can be fully determined, completing the algorithm. Finally, the local
interaction energy can then be written using the tensor product representation
as
\begin{equation}
E_{loc}\left(\left\{ \mathcal{G}_{12,\ell}\right\} ,\{u_{\Gamma}\},\left\{ \mathcal{A}_{<\ell}\right\} ,\left\{ \mathcal{A}_{>\ell}\right\} \right)=\frac{u^{T}\left(\barhat[H]_{loc;i}^{\left(3\right)}\right)_{u}u}{u^{T}\left(\barhat[1]\right)_{u}u},\label{eq:E for n=00003D3-1}
\end{equation}
while the kinetic energy is $\sum_{\ell}\int dk\epsilon_{k\ell}n_{k\ell}$.
The $n_{k\ell}$ are subject to the two linear constraints given in
Eqns. (\ref{eq:linearc2}) and (\ref{eq:linearc3}), which can be
implemented in the $u$-representation using

\begin{align}
 & n_{\ell}\left(\left\{ \mathcal{G}_{12;\ell}\right\} ,\{u_{\Gamma}\}\right)=\frac{u^{T}\left(\barhat[a]_{\ell}^{\dagger\left(1\right)}\barhat[a]_{\ell}^{\left(1\right)}\right)_{u}u}{u^{T}\left(\barhat[1]\right)_{u}u},\label{eq:nell}\\
 & \Delta_{\ell}\left(\left\{ \mathcal{G}_{12;\ell}\right\} ,\{u_{\Gamma}\}\right)=\frac{S_{\ell,12}}{S_{\ell,11}}\frac{u^{T}\left(\barhat[a]_{\ell}^{\dagger\left(1\right)}\barhat[a]_{\ell}^{\left(2\right)}\right)_{u}u}{u^{T}\left(\barhat[1]\right)_{u}u},\label{eq:delta}
\end{align}
while the $n_{k\ell;0}$ are subject to $\int_{<}dk=n_{\ell}$.

It is useful to appreciate what can be gleaned from the form of the
preceding equations. The $g'_{\ell,A}$ block is reminiscent of the
$\hat{\rho}_{G2}$ case where $\Delta_{\ell}$ captures the quasi-particle
renormalization, while $g'_{\ell,B}$ and $g'_{\ell,C}$ are reminiscent
of the $\hat{\rho}_{B2}$ case where $\mathcal{A}_{<\ell}$ and $\mathcal{A}_{>\ell}$
capture the super-exchange effects. This observation helps illustrate
how the $\hat{\rho}_{G3}$ simultaneously captures the physics of
both $\hat{\rho}_{G2}$ and $\hat{\rho}_{B2}$. 

In the following, we further elaborate on two points which were not
fully elucidated in Ref. \cite{Cheng2023035127}. First, we provide
explicit expressions for the tensor product representation in the
case of $\mathcal{N}=3$. Second, we explore various relations derived
using the block structure of the integer time Dyson equation.

\subsubsection{Evaluating observables under $\barhat[\rho]_{loc}$ using the tensor
product representation}

Consider the tensor product representation of $\barhat[O]$ evaluated
under $\barhat[\rho]_{loc}$. Given that the interacting projector
only acts on the first and second time step, the 3-dimensional tensor
representation may be reduced to a 2-dimensional tensor representation
as 
\begin{equation}
\left[\left(\barhat[O]\right)_{u}\right]_{\Gamma_{1}\Gamma_{2}}=\left\langle \barhat[X]_{i\Gamma_{1}}^{\left(1\right)}\barhat[X]_{i\Gamma_{2}}^{\left(2\right)}\barhat[O]\right\rangle _{\barhat[\rho]_{loc,i;0}}.
\end{equation}
Furthermore, the tensor contraction can be simplified as $(\barhat[O])_{u}\cdot\left(u_{1}\otimes u_{2}\right)=u^{T}(\barhat[O])_{u}u$
with $u^{T}=u_{1}=u_{2}$. Therefore, even though the compound space
of $\mathcal{N}=3$ is larger than $\mathcal{N}=2$ for a given original
Hilbert space, the computational cost in the tensor representation
for $\hat{\rho}_{G3}$ is similar to that of $\hat{\rho}_{G2}$ due
to this dimension reduction. Similar to the case of the $\hat{\rho}_{G2}$,
$\left(\barhat[O]\right)_{u}$ has a direct product structure given
by Eq. (\ref{eq:direct_product_structure}). We first specify the
components of the$A$ block quantities at a given spin orbital $\ell$
\begin{align}
 & \left(\barhat[1]\right)_{u;\ell}=\left(\begin{array}{cc}
\mathcal{G}_{\ell,12}^{2}+\frac{1}{4} & \frac{1}{4}-\mathcal{G}_{\ell,12}^{2}\\
\frac{1}{4}-\mathcal{G}_{\ell,12}^{2} & \mathcal{G}_{\ell,12}^{2}+\frac{1}{4}
\end{array}\right),\\
 & \left(\barhat[a]_{\ell}^{\dagger\left(1\right)}\barhat[a]_{\ell}^{\left(1\right)}\right)_{u;\ell,S}=\left(\barhat[a]_{\ell}^{\dagger\left(2\right)}\barhat[a]_{\ell}^{\left(2\right)}\right)_{u;\ell,S}\\
 & =\left(\begin{array}{cc}
0 & \frac{1}{8}\left(1-4\mathcal{G}_{\ell,12}^{2}\right)\\
\frac{1}{8}\left(1-4\mathcal{G}_{\ell,12}^{2}\right) & \mathcal{G}_{\ell,12}^{2}+\frac{1}{4}
\end{array}\right),\\
 & \left(\barhat[a]_{\ell}^{\dagger\left(1\right)}\barhat[a]_{\ell}^{\left(2\right)}\right)_{u;\ell,S}=-\left(\barhat[a]_{\ell}^{\dagger\left(2\right)}\barhat[a]_{\ell}^{\left(1\right)}\right)_{u;\ell,S}\\
 & =\left(\begin{array}{cc}
0 & \frac{\mathcal{G}_{\ell,12}}{2}\\
\frac{\mathcal{G}_{\ell,12}}{2} & 0
\end{array}\right),
\end{align}
where $S$ indicates the symmetric part of the matrix, which is defined
as 
\begin{equation}
\left(\barhat[O]\right)_{u,S}=\frac{1}{2}\left(\left(\barhat[O]\right)_{u}+\left(\barhat[O]\right)_{u}^{T}\right),\label{eq:S in u}
\end{equation}
which is useful given that $u^{T}\left(\barhat[O]\right)_{u}u=u^{T}\left(\barhat[O]\right)_{u,S}u$.
Similarly, the $w$-representation can be defined using $u=Vw$, where
$V=\otimes_{\ell}V_{\ell}$ with 
\begin{equation}
V_{\ell}=\frac{1}{\sqrt{2}}\left(\begin{array}{cc}
\frac{1}{2\mathcal{G}_{\ell,12}}+1 & 1-\frac{1}{2\mathcal{G}_{\ell,12}}\\
1-\frac{1}{2\mathcal{G}_{\ell,12}} & \frac{1}{2\mathcal{G}_{\ell,12}}+1
\end{array}\right),
\end{equation}
such that $\left(\barhat[1]\right)_{w;\ell}$ is the identity matrix
and $\left(\barhat[a]_{\ell}^{\dagger\left(1\right)}\barhat[a]_{\ell}^{\left(1\right)}\right)_{w;\ell,S}$
is diagonal, resulting in 
\begin{align}
 & \left(\barhat[1]\right)_{w;\ell}=\left(\begin{array}{cc}
1 & 0\\
0 & 1
\end{array}\right),\label{eq:1w}\\
 & \left(\barhat[a]_{\ell}^{\dagger\left(1\right)}\barhat[a]_{\ell}^{\left(1\right)}\right)_{w;\ell,S}=\left(\barhat[a]_{\ell}^{\dagger\left(2\right)}\barhat[a]_{\ell}^{\left(2\right)}\right)_{w;\ell,S},\label{eq:n1wS}\\
 & =\left(\begin{array}{cc}
-\frac{\left(1-2\mathcal{G}_{\ell,12}\right){}^{2}}{8\mathcal{G}_{\ell,12}} & 0\\
0 & \frac{\left(2\mathcal{G}_{\ell,12}+1\right){}^{2}}{8\mathcal{G}_{\ell,12}}
\end{array}\right),\label{eq:n1w}
\end{align}
and correspondingly, we have 
\begin{align}
 & \left(\barhat[a]_{\ell}^{\dagger\left(1\right)}\barhat[a]_{\ell}^{\left(2\right)}\right)_{w;\ell,S}=-\left(\barhat[a]_{\ell}^{\dagger\left(2\right)}\barhat[a]_{\ell}^{\left(1\right)}\right)_{w;\ell,S}\\
 & =\left(\begin{array}{cc}
\frac{1}{2}\left(\mathcal{G}_{\ell,12}-\frac{1}{4\mathcal{G}_{\ell,12}}\right) & \frac{\mathcal{G}_{\ell,12}}{2}+\frac{1}{8\mathcal{G}_{\ell,12}}\\
\frac{\mathcal{G}_{\ell,12}}{2}+\frac{1}{8\mathcal{G}_{\ell,12}} & \frac{1}{2}\left(\mathcal{G}_{\ell,12}-\frac{1}{4\mathcal{G}_{\ell,12}}\right)
\end{array}\right).
\end{align}
 Finally, we provide explicit expressions for $\left(\barhat[a]_{\ell}^{\dagger\left(3\right)}\barhat[a]_{\ell}^{\left(3\right)}\right)_{w;\ell}$,
with the components given by

\begin{align}
 & \left[\left(\barhat[a]_{\ell}^{\dagger\left(3\right)}\barhat[a]_{\ell}^{\left(3\right)}\right)_{w;\ell}\right]_{11}=\frac{1}{2\mathcal{G}_{\ell,12}}\big(\mathcal{G}_{\ell,13}\left(\mathcal{G}_{\ell,31}-\mathcal{G}_{\ell,32}\right)\nonumber \\
 & +\mathcal{G}_{\ell,23}\left(\mathcal{G}_{\ell,31}+\mathcal{G}_{\ell,32}\right)+2\mathcal{G}_{\ell,12}\mathcal{G}_{\ell,33}\big),\label{eq:n3w11}\\
 & \left[\left(\barhat[a]_{\ell}^{\dagger\left(3\right)}\barhat[a]_{\ell}^{\left(3\right)}\right)_{w;\ell}\right]_{12}=\frac{\left(\mathcal{G}_{\ell,13}-\mathcal{G}_{\ell,23}\right)\left(\mathcal{G}_{\ell,31}+\mathcal{G}_{\ell,32}\right)}{2\mathcal{G}_{\ell,12}},\\
 & \left[\left(\barhat[a]_{\ell}^{\dagger\left(3\right)}\barhat[a]_{\ell}^{\left(3\right)}\right)_{w;\ell}\right]_{21}=-\frac{\left(\mathcal{G}_{\ell,13}+\mathcal{G}_{\ell,23}\right)\left(\mathcal{G}_{\ell,31}-\mathcal{G}_{\ell,32}\right)}{2\mathcal{G}_{\ell,12}},\\
 & \left[\left(\barhat[a]_{\ell}^{\dagger\left(3\right)}\barhat[a]_{\ell}^{\left(3\right)}\right)_{w;\ell}\right]_{22}=\frac{1}{2\mathcal{G}_{\ell,12}}\big(\mathcal{G}_{\ell,23}\left(\mathcal{G}_{\ell,31}-\mathcal{G}_{\ell,32}\right)\nonumber \\
 & -\mathcal{G}_{\ell,13}\left(\mathcal{G}_{\ell,31}+\mathcal{G}_{\ell,32}\right)+2\mathcal{G}_{\ell,12}\mathcal{G}_{\ell,33}\big),\label{eq:n3w22}
\end{align}
which can be used to compute the local interaction energy.

\subsubsection{Block structure of the integer time Dyson equation\label{subsec:Block-structure-of-integer-time-Dyson}}

Here we derive various useful equations using the block form of the
integer time Dyson equation. To make our discussion general, we assume
that the integer time Green's functions are not diagonal in the orbital
index, resulting in the following block matrix equation 

\begin{equation}
\begin{pmatrix}S_{F,A} & S_{F,B}\\
S_{F,C} & S_{F,D}
\end{pmatrix}=\begin{pmatrix}S_{0,A} & S_{0,B}\\
S_{0,C} & S_{0,D}
\end{pmatrix}\begin{pmatrix}S_{A} & 0\\
0 & 1
\end{pmatrix},
\end{equation}
where the blocks for $S_{F}$ and $S_{0}$ can be explicitly expressed
in terms of the blocks of $g$ and $\mathcal{G}$ using the inverse
formula for a $2\times2$ block matrix, resulting in the following
relations 
\begin{align}
 & 1+S_{F,D}=\frac{1}{g_{D}-g_{C}g_{A}^{-1}g_{B}}\label{eq:sD}\\
 & =1+S_{0,D}=\frac{1}{\mathcal{G}_{D}-\mathcal{G}_{C}\mathcal{G}_{A}^{-1}\mathcal{G}_{B}},\label{eq:sD2}\\
 & S_{F,B}=-g_{A}^{-1}g_{B}\left(1+S_{F,D}\right)\label{eq:sB}\\
 & =S_{0,B}=-\mathcal{G}_{A}^{-1}\mathcal{G}_{B}\left(1+S_{0,D}\right),\label{eq:sB2}\\
 & S_{F,C}=-\left(1+S_{F,D}\right)g_{C}g_{A}^{-1}\label{eq:sC}\\
 & =S_{0,C}S_{A}=\left(1+S_{0,D}\right)\mathcal{G}_{C}\mathcal{G}_{A}^{-1}S_{A},\label{eq:sC2}\\
 & S_{F,A}=S_{F}^{\left(A\right)}+g_{A}^{-1}g_{B}\left(1+S_{F,D}\right)g_{C}g_{A}^{-1}\label{eq:sA}\\
 & =S_{0,A}S_{A}=S_{0}^{\left(A\right)}S_{A}+\mathcal{G}_{A}^{-1}\mathcal{G}_{B}\left(1+S_{0,D}\right)\mathcal{G}_{C}\mathcal{G}_{A}^{-1}S_{A}.\label{eq:sA1}
\end{align}
Using Eqns. (\ref{eq:sD}), (\ref{eq:sD2}), (\ref{eq:sB}), and (\ref{eq:sB2}),
we obtain 
\begin{equation}
g_{A}^{-1}g_{B}=\mathcal{G}_{A}^{-1}\mathcal{G}_{B}.\label{eq:gagbGAGA}
\end{equation}
Similarly, using Eqns. (\ref{eq:sD}), (\ref{eq:sD2}), and (\ref{eq:sC}),
we obtain 
\begin{equation}
g_{C}g_{A}^{-1}=\mathcal{G}_{C}\mathcal{G}_{A}^{-1}S_{A}.\label{eq:gc}
\end{equation}
Using Eqns. (\ref{eq:gagbGAGA}), (\ref{eq:gc}), (\ref{eq:sA}),
and (\ref{eq:sA1}), we verify that $\mathcal{G}_{A}$ and $g_{A}$
can be used to determine $S_{A}$ as 
\begin{equation}
S_{F}^{\left(A\right)}=g_{A}^{-1}-1=S_{0}^{\left(A\right)}S_{A}=\left(\mathcal{G}_{A}^{-1}-1\right)S_{A}.\label{eq:sA0local}
\end{equation}
Using Eqns. (\ref{eq:gc}) and (\ref{eq:sA0local}), we have 
\begin{equation}
g_{C}\left(1-g_{A}\right)^{-1}=\mathcal{G}_{C}\left(1-\mathcal{G}_{A}\right)^{-1}.\label{eq:gconega}
\end{equation}
Finally, using Eq. (\ref{eq:sA0local}), we have 
\begin{equation}
g_{D}-g_{C}g_{A}^{-1}g_{B}=\mathcal{G}_{D}-\mathcal{G}_{C}\mathcal{G}_{A}^{-1}\mathcal{G}_{B},
\end{equation}
which can be used to solve for $\mathcal{G}_{A}$. Additionally, it
is useful to explicitly write expressions for $\mathcal{G}_{B}$ ,
$\mathcal{G}_{C}$, and $\mathcal{G}_{D}$ as 
\begin{align}
\mathcal{G}_{B} & =\mathcal{G}_{A}g_{A}^{-1}g_{B},\label{eq:g0B}\\
\mathcal{G}_{C} & =g_{C}\left(1-g_{A}\right)^{-1}\left(1-\mathcal{G}_{A}\right),\label{eq:g0C}\\
\mathcal{G}_{D} & =g_{D}+g_{C}\left(1-g_{A}\right)^{-1}\left(g_{A}-\mathcal{G}_{A}\right)g_{A}^{-1}g_{B}.\label{eq:g0D}
\end{align}

\subsection{Derivation of the qubit energy form\label{subsec:Derivation-of-qubit-N3}}

In this section, we derive the qubit energy form, given in Eq. (\ref{eq:energy_n3_qubit}),
in several steps. First, in Section \ref{subsec:Representation-of-A-block_rhog3},
we introduce a polar representation for the $A$ block. Second, in
Section \ref{subsec:Resolving-the-self-consistency-A}, we introduce
$\rhoeff$ as a unitary transformation of $ww^{T}$, and we solve
the self-consistency condition for the $A$ block using $n_{\ell}$,
$\xi_{\ell}$, and $\Delta_{\ell}$. Third, in Section \ref{subsec:Resolving-the-BCD},
we resolve the self-consistency of the $B$, $C$, and $D$ blocks
using the extra information provided by $\mathcal{A}_{<\ell}$ and
$\mathcal{A}_{>\ell}$.

\subsubsection{Polar Representation of the A block\label{subsec:Representation-of-A-block_rhog3}}

We first review some mathematical properties of the matrix group with
following form 
\begin{equation}
\mathcal{S}\left(c,\phi\right)=c\left(\begin{array}{cc}
\cos(\phi) & \sin(\phi)\\
-\sin(\phi) & \cos(\phi)
\end{array}\right),\label{eq:scphi}
\end{equation}
where $c>0$ and $\phi\in\left[0,2\pi\right]$ and the group multiplication
is 
\begin{equation}
\mathcal{S}\left(c_{1},\phi_{1}\right)\mathcal{S}\left(c_{2},\phi_{2}\right)=\mathcal{S}\left(c,\phi\right).
\end{equation}
There is an isomorphism for this matrix group to $\mathbb{R}^{+}\times S^{1}$
with $\mathcal{S}\left(c,\phi\right)\rightarrow\left(c,\phi\right)$,
with the group product taken as 
\begin{align}
c_{1}c_{2}=c,\hspace{1em}\phi_{1}+\phi_{2}=\phi.
\end{align}
Interpreting $\mathcal{S}\left(c,\phi\right)$ as the integer time
self-energy, the corresponding integer time Green's function is given
as 
\begin{equation}
g\left(c,\phi\right)=\frac{1}{1+\mathcal{S}}=\left(\begin{array}{cc}
\frac{c\cos(\phi)+1}{c^{2}+2c\cos(\phi)+1} & -\frac{c\sin(\phi)}{c^{2}+2c\cos(\phi)+1}\\
\frac{c\sin(\phi)}{c^{2}+2c\cos(\phi)+1} & \frac{c\cos(\phi)+1}{c^{2}+2c\cos(\phi)+1}
\end{array}\right),\label{eq:gcphi}
\end{equation}
and it can be seen that 
\begin{align}
g_{11} & \equiv\left[g\left(c,\phi\right)\right]_{11}=\left[g\left(c,\phi\right)\right]_{22},\\
g_{12} & \equiv\left[g\left(c,\phi\right)\right]_{12}=-\left[g\left(c,\phi\right)\right]_{21}.
\end{align}
It is also useful to express $c$ and $\text{\ensuremath{\phi}}$
in terms of $g_{11}$ and $g_{12}$ as 
\begin{align}
c & =\sqrt{\frac{1-2g_{11}}{g_{11}^{2}+g_{12}^{2}}+1},\label{eq:cing11g12}\\
\tan\left(\phi\right) & =\frac{g_{12}}{g_{12}^{2}+\left(g_{11}-1\right)g_{11}}.\label{eq:tanphigeneral}
\end{align}

We now use the preceding results to study the $A$ block of the integer
time Green's function. Given that $g_{\ell,A}$ and $\mathcal{G}_{\ell,A}$
have the form given by Eq. (\ref{eq:gcphi}), the $S_{F,\ell}^{\left(A\right)}$
and $S_{0,\ell}^{\left(A\right)}$ also have the form of Eq. (\ref{eq:scphi}),
and therefore $S_{\ell,A}$ also has the form of Eq. (\ref{eq:scphi}).
We can then use the polar representation to express the following
quantities 
\begin{align}
 & S_{0,\ell}^{\left(A\right)}=\mathcal{G}_{\ell;A}^{-1}-1=\mathcal{S}\left(1,\phi_{\ell,0}\right),\\
 & S_{F,\ell}^{\left(A\right)}=g_{\ell;A}^{-1}-1=\mathcal{S}\left(c_{\ell},\phi_{\ell}\right),\\
 & S_{\ell,A}=\left(S_{0,\ell}^{\left(A\right)}\right)^{-1}S_{F,\ell}^{\left(A\right)}=\mathcal{S}\left(c_{\ell},\theta_{\ell}\right),\label{eq:sell}
\end{align}
allowing for the integer time Dyson equation to be recast as
\begin{equation}
\theta_{\ell}=\phi_{\ell}-\phi_{\ell,0},\label{eq:theta_dyson-1}
\end{equation}
which yields Eq. (\ref{eq:phiell0}). Using $g_{11}\rightarrow n_{\ell}$,
$g_{12}\rightarrow g_{\ell,12}$, $c\rightarrow c_{\ell},$ and $\phi\rightarrow\phi_{\ell}$
in Eqns. (\ref{eq:cing11g12}) and (\ref{eq:tanphigeneral}), we obtain
Eqns. (\ref{eq:cell}) and (\ref{eq:phiell}). Using $g_{11}\rightarrow1/2$
, $g_{12}\rightarrow\mathcal{G}_{\ell,12}$, $c\rightarrow1$, and
$\phi\rightarrow\phi_{\ell,0}$, we obtain 
\begin{equation}
\tan\left(\phi_{\ell,0}\right)=\frac{\mathcal{G}_{\ell,12}}{\mathcal{G}_{\ell,12}^{2}-\frac{1}{4}},
\end{equation}
which yields Eq. (\ref{eq:g012}) by assuming $\phi_{\ell,0}\in\left[-\pi/2,0\right]$.
Using Eq. (\ref{eq:sell}), we can rewrite Eq. (\ref{eq:gllattice})
in terms of $c_{\ell}$, $\theta_{\ell}$ and $n_{k\ell}$ as

\begin{align}
 & g'_{\ell,A}=\left(\begin{array}{cc}
n_{\ell} & \cot\left(\theta_{\ell}\right)\Delta_{\ell}\\
-\cot\left(\theta_{\ell}\right)\Delta_{\ell} & n_{\ell}
\end{array}\right),\label{eq:gAintheta-1}\\
 & g'_{\ell,B}=\sqrt{\frac{1}{c_{\ell}\sin\left(\theta_{\ell}\right)}}\left(\begin{array}{c}
\cos\left(\theta_{\ell}\right)\mathcal{A}_{>,\ell}\\
\mathcal{A}_{<,\ell}+\sin\left(\theta_{\ell}\right)\mathcal{A}_{>,\ell}
\end{array}\right),\\
 & g'_{\ell,C}=-\sqrt{\frac{c_{\ell}}{\sin\left(\theta_{\ell}\right)}}\left(\begin{array}{c}
\cos\left(\theta_{\ell}\right)\mathcal{A}_{<,\ell}\\
\sin\left(\theta_{\ell}\right)\mathcal{A}_{<,\ell}+\mathcal{A}_{>,\ell}
\end{array}\right)^{T}.
\end{align}
The matrix elements in $w$ representation are given by 

\begin{align}
 & \left[\left(\barhat[a]_{\ell}^{\dagger\left(1\right)}\barhat[a]_{\ell}^{\left(1\right)}\right)_{w;\ell,S}\right]_{11}=\frac{1}{2}\left(\csc\left(\phi_{\ell,0}\right)+1\right),\label{eq:g11ws}\\
 & \left[\left(\barhat[a]_{\ell}^{\dagger\left(1\right)}\barhat[a]_{\ell}^{\left(1\right)}\right)_{w;\ell,S}\right]_{22}=\frac{1}{2}\left(1-\csc\left(\phi_{\ell,0}\right)\right),\\
 & \left[\left(\barhat[a]_{\ell}^{\dagger\left(1\right)}\barhat[a]_{\ell}^{\left(1\right)}\right)_{w;\ell,S}\right]_{12}=\left[\left(\barhat[a]_{\ell}^{\dagger\left(1\right)}\barhat[a]_{\ell}^{\left(1\right)}\right)_{w;\ell,S}\right]_{21}=0,\\
 & \left(\barhat[a]_{\ell}^{\dagger\left(1\right)}\barhat[a]_{\ell}^{\left(2\right)}\right)_{w;\ell,S}=\left(\begin{array}{cc}
\frac{1}{2}\cot\left(\phi_{\ell,0}\right) & -\frac{1}{2}\csc\left(\phi_{\ell,0}\right)\\
-\frac{1}{2}\csc\left(\phi_{\ell,0}\right) & \frac{1}{2}\cot\left(\phi_{\ell,0}\right)
\end{array}\right).\label{eq:g12ws}
\end{align}

\subsubsection{Resolving the self-consistency in the $A$-block\label{subsec:Resolving-the-self-consistency-A}}

\global\long\def\eff{q}%
A key ingredient of the qubit parametrization is the introduction
of the qubit representation, defined by
\begin{align}
 & \left(\barhat[O]\right)_{\eff}=\mathcal{U}^{\dagger}\left(\barhat[O]\right)_{w}\mathcal{U},\label{eq:Oeff}\\
 & \rhoeff=\mathcal{U}^{\dagger}ww^{T}\mathcal{U},
\end{align}
where $\mathcal{U}$ is a unitary matrix defined as $\mathcal{U}=\mathcal{U}_{1}\otimes\dots\otimes\mathcal{U}_{2N_{orb}}$
and 
\begin{equation}
\mathcal{U}_{\ell}=\left(\begin{array}{cc}
\cos\left(\psi_{\ell}\right) & \sin\left(\psi_{\ell}\right)\\
-\sin\left(\psi_{\ell}\right) & \cos\left(\psi_{\ell}\right)
\end{array}\right),\label{eq:Uell}
\end{equation}
where $\psi_{\ell}$ is a parameter that will be determined by demanding
that $\rhoeff$ yields the physical density. This should be contrasted
to the $w$ representation, where the local density of $ww^{T}$ is
generally different from the physical density. 

We begin by deriving Eq. (\ref{eq:thetaell}), which determines $\theta_{\ell}$
for a given $n_{\ell},$ $\xi_{\ell}$, and $\Delta_{\ell}$. Using
Eq. (\ref{eq:Oeff}), Eqns. (\ref{eq:g11ws})-(\ref{eq:g12ws}) are
transformed to 
\begin{align}
 & \left(\barhat[a]_{\ell}^{\dagger\left(1\right)}\barhat[a]_{\ell}^{\left(1\right)}\right)_{\eff;\ell,S}=\left(\begin{array}{cc}
\frac{1}{2}\left(p_{\ell}+1\right) & \frac{q_{\ell}}{2}\\
\frac{q_{\ell}}{2} & \frac{1}{2}\left(1-p_{\ell}\right)
\end{array}\right),\\
 & \left[\left(\barhat[a]_{\ell}^{\dagger\left(1\right)}\barhat[a]_{\ell}^{\left(2\right)}\right)_{\eff;\ell,S}\right]_{11}=\frac{1}{2}\left(\cot\left(\phi_{\ell,0}\right)+q_{\ell}\right),\\
 & \left[\left(\barhat[a]_{\ell}^{\dagger\left(1\right)}\barhat[a]_{\ell}^{\left(2\right)}\right)_{\eff;\ell,S}\right]_{22}=\frac{1}{2}\left(\cot\left(\phi_{\ell,0}\right)-q_{\ell}\right),\\
 & \left[\left(\barhat[a]_{\ell}^{\dagger\left(1\right)}\barhat[a]_{\ell}^{\left(2\right)}\right)_{\eff;\ell,S}\right]_{12}=\left[\left(\barhat[a]_{\ell}^{\dagger\left(1\right)}\barhat[a]_{\ell}^{\left(2\right)}\right)_{\eff;\ell,S}\right]_{21}=-\frac{p_{\ell}}{2},
\end{align}
where 
\begin{align}
 & p_{\ell}=\cos\left(2\psi_{\ell}\right)\csc\left(\phi_{\ell,0}\right),\label{eq:plphi0psi}\\
 & q_{\ell}=\sin\left(2\psi_{\ell}\right)\csc\left(\phi_{\ell,0}\right).\label{eq:qlphi0psi}
\end{align}
 Using the self-consistency condition for the 1,1 and 1,2 entries,
we have 
\begin{align}
 & \text{Tr}\left(\left(\barhat[a]_{\ell}^{\dagger\left(1\right)}\barhat[a]_{\ell}^{\left(1\right)}\right)_{\eff;\ell,S}\rhoeff_{\ell}\right)=n_{\ell},\label{eq:11effn}\\
 & \text{Tr}\left(\left(\barhat[a]_{\ell}^{\dagger\left(1\right)}\barhat[a]_{\ell}^{\left(2\right)}\right)_{\eff;\ell,S}\rhoeff_{\ell}\right)=g_{\ell,12},\label{12effg}\\
 & \rhoeff_{\ell}=\textrm{Tr}_{/\ell}\rhoeff=\left(\begin{array}{cc}
1-n_{\ell} & \xi_{\ell}\\
\xi_{\ell} & n_{\ell}
\end{array}\right).
\end{align}
 Eqns. (\ref{eq:11effn}) and (\ref{12effg}) reduce to two linear
equations in $p_{\ell}$ and $q_{\ell}$ as 
\begin{align}
\xi_{\ell}q_{\ell}-\delta n_{\ell}p_{\ell} & =\delta n_{\ell},\\
\xi_{\ell}p_{\ell}+\delta n_{\ell}q_{\ell} & =\frac{\cot\left(\phi_{\ell,0}\right)}{2}-g_{\ell,12},
\end{align}
and $p_{\ell}$ and $q_{\ell}$ can be determined from Eqns. (\ref{eq:pl})
and (\ref{eq:ql}). Using Eqns. (\ref{eq:plphi0psi}) and (\ref{eq:qlphi0psi}),
we obtain 
\begin{equation}
p_{\ell}^{2}+q_{\ell}^{2}=\csc^{2}\left(\phi_{\ell,0}\right)=\cot^{2}\left(\phi_{\ell,0}\right)+1.\label{eq:p2plusq2}
\end{equation}
Both $\cot\left(\phi_{\ell}\right)$ and $\cot\left(\phi_{\ell,0}\right)$
can be expressed in terms of $\cot\left(\theta_{\ell}\right)$ as
\begin{align}
 & \cot\left(\phi_{\ell}\right)=\Delta_{\ell}\cot\left(\theta_{\ell}\right)-\frac{\xi_{\ell,0}^{2}}{\Delta_{\ell}\cot\left(\theta_{\ell}\right)},\\
 & \cot\left(\phi_{\ell;0}\right)=\cot\left(\phi_{\ell}-\theta_{\ell}\right)=\frac{\cot\left(\phi_{\ell}\right)\cot\left(\theta_{\ell}\right)+1}{\cot\left(\theta_{\ell}\right)-\cot\left(\phi_{\ell}\right)}\\
 & =-\frac{\cot\left(\theta_{\ell}\right)\left(-\xi_{\ell,0}^{2}+\Delta_{\ell}^{2}\cot^{2}\left(\theta_{\ell}\right)+\Delta_{\ell}\right)}{\left(\Delta_{\ell}-1\right)\Delta_{\ell}\cot^{2}\left(\theta_{\ell}\right)-\xi_{\ell,0}^{2}}.\label{eq:phiell0Intheta}
\end{align}
Substituting Eqns. (\ref{eq:pl}) and (\ref{eq:ql}) into Eq. (\ref{eq:p2plusq2})
and using Eq. (\ref{eq:phiell0Intheta}), we obtain a sixth order
equation in $\cot\left(\theta_{\ell}\right)$ which can be factored
into the following form
\begin{align}
 & \left(\Delta_{\ell}^{2}\left(1-2\xi_{\ell,0}^{2}\right)\cot^{2}\left(\theta_{\ell}\right)+\xi_{\ell,0}^{4}+\Delta_{\ell}^{4}\cot^{4}\left(\theta_{\ell}\right)\right)\nonumber \\
 & \times\left(\cot^{2}\left(\theta_{\ell}\right)\left(\xi_{\ell,0}^{2}+\Delta_{\ell}^{2}-\Delta_{\ell}\right)-\xi_{\ell}^{2}\csc^{2}\left(\theta_{\ell}\right)\right)=0.\label{eq:equaiton for cot theta}
\end{align}
Notice that the first factor in Eq. (\ref{eq:equaiton for cot theta})
is positive, implying that the second factor is zero, and $\theta_{\ell}$
may be obtained as 

\begin{equation}
\cos\left(\theta_{\ell}\right)=\frac{\xi_{\ell}}{\sqrt{\xi_{\ell,0}^{2}+\left(\Delta_{\ell}-1\right)\Delta_{\ell}}},
\end{equation}
which yields Eq. (\ref{eq:thetaell}). Notice that $\left|\cos\left(\theta_{\ell}\right)\right|\leq1$,
yielding a constraint on $\xi_{\ell}$ as 
\begin{equation}
\left|\xi_{\ell}\right|\leq\sqrt{\xi_{\ell,0}^{2}+\left(\Delta_{\ell}-1\right)\Delta_{\ell}}.\label{eq:xiellmax}
\end{equation}

\subsubsection{Resolving the $B$, $C$, and $D$ blocks\label{subsec:Resolving-the-BCD}}

We begin by deriving a subtle symmetry between $\mathcal{G}_{B}$
and $\mathcal{G}_{C}$, described by Eq. (\ref{eq:g013}) and Eq.
(\ref{eq:g023}), which can be rewritten as 
\begin{equation}
\widetilde{\mathcal{G}_{C}^{T}}=-\mathcal{G}_{B},\label{eq:gcgbrelation}
\end{equation}
where the tilde is defined by the following rules. For a $2\times1$
matrix $m$, we have $\left[\tilde{m}\right]_{21}=\left[m\right]_{11}$
and $\left[\tilde{m}\right]_{11}=\left[m\right]_{21}$. For a $1\times2$
matrix $m$, we have $\left[\tilde{m}\right]_{12}=\left[m\right]_{11}$
and $\left[\tilde{m}\right]_{11}=\left[m\right]_{12}$. For a $2\times2$
matrix $m$, we have $\left[\tilde{m}\right]_{11}=\left[m\right]_{22}$,
$\left[\tilde{m}\right]_{22}=\left[m\right]_{11}$, $\left[\tilde{m}\right]_{12}=\left[m\right]_{21}$,
and $\left[\tilde{m}\right]_{21}=\left[m\right]_{12}$. In order to
prove Eq. (\ref{eq:gcgbrelation}), we use Eq. (\ref{eq:gc}) to obtain
\begin{align}
 & \mathcal{G}_{C}=g_{C}g_{A}^{-1}S_{A}^{-1}\mathcal{G}_{A},\\
 & \widetilde{\mathcal{G}_{C}^{T}}=\widetilde{\mathcal{G}_{A}^{T}}\widetilde{\left(S_{A}^{-1}\right)^{T}}\widetilde{\left(g_{A}^{-1}\right)^{T}}\widetilde{g_{C}^{T}}=\mathcal{G}_{A}S_{A}^{-1}g_{A}^{-1}\widetilde{g_{C}^{T}}.
\end{align}
Given that $\mathcal{G}_{B}=\mathcal{G}_{A}g_{A}^{-1}g_{B}$ (see
Eq. (\ref{eq:g0B})) and that $S_{A}$ commutes with $g_{A}$ (see
Section \ref{subsec:Resolving-the-self-consistency-A}), Eq. (\ref{eq:gcgbrelation})
can be proven by verifying that 
\begin{align}
 & -S_{A}g_{B}=\widetilde{g_{C}^{T}}\\
 & =-\sqrt{\frac{c_{\ell}}{\sin\left(\theta_{\ell}\right)}}\left(\begin{array}{c}
\sin\left(\theta_{\ell}\right)\mathcal{A}_{<,\ell}+\mathcal{A}_{>,\ell}\\
\cos\left(\theta_{\ell}\right)\mathcal{A}_{<,\ell}
\end{array}\right).
\end{align}
Using Eq. (\ref{eq:g0B}), we obtain 
\begin{align}
 & \mathcal{G}_{\ell,13}=-\mathcal{G}_{\ell,32}=\frac{1}{2\left(g_{\ell,12}^{2}+n_{\ell}^{2}\right)\sqrt{c_{\ell}\sin\left(\theta_{\ell}\right)}}\nonumber \\
 & \times\big(\left(2g_{\ell,12}\mathcal{G}_{\ell,12}+n_{\ell}\right)\cos\left(\theta_{\ell}\right)\mathcal{A}_{>,\ell}\nonumber \\
 & -\left(g_{\ell,12}-2n_{\ell}\mathcal{G}_{\ell,12}\right)\left(\mathcal{A}_{<,\ell}+\sin\left(\theta_{\ell}\right)\mathcal{A}_{>,\ell}\right)\big),\\
 & \mathcal{G}_{\ell,23}=-\mathcal{G}_{\ell,31}=\frac{1}{2\left(g_{\ell,12}^{2}+n_{\ell}^{2}\right)\sqrt{c_{\ell}\sin\left(\theta_{\ell}\right)}}\nonumber \\
 & \times\big(\left(2g_{\ell,12}\mathcal{G}_{\ell,12}+n_{\ell}\right)\left(\mathcal{A}_{<,\ell}+\sin\left(\theta_{\ell}\right)\mathcal{A}_{>,\ell}\right)\nonumber \\
 & +\left(g_{\ell,12}-2n_{\ell}\mathcal{G}_{\ell,12}\right)\cos\left(\theta_{\ell}\right)\mathcal{A}_{>,\ell}\big).
\end{align}
To simplify these two equations, we introduce $\mathcal{I}_{\ell}$,
$\mathcal{J}_{\ell}$, $\mathcal{A}'_{<,\ell}$, and $\mathcal{A}'_{>,\ell}$
(defined in Eqns. (\ref{eq:Iell})-(\ref{eq:Aprime>ell})), which
yields Eqns. (\ref{eq:g013}) and (\ref{eq:g023}). To compute $\mathcal{G}_{\ell,33}$,
Eqns. (\ref{eq:g0D}), (\ref{eq:gagbGAGA}), and (\ref{eq:gconega})
can be used to obtain 
\begin{align}
 & \mathcal{G}_{D}=g_{D}+\mathcal{G}_{C}\left(1-\mathcal{G}_{A}\right)^{-1}\left(g_{A}-\mathcal{G}_{A}\right)\mathcal{G}_{A}^{-1}\mathcal{G}_{B}\label{eq:g033}\\
 & =n_{\ell}+\frac{4}{4\mathcal{G}_{\ell,12}^{2}+1}\nonumber \\
 & \times\left(\left(\mathcal{G}_{\ell,13}^{2}-\mathcal{G}_{\ell,23}^{2}\right)\left(g_{\ell,12}-\mathcal{G}_{\ell,12}\right)-2\delta n_{\ell}\mathcal{G}_{\ell,13}\mathcal{G}_{\ell,23}\right).\label{eq:Gell33alt}
\end{align}
Introducing $i_{\ell}$ and $j_{\ell}$, defined in Eqns. (\ref{eq:iell})
and (\ref{eq:jell}), we can simplify Eq. (\ref{eq:Gell33alt}) to
(\ref{eq:Gell33}). The matrix elements of $\left(\barhat[a]_{\ell}^{\dagger\left(3\right)}\barhat[a]_{\ell}^{\left(3\right)}\right)_{w,\ell}$
can then be determined as 
\begin{align}
 & \left(\barhat[a]_{\ell}^{\dagger\left(3\right)}\barhat[a]_{\ell}^{\left(3\right)}\right)_{w,\ell}=\left(\begin{array}{cc}
h_{z}+h_{0} & h_{x}\\
h_{x} & h_{0}-h_{z}
\end{array}\right),\\
 & h_{0}=\mathcal{G}_{\ell,33}-\frac{1}{2}j_{\ell}\csc\left(\phi_{\ell,0}\right),\\
 & h_{z}=\frac{1}{2}i_{\ell}\csc\left(\phi_{\ell,0}\right),\\
 & h_{x}=\frac{1}{2}j_{\ell}\csc\left(\phi_{\ell,0}\right).
\end{align}
Eqns. (\ref{eq:Oeff}) and (\ref{eq:Uell}) may be used to evaluate
\begin{align}
\left(\barhat[a]_{\ell}^{\dagger\left(3\right)}\barhat[a]_{\ell}^{\left(3\right)}\right)_{\eff,\ell} & =\left(\begin{array}{cc}
f_{\ell,z}+f_{\ell,0} & f_{\ell,x}\\
f_{\ell,x} & f_{\ell,0}-f_{\ell,z}
\end{array}\right)\\
 & =f_{\ell,0}+f_{\ell,x}\hat{\sigma}^{x}+f_{\ell,z}\hat{\sigma}^{z},
\end{align}
and we define
\begin{align}
\hat{n}_{eff,\ell} & \equiv\left(\barhat[a]_{\ell}^{\dagger\left(3\right)}\barhat[a]_{\ell}^{\left(3\right)}\right)_{\eff}\\
 & =I_{2^{\ell-1}}\otimes\left(\barhat[a]_{\ell}^{\dagger\left(3\right)}\barhat[a]_{\ell}^{\left(3\right)}\right)_{\eff,\ell}\otimes I_{2^{2N_{orb}-\ell}}\\
 & =f_{\ell,0}+f_{\ell,x}\hat{\sigma}_{\ell}^{x}+f_{\ell,z}\hat{\sigma}_{\ell}^{z},
\end{align}
where $f_{\ell,0}$, $f_{\ell,x}$, and $f_{\ell,z}$ are given in
Eqns. (\ref{eq:fell0}), (\ref{eq:fellx}), and (\ref{eq:fellz}).
The derivations of Eqns. (\ref{eq:algolist1})-(\ref{eq:fellx}) are
now complete.

\subsection{Examining the qubit energy form in special cases\label{subsec:Examining-special-cases}}

In this section, we showcase the qubit energy form for the special
case of half-filled orbitals. Additionally, we examine how the qubit
energy form for $\hat{\rho}_{G3}$ recovers the qubit energy forms
for $\hat{\rho}_{G2}$ and $\hat{\rho}_{B2}$.

\subsubsection{The case of half-filled orbitals\label{subsec:The-case-of-half-filled-orbs}}

In this section, we examine the case of half-filling with particle-hole
symmetry where $n_{\ell}=1/2$ and $\mathcal{A}_{<\ell}=\mathcal{A}_{>\ell}=\frac{1}{2}A_{\ell}$.
Using the general algorithm given in Eqs. (\ref{eq:algolist1})-(\ref{eq:fellx}),
we provide corresponding results. Starting with $\xi_{\ell,0}=\frac{1}{2}$
and $\delta n_{\ell}=0$, we obtain 
\begin{align}
 & \theta_{\ell}=\cos^{-1}\left(\frac{2\xi_{\ell}}{1-2\Delta_{\ell}}\right),\\
 & c_{\ell}=1,\\
 & \cot\left(\phi_{\ell,0}\right)=\frac{\cot\left(\theta_{\ell}\right)\left(1-4\Delta_{\ell}\left(\Delta_{\ell}\cot^{2}\left(\theta_{\ell}\right)+1\right)\right)}{4\left(\Delta_{\ell}-1\right)\Delta_{\ell}\cot^{2}\left(\theta_{\ell}\right)-1},\\
 & \mathcal{G}_{\ell,12}=\frac{\cot\left(\frac{\theta_{\ell}}{2}\right)\left(-2\Delta_{\ell}\cos\left(\theta_{\ell}\right)+\cos\left(\theta_{\ell}\right)-1\right)}{2\left(2\Delta_{\ell}-1\right)\cos\left(\theta_{\ell}\right)-2},\\
 & \mathcal{I}_{\ell}=\frac{4\Delta_{\ell}^{2}\cos^{2}\left(\theta_{\ell}\right)+\sin^{2}\left(\theta_{\ell}\right)}{4\left(\cos\left(\theta_{\ell}\right)-1\right)\left(\left(2\Delta_{\ell}-1\right)\cos\left(\theta_{\ell}\right)-1\right)},\\
 & \mathcal{J}_{\ell}=\frac{4\Delta_{\ell}^{2}\cos\left(\theta_{\ell}\right)\cot\left(\theta_{\ell}\right)+\sin\left(\theta_{\ell}\right)}{4\left(2\Delta_{\ell}-1\right)\cos\left(\theta_{\ell}\right)-4},\\
 & \mathcal{A}'_{<,\ell}=\frac{1}{2}A_{\ell}\left(\sin\left(\theta_{\ell}\right)+1\right),\\
 & \mathcal{A}'_{>,\ell}=\frac{1}{2}A_{\ell}\cos\left(\theta_{\ell}\right),\\
 & \mathcal{G}_{\ell,13}=\mathcal{G}_{\ell,23}=-\frac{A_{\ell}\left(\sin\left(\theta_{\ell}\right)+\cos\left(\theta_{\ell}\right)+1\right)}{2\sqrt{\sin\left(\theta_{\ell}\right)}\left(\left(2\Delta_{\ell}-1\right)\cos\left(\theta_{\ell}\right)-1\right)},\\
 & i_{\ell}=\frac{2A_{\ell}^{2}\left(\csc\left(\theta_{\ell}\right)+1\right)}{4\Delta_{\ell}^{2}\cot^{2}\left(\theta_{\ell}\right)+1},
\end{align}
and $j_{\ell}=0$, $\mathcal{G}_{\ell,33}=\frac{1}{2}$, and 
\begin{align}
 & p_{\ell}=\frac{\cot\left(\theta_{\ell}\right)}{2\xi_{\ell}}\left(\frac{1-4\Delta_{\ell}\left(\Delta_{\ell}\cot^{2}\left(\theta_{\ell}\right)+1\right)}{4\left(\Delta_{\ell}-1\right)\Delta_{\ell}\cot^{2}\left(\theta_{\ell}\right)-1}-2\Delta_{\ell}\right),
\end{align}
and $q_{\ell}=0$, $f_{\ell,0}=\frac{1}{2}$, $f_{\ell,x}=0$, and
\begin{align}
 & f_{\ell,z}=-\frac{A_{\ell}^{2}\left(2\Delta_{\ell}-1\right)\cot\left(\theta_{\ell}\right)\left(\csc\left(\theta_{\ell}\right)+1\right)}{2\xi_{\ell}\left(4\left(\Delta_{\ell}-1\right)\Delta_{\ell}\cot^{2}\left(\theta_{\ell}\right)-1\right)}\\
 & =\frac{A_{\ell}^{2}}{4\xi_{\ell}^{2}-1}\left(\sqrt{1-\frac{4\xi_{\ell}^{2}}{\left(1-2\Delta_{\ell}\right){}^{2}}}+1\right).
\end{align}
It is also useful to define 
\begin{equation}
\mathcal{F}_{\ell}\equiv-2f_{\ell,z}=\frac{2A_{\ell}^{2}}{1-4\xi_{\ell}^{2}}\left(\sqrt{1-\frac{4\xi_{\ell}^{2}}{\left(1-2\Delta_{\ell}\right){}^{2}}}+1\right),\label{eq:F_half_filling}
\end{equation}
yielding 
\begin{align}
 & \left(\barhat[n]_{\ell}^{\left(3\right)}\right)_{\eff,\ell}=\left(\begin{array}{cc}
\frac{1}{2}\left(1-\mathcal{F}_{\ell}\right) & 0\\
0 & \frac{1}{2}\left(\mathcal{F}_{\ell}+1\right)
\end{array}\right),\\
 & \hat{n}_{eff,\ell}=\left(\barhat[n]_{\ell}^{\left(3\right)}\right)_{\eff}=\frac{1}{2}+\left(\hat{n}_{\ell}-\frac{1}{2}\right)\mathcal{F}_{\ell}.\label{eq:neff}
\end{align}
Eq. (\ref{eq:neff}) can be plugged into Eq. (\ref{eq:energy_n3_qubit})
to obtain the total energy. This result will be applied to the multiorbital
Hubbard model in Section \ref{sec:Applications:-Multi-orbital-Hubb}. 

\subsubsection{Recovering the qubit energy form for $\hat{\rho}_{G2}$ \label{subsec:Recovering-the-N=00003D2_gtype}}

Given that $\hat{\rho}_{G2}$ is a special case of $\hat{\rho}_{G3}$,
it is clear that the former can be obtained by constraining the latter.
We previously demonstrated that restricting the momentum density distribution
to be flat in each region and taking $\mathcal{G}_{\ell,12}=1/2$
within $\hat{\rho}_{G3}$ will recover $\hat{\rho}_{G2}$ in the case
where $\{n_{kl;0}\}$ corresponds to a pure state \cite{Cheng2023035127}.
Here we illustrate this fact using the qubit parametrization. We begin
by enforcing $\mathcal{G}_{\ell,12}=1/2$, and Eq. (\ref{eq:g012})
yields $\phi_{\ell,0}=-\pi/2$. Using Eq. (\ref{eq:phiell0Intheta})
and $\phi_{\ell,0}=-\pi/2$, we obtain 
\begin{equation}
-\xi_{\ell,0}^{2}+\Delta_{\ell}^{2}\cot^{2}\left(\theta_{\ell}\right)+\Delta_{\ell}=0.\label{eq:gutzeq}
\end{equation}
Using Eqns. (\ref{eq:thetaell}) and (\ref{eq:gutzeq}), we obtain
\begin{align}
 & \Delta_{\ell}=\xi_{\ell,0}^{2}-\xi_{\ell}^{2},\label{eq:deltagutz}\\
 & \cot\left(\theta_{\ell}\right)=\frac{\xi_{\ell}}{\xi_{\ell,0}^{2}-\xi_{\ell}^{2}},\\
 & g_{\ell,12}=\xi_{\ell},\\
 & c_{\ell}=\sqrt{1-\frac{2\text{\ensuremath{\delta}n}_{\ell}}{n_{\ell}^{2}+\xi_{\ell}^{2}}},
\end{align}
yielding $\mathcal{I}_{\ell}=\frac{1}{2}\left(n_{\ell}+\xi_{\ell}\right)$
and $\mathcal{J}_{\ell}=\frac{1}{2}\left(\xi_{\ell}-n_{\ell}\right)$.
Using the assumption of a flat momentum density distribution, given
by $n_{k\ell}|_{k\in<}=\frac{\xi_{\ell}^{2}}{n_{\ell}}+n_{\ell}$
and $n_{k\ell}|_{k\in>}=\frac{\xi_{\ell}^{2}}{n_{\ell}-1}+n_{\ell}$,
which yields a quasi-particle weight of 
\begin{equation}
n_{k\ell}|_{k\in<}-n_{k\ell}|_{k\in>}=\frac{\xi_{\ell}^{2}}{\xi_{\ell,0}^{2}},
\end{equation}
we obtain 
\begin{align}
 & \mathcal{A}_{<,\ell}=\sqrt{\left(\xi_{\ell,0}^{2}-\xi_{\ell}^{2}\right)\left(n_{\ell}^{2}+\xi_{\ell}^{2}\right)},\\
 & \mathcal{A}_{>,\ell}=\sqrt{\left(\xi_{\ell,0}^{2}-\xi_{\ell}^{2}\right)\left(\left(n_{\ell}-1\right){}^{2}+\xi_{\ell}^{2}\right)},\\
 & \mathcal{A}'_{<,\ell}=n_{\ell}\sqrt{\frac{\xi_{\ell,0}^{2}-\xi_{\ell}^{2}}{n_{\ell}^{2}+\xi_{\ell}^{2}}},\\
 & \mathcal{A}'_{>,\ell}=\xi_{\ell}\sqrt{\frac{\xi_{\ell,0}^{2}-\xi_{\ell}^{2}}{n_{\ell}^{2}+\xi_{\ell}^{2}}},\\
 & \mathcal{G}_{\ell,13}=\mathcal{G}_{\ell,23}=-\mathcal{G}_{\ell,31}=-\mathcal{G}_{\ell,32}=\frac{1}{2},
\end{align}
and $i_{\ell}=1$, $j_{\ell}=0$, $\mathcal{G}_{\ell,33}=\frac{1}{2}$,
$p_{\ell}=-1$, $q_{\ell}=0$, $f_{\ell,0}=\frac{1}{2}$, $f_{\ell,z}=-\frac{1}{2}$,
$f_{\ell,x}=0$, and 
\begin{equation}
\left(\barhat[n]_{\ell}^{\left(3\right)}\right)_{\eff,\ell}=\left(\begin{array}{cc}
0 & 0\\
0 & 1
\end{array}\right),
\end{equation}
yielding $\hat{n}_{eff,\ell}=\hat{n}_{\ell}$, recovering the qubit
energy form obtained from $\hat{\rho}_{G2}$. 

\subsubsection{Recovering the qubit energy form for $\hat{\rho}_{B2}$\label{subsec:Recovering-the-N=00003D2-bytpe}}

Given that $\hat{\rho}_{B2}$ is a special case of $\hat{\rho}_{G3}$,
it is clear that the former can be obtained by constraining the latter.
However, it should be noted that in the qubit parametrization, $\hat{K}_{2}$
in $\hat{\rho}_{G3}$ is assumed to correspond to a Slater determinant,
but $\hat{K}_{2}$ must be the identity to recover $\hat{\rho}_{B2}$.
Nonetheless, we demonstrate that the qubit energy form of $\hat{\rho}_{G3}$
can still be constrained to recover the qubit energy form of $\hat{\rho}_{B2}$.
The solution is to restrict $\rhoeff$ to be a diagonal matrix, implying
that $\xi_{\ell}=0$, yielding $\theta_{\ell}=\frac{\pi}{2}$, $g_{\ell,12}=0$,
$c_{\ell}=\frac{1}{n_{\ell}}-1$, $\phi_{\ell}=0$, $\phi_{\ell,0}=-\frac{\pi}{2}$,
$\mathcal{G}_{\ell,12}=\frac{1}{2}$, $\mathcal{I}_{\ell}=\frac{n_{\ell}}{2}$,
$\mathcal{J}_{\ell}=-\frac{n_{\ell}}{2}$, $\mathcal{A}'_{<,\ell}=\mathcal{A}_{<,\ell}+\mathcal{A}_{>,\ell}=A_{\ell}$,
$\mathcal{A}'_{>,\ell}=0$, $\mathcal{G}_{\ell,13}=\mathcal{G}_{\ell,23}=\frac{A_{\ell}}{2\xi_{\ell,0}}$,
$i_{\ell}=\frac{A_{\ell}^{2}}{\xi_{\ell,0}^{2}}$, $j_{\ell}=0$,
$\mathcal{G}_{\ell,33}=n_{\ell}-\frac{A_{\ell}^{2}\delta n_{\ell}}{\xi_{\ell,0}^{2}}$,
$p_{\ell}=-1$, $q_{\ell}=0$, $f_{\ell,0}=n_{\ell}-\frac{A_{\ell}^{2}\text{\ensuremath{\delta}n}_{\ell}}{\xi_{\ell,0}^{2}}$,
$f_{\ell,z}=-\frac{A_{\ell}^{2}}{2\xi_{\ell,0}^{2}}$, and $f_{\ell,x}=0$.
Finally, we have

\begin{align}
 & \left(\barhat[n]_{\ell}^{\left(3\right)}\right)_{\eff,\ell}=n_{\ell}+\frac{A_{\ell}^{2}}{\xi_{\ell,0}^{2}}\left(\begin{array}{cc}
-n_{\ell} & 0\\
0 & 1-n_{\ell}
\end{array}\right),\\
 & \hat{n}_{eff,\ell}=n_{\ell}+\frac{A_{\ell}^{2}}{\xi_{\ell,0}^{2}}\left(\hat{n}_{\ell}-n_{\ell}\right),
\end{align}
recovering the qubit energy form of $\hat{\rho}_{B2}$. The preceding
result demonstrates that when all orbitals have $\xi_{\ell}=0$, implying
that the system is a Mott insulator, the $\hat{\rho}_{G3}$ solution
can be recovered by $\hat{\rho}_{B2}$, demonstrating the power of
$\hat{\rho}_{B2}$ in the Mott phase.

\section{Applications: multiorbital Hubbard model at half filling with particle-hole
symmetry in $d=\infty$ \label{sec:Applications:-Multi-orbital-Hubb}}

Here we showcase how to use the qubit energy form for $\hat{\rho}_{G3}$
to study the multiorbital Hubbard model in $d=\infty$ with density-density
interactions. While the qubit energy form can be applied to arbitrary
densities, here we study the case of half-filling where the local
interaction energy takes a simple form. In Section \ref{subsec:Numerical-minimization-of},
we explicitly evaluate the qubit energy form for half-filling and
demonstrate how to analytically minimize over the momentum density
distribution, yielding a final energy form which can straightforwardly
be numerically minimized. In Section \ref{subsec:Understanding-the-Mott-single-band-Delta},
we demonstrate that $\Delta_{\ell}$, defined in Eq. (\ref{eq:delta definition}),
is a key variable for understanding the Mott transition. For a special
density-of-states, the minimization can be analytically performed,
yielding an analytical relation between the ground state energy and
the Hubbard $U$ via $\Delta_{\ell}$. In Section \ref{subsec:Understanding-the-effect-J-two-band},
we study the effect of the Hund coupling $J$ on the nature of the
Mott transition in the multiorbital Hubbard model. 

\subsection{Numerical minimization of the qubit trial energy\label{subsec:Numerical-minimization-of}}

\subsubsection{Evaluating the qubit trial energy\label{subsec:Minimizing-over-local-param} }

In order to elucidate the results of the multiorbital Hubbard model,
we first begin by considering the single orbital case with density
of states $D\left(\epsilon\right)$ having particle-hole symmetry.
Due to spin symmetry, we omit the spin-orbital index $\ell$ in $\xi_{\ell},$$\Delta_{\ell}$,
and $A_{\ell}=\mathcal{A}_{<\ell}+\mathcal{A}_{>\ell}$. The qubit
energy form for $\hat{\rho}_{G3}$ is given as (see Section \ref{subsec:The-case-of-half-filled-orbs}
for derivation)
\begin{align}
 & E(n(\epsilon),\rhoeff)=2\int_{-\infty}^{\infty}d\epsilon D\left(\epsilon\right)\epsilon n\left(\epsilon\right)+U\langle\hat{n}_{eff,\uparrow}\hat{n}_{eff,\downarrow}\rangle_{\rhoeff},\\
 & \hat{n}_{eff,\ell}=\frac{1}{2}+\mathcal{F}(\Delta,A,\xi)\left(\hat{n}_{\ell}-\frac{1}{2}\right),\\
 & \mathcal{F}(\Delta,A,\xi)=\frac{2A^{2}}{1-4\xi^{2}}\left(\sqrt{1-\frac{4\xi^{2}}{\left(1-2\Delta\right){}^{2}}}+1\right),\label{eq:F_half_filling-1}\\
 & A=\int_{-\infty}^{\infty}d\epsilon D(\epsilon)\sqrt{n(\epsilon)\left(1-n(\epsilon)\right)},\label{eq:A-half-filling}\\
 & \Delta=\int_{0}^{\infty}d\epsilon D(\epsilon)n(\epsilon),\label{eq:delta-half-filling}\\
 & \xi=\frac{1}{2}\langle\hat{\sigma}_{\ell}^{x}\rangle_{\rhoeff},
\end{align}
where particle-hole symmetry implies $n\left(\epsilon\right)+n\left(-\epsilon\right)=1$,
and in order to ensure that $\mathcal{F}(\Delta,A,\xi)$ is real,
$\xi$ must satisfy the following constraint
\begin{align}
 & \left|\xi\right|\leq\frac{1}{2}-\Delta.\label{eq:xi_constraint}
\end{align}
\global\long\def\bared{\mathcal{D}}%
The effective many-body density matrix $\rhoeff$ can be encoded as
$\rhoeff=|\Psi\rangle\langle\Psi|$ where $|\Psi\rangle$ can be parametrized
using a single parameter $\bared\in[0,1/2]$ as
\begin{equation}
|\Psi\rangle=\left(\begin{array}{cccc}
\sqrt{\bared} & \sqrt{\frac{1}{2}-\bared} & \sqrt{\frac{1}{2}-\bared} & \sqrt{\bared}\end{array}\right)^{T}.
\end{equation}
This parametrization ensures that $n_{\ell}=1/2$ and $\xi$ can be
determined as 
\begin{equation}
\xi=2\sqrt{\left(\frac{1}{2}-\bared\right)\bared}.\label{eq:xie_ell_deff}
\end{equation}
The double occupancy is given as 
\begin{align}
 & d(\Delta,A,\bared)=\left\langle \hat{n}_{eff,\uparrow}\hat{n}_{eff,\downarrow}\right\rangle _{\rhoeff}=\left(\bared-\frac{1}{4}\right)\mathcal{F}^{2}+\frac{1}{4}\\
 & =\frac{1}{4}\left(\frac{4A^{4}(1-2\Delta+h\left(\Delta,\bared\right))^{2}}{(1-2\Delta)^{2}\left(4\bared-1\right){}^{3}}+1\right),\label{eq:d_f_of_A_delta_delta_eff}\\
 & h(\Delta,\bared)=\sqrt{8\bared\left(2\bared-1\right)+(1-2\Delta)^{2}}
\end{align}
and to have a real $d(\Delta,A,\bared)$, the $h(\Delta,\bared)$
must be real which requires $\bared$ to satisfy the following constraint
\begin{align}
 & \bared\in[0,\bared_{-}]\cup[\bared_{+},\frac{1}{2}],\label{eq:deff-range}\\
 & \bared_{\pm}=\frac{1}{4}\pm\frac{1}{2}\sqrt{(1-\Delta)\Delta}.\label{eq:deffmax-pm}
\end{align}
Finally, the trial energy for the single orbital case is a function
only of $n(\epsilon)$ and $\bared$, given as

\begin{equation}
E(n(\epsilon),\bared)=2\int_{-\infty}^{\infty}d\epsilon D\left(\epsilon\right)\epsilon n\left(\epsilon\right)+Ud(\Delta,A,\bared).
\end{equation}

We now proceed to the multiorbital case, with a local interaction
given as
\begin{equation}
\hat{H}_{loc}=\sum_{\ell<\ell'}U_{\ell\ell'}\delta\hat{n}_{\ell}\delta\hat{n}_{\ell'},
\end{equation}
where $\delta\hat{n}_{\ell}=\hat{n}_{\ell}-\frac{1}{2}$. The qubit
trial energy is given as
\begin{align}
 & E(\{n_{\ell}(\epsilon)\},\rhoeff)=\sum_{\ell}\int_{-\infty}^{\infty}d\epsilon D_{\ell}\left(\epsilon\right)\epsilon n_{\ell}\left(\epsilon\right)\nonumber \\
 & +\sum_{\ell<\ell'}\mathcal{F}(\Delta_{\ell},A_{\ell},\xi_{\ell})\mathcal{F}(\Delta_{\ell'},A_{\ell'},\xi_{\ell'})U_{\ell\ell'}\langle\delta\hat{n}_{\ell}\delta\hat{n}_{\ell'}\rangle_{\rhoeff},
\end{align}
where $\rhoeff$ is a pure state that is restricted to $\langle\hat{n}_{\ell}\rangle_{\rhoeff}=\frac{1}{2}$,
particle-hole symmetry implies $n_{\ell}\left(\epsilon\right)+n_{\ell}\left(-\epsilon\right)=1$,
and $\xi_{\ell}$ must satisfy the following constraint
\begin{align}
 & \left|\xi_{\ell}\right|\leq\frac{1}{2}-\Delta_{\ell}.\label{eq:xi_constraint_ell}
\end{align}
The expression for $\mathcal{F}(\Delta_{\ell},A_{\ell},\xi_{\ell})$
is given in Eq. (\ref{eq:F_half_filling-1}). The most straightforward
approach to satisfying Eq. (\ref{eq:xi_constraint_ell}) is to first
choose $\rhoeff$, and then Eq. (\ref{eq:xi_constraint_ell}) becomes
a linear constraint on $n_{\ell}(\epsilon)$ (see Section \ref{sec:Overview-of-the}
for further discussion). 

\subsubsection{Numerical minimization of the qubit trial energy\label{subsec:Numerical-minimization-of-qubit-trial}}

\begin{figure}
\includegraphics[width=0.99\columnwidth]{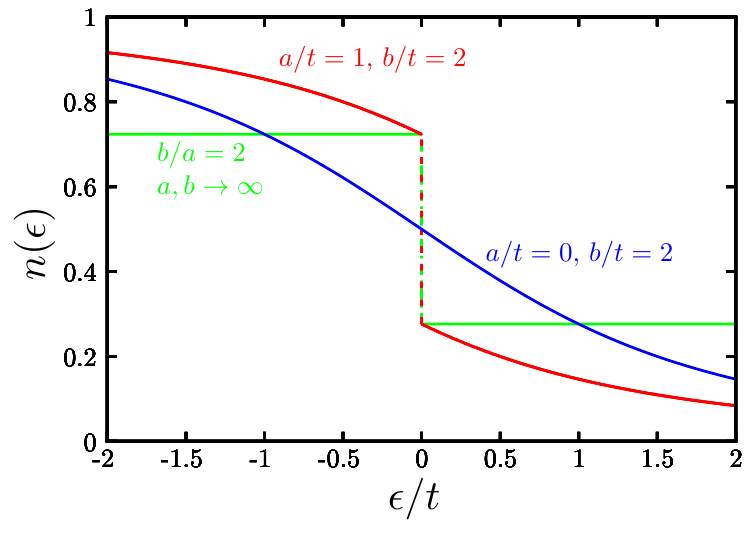}\caption{\label{fig:momentum-density-distribution}A plot of the partially
optimized momentum density distribution $n\left(\epsilon,a,b\right)$
for various values of $a$ and $b$. }
\end{figure}

We now proceed to minimizing the qubit trial energy. As before, we
first focus on the single orbital model for clarity, and then consider
the multiorbital case. The general numerical minimization has been
described in Section \ref{sec:Overview-of-the}, which consists of
two steps: First, the momentum density distribution is partially optimized
under the constraint of $n_{\ell}$, $\Delta_{\ell}$, $\mathcal{A}_{<\ell}$,
and $\mathcal{A}_{>\ell}$, or through four Lagrange multiplier $a_{<\ell}$,
$a_{>\ell},$ $b_{<\ell}$, and $b_{>\ell}$. Second, one needs to
minimize over the remaining $2^{2N_{orb}}+3\times2N_{orb}-1$ independent
variational parameters subjected to the inequality constraint given
by Eq. (\ref{eq:general_constraint}). For the half-filled, particle-hole
symmetric case, there are several simplifications. First, the optimized
momentum density distribution now only depends on $\Delta_{\ell}$
and $A_{\ell}=\mathcal{A}_{<\ell}+\mathcal{A}_{>\ell}$, or just two
Lagrange multipliers $a=a_{<\sigma}=-a_{>\sigma}$ and $b=b_{>\sigma}=b_{<\sigma}$,
and the partially optimized momentum density distribution is given
as 
\begin{equation}
n\left(\epsilon,a,b\right)=\begin{cases}
\frac{1}{2}\left(1-\frac{a+\epsilon}{\sqrt{(a+\epsilon)^{2}+b^{2}}}\right) & \epsilon>0\\
\frac{1}{2}\left(1+\frac{a-\epsilon}{\sqrt{(a-\epsilon)^{2}+b^{2}}}\right) & \epsilon<0
\end{cases},\label{eq:n_epsilon_a_b}
\end{equation}
where the Lagrange multipliers $a$ and $b$ are determined by $\Delta$
and $A$ via inverting the following relation
\begin{align}
 & A(a,b)=\int_{-\infty}^{\infty}d\epsilon D(\epsilon)\sqrt{n(\epsilon,a,b)\left(1-n(\epsilon,a,b)\right)},\label{eq:A-a-b}\\
 & \Delta(a,b)=\int_{0}^{\infty}d\epsilon D(\epsilon)n(\epsilon,a,b),\label{eq:delta-a-b}
\end{align}
which yields $a\left(\Delta,A\right)$ and $b\left(\Delta,A\right)$.
In order to better appreciate the flexibility of $n\left(\epsilon,a,b\right)$,
it is useful to examine the behavior for various choices of $a$ and
$b$ (see Figure \ref{fig:momentum-density-distribution}). For $a>0$
and $b>0$, the distribution corresponds to a Fermi liquid, given
that the discontinuity of $n\left(\epsilon,a,b\right)$ at $\epsilon=0$
yields a quasi-particle weight $Z$ of 
\begin{equation}
Z=\frac{a}{\sqrt{a^{2}+b^{2}}},\label{eq:Z_a_b}
\end{equation}
which will be between zero and one. For the special case where $a\rightarrow\infty$
and $b\rightarrow\infty$ with $a/b$ remaining finite, the distribution
recovers a flat distribution, which is obtained for an optimized $\hat{\rho}_{G2}$.
Finally, when $a=0$ and $b>0$, the system is in the Mott phase where
$Z=0$. 

The second simplification for half-filling and particle-hole symmetry
is that the number of independent parameters is $2^{2N_{orb}}+2N_{orb}-1$,
given that $n_{\ell}=1/2$ and $\mathcal{A}_{<\ell}=\mathcal{A}_{>\ell}$.
If we have further symmetry between different spin orbitals, the number
of independent parameters can further be reduced. For example, for
the single-orbital case with spin symmetry, we have $A_{\ell}=A$
and $\Delta_{\ell}=\Delta$, and therefore the number of independent
parameters is $2^{2}+2-1-2=3$. As discussed in Section \ref{sec:Overview-of-the},
there are two possible strategies. The first strategy starts from
$a,b$, which determines $n\left(\epsilon\right)$, and $\rhoeff$
is specified via $\bared$, which will have a range given by Eq. (\ref{eq:deff-range}).
Mathematically, this qubit trial energy is given by 
\begin{align}
E(a,b,\bared)= & 2\int_{-\infty}^{\infty}d\epsilon D\left(\epsilon\right)\epsilon n\left(\epsilon,a,b\right)\nonumber \\
 & +Ud(\Delta(a,b),A(a,b),\bared).\label{eq:single-orbital-trial-1}
\end{align}
This strategy allows the energy to be explicitly written in terms
of the variational parameters, and thus it is straightforward to compute
the derivatives via automatic differentiation. For the multiorbital
case, the constraint between $\rhoeff$ and $\{\Delta_{\ell}\}$ is
given by Eq. (\ref{eq:xi_constraint_ell}), which is not straightforward
to implement for a given $\{\Delta_{\ell}\}$. Therefore, it is useful
to pursue a second strategy.

We begin by illustrating the second strategy in the single-orbital
case. This strategy begins by specifying $\rhoeff$ via $\bared$,
then the range of $\Delta$ can be determined from the constraint
given by Eq. (\ref{eq:xi_constraint}), and then $\Delta$ and $b$
can be specified, which determines $a$ via inverting Eq. (\ref{eq:delta-a-b}),
denoted as $a\left(\Delta,b\right)$. Mathematically, the resulting
qubit trial energy is given by 

\begin{align}
E(\Delta,b,\bared)= & 2\int_{-\infty}^{\infty}d\epsilon D\left(\epsilon\right)\epsilon n\left(\epsilon,a(\Delta,b),b\right)\nonumber \\
 & +Ud(\Delta,A(a(\Delta,b),b),\bared).\label{eq:single-orbital-trial-2}
\end{align}
For the multiorbital case, the qubit trial energy is given as 
\begin{align}
 & E(\{\Delta_{\ell}\},\{b_{\ell}\},\rhoeff)\nonumber \\
 & =\sum_{\ell}\int_{-\infty}^{\infty}d\epsilon D_{\ell}\left(\epsilon\right)\epsilon n\left(\epsilon,a_{\ell}(\Delta_{\ell},b_{\ell}),b_{\ell}\right)\nonumber \\
 & +\sum_{\ell<\ell'}\mathcal{F}_{\ell}\mathcal{F}_{\ell'}U_{\ell\ell'}\langle\delta\hat{n}_{\ell}\delta\hat{n}_{\ell'}\rangle_{\rhoeff},\\
 & \mathcal{F}_{\ell}=\mathcal{F}(\Delta_{\ell},A_{\ell}(a_{\ell}(\Delta_{\ell},b_{\ell}),b_{\ell}),\frac{1}{2}\langle\hat{\sigma}_{\ell}^{x}\rangle_{\rhoeff}),
\end{align}
where $a_{\ell}(\Delta_{\ell},b_{\ell})$ and $A_{\ell}(a_{\ell},b_{\ell})$
are the multiorbital versions of $a(\Delta,b)$ and $A(a,b)$. The
second strategy allows one to automatically implement all of the constraints
in the multiorbital case, and the only downside is that $a_{\ell}(\Delta_{\ell},b_{\ell})$
must be numerically evaluated, though this is a trivial task.

\subsection{Understanding the Mott transition in the single orbital model\label{subsec:Understanding-the-Mott-single-band-Delta}}

In the preceding section, the resulting qubit trial energy for the
single orbital model at half-filling is either parametrized by $a$,
$b$, and $\bared$ (i.e. Eq. (\ref{eq:single-orbital-trial-1})),
or $\Delta$, $b$ and $\bared$ (i.e. Eq. (\ref{eq:single-orbital-trial-2})),
which both allow for straightforward numerical minimization for a
given $U$ and $D(\epsilon)$. In this section, we explore a different
perspective based on the one body reduced density matrix viewpoint
(see Ref. \cite{companion}), which provides a clearer understanding
of the Mott transition. The first step begins with the parametrization
of the qubit trial energy using $\Delta$, $A$, and $\bared$, where
the local interaction energy is $Ud(\Delta,A,\bared)$, where $d(\Delta,A,\bared)$
is defined in Eq. (\ref{eq:d_f_of_A_delta_delta_eff}). It should
be noted that $d(\Delta,A,\bared)$ has the form $1/4+A^{4}f(\Delta,\bared)$,
and therefore we can optimize $d(\Delta,A,\bared)$ with respect to
$\bared$ for fixed $\Delta$ and $A$, yielding an optimized $\bared$
as a function of $\Delta$, denoted as $\bared\left(\Delta\right)$.
Therefore, the interaction energy is purely a function of $\Delta$
and $A$, which can be viewed as a one body reduced density matrix
functional, yielding a simple picture of the Mott transition. Two
key points should be appreciated. First, $\bared\left(\Delta\right)$
has a critical value of $\Delta$, denoted $\Delta_{c}$, and $\bared\left(\Delta\right)=0$
for $\Delta\geq\Delta_{c}$, which corresponds to the Mott insulator.
The second point is that $\Delta_{c}$ is independent of $D(\epsilon)$.
While the total energy can be obtained by numerically minimizing over
$\Delta$ and $A$, further insight can be obtained by expressing
$U$ and the total energy as functions of $\Delta$.

\subsubsection{Qubit trial energy in terms of $\Delta$ and $A$\label{subsec:Qubit-energy-form-delta-A}}

\begin{figure}
\includegraphics[width=0.99\columnwidth]{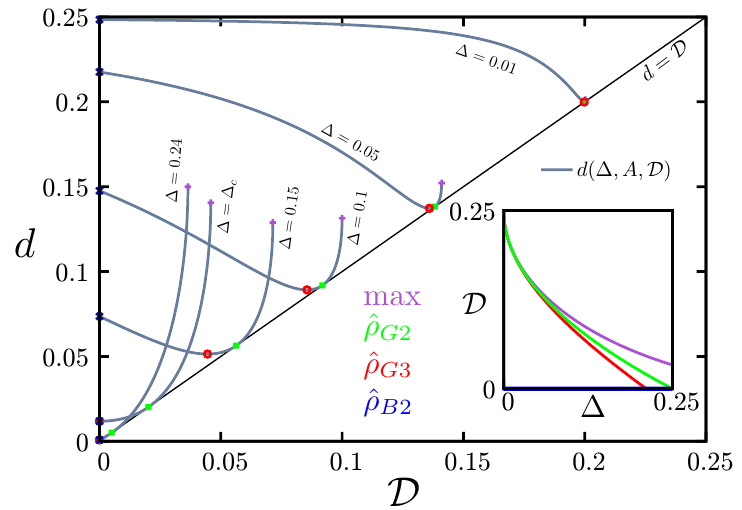}\caption{\label{fig:visualize_d_deff} A plot of the double occupancy $d$
given in Eq. (\ref{eq:d_f_of_A_delta_delta_eff}) as a function of
$\protect\bared$ for various $\Delta$ and $A=\sqrt{\left(2-4\Delta\right)\Delta}$.
The points labeled $\hat{\rho}_{G3}$ provide the optimized value
for $\protect\bared$ given by Eq. (\ref{eq:deff_with_bounds}), while
the $\hat{\rho}_{G2}$ and $\hat{\rho}_{B2}$ points provide the $\protect\bared$
values for the $\hat{\rho}_{G2}$ and $\hat{\rho}_{B2}$ given by
Eq. (\ref{eq:dG=00003Dd}) and $\protect\bared=0$, respectively.
The purple points labeled $max$ provide the maximum value for $\protect\bared$,
given as $\protect\bared_{-}$ defined in Eq. (\ref{eq:deffmax-pm}).
The inset plots $\protect\bared$ as a function of $\Delta$ for $\hat{\rho}_{G3}$,
$\hat{\rho}_{G2}$, and $\hat{\rho}_{B2}$, in addition to the maximum
value. }
\end{figure}

As outlined in Section \ref{subsec:Numerical-minimization-of-qubit-trial},
the qubit trial energy can be parametrized as
\begin{align}
 & E(\Delta,A,\bared)=K(\Delta,A)+Ud(\Delta,A,\bared),\\
 & K(\Delta,A)=2\int_{-\infty}^{\infty}d\epsilon D\left(\epsilon\right)\epsilon n\left(\epsilon,a(\Delta,A),b(\Delta,A)\right).
\end{align}
For simplicity, we study the case where $U\ge0$. We now consider
how to minimize the total energy over $\bared$, which amounts to
minimizing $d(\Delta,A,\bared)$ over $\bared$. In order to visualize
the minimization, we plot $d$ as a function of $\bared$ for a given
$\Delta$ and $A$. Given that $A$ will not influence the minimization
over $\bared$, we choose $A=\sqrt{\left(2-4\Delta\right)\Delta}$
from a flat momentum density distribution (see Figure \ref{fig:visualize_d_deff}).
For a given $\Delta$ curve, there are four distinct sets of points
denoted as $\hat{\rho}_{G2}$, $\hat{\rho}_{B2}$, $\hat{\rho}_{G3}$,
and $max$, where the $\hat{\rho}_{G3}$ points provide the optimized
values for $\bared$, the $\hat{\rho}_{G2}$ and $\hat{\rho}_{B2}$
points provide the $\bared$ values for $\hat{\rho}_{G2}$ and $\hat{\rho}_{B2}$,
respectively, and a $max$ point provides the maximum value for $\bared$
given by $\bared_{-}$ in Eq. (\ref{eq:deffmax-pm}). For small values
of $\Delta$, the optimized value of $\bared$ is nonzero, and $\bared$
monotonically decreases with increasing $\Delta$. For $\Delta$ larger
than some critical value, the optimized value for $\bared$ is zero.
Having obtained a graphical understanding of this function, we proceed
to mathematically minimize the $d$ over $\bared$ for a given $\Delta$
and $A$, which is a constrained minimization given that $\bared\in[0,\bared_{-}]$.
Solving $\partial d/\partial\bared=0$ yields 
\begin{align}
 & \bared^{\star}(\Delta)=\frac{1}{4}-\frac{1}{4}\sqrt{6(\Delta-1)\Delta+\left(3\Delta-\frac{3}{2}\right)h_{1}+\frac{9}{2}},\label{eq:deff_star}\\
 & h_{1}=\sqrt{4(\Delta-1)\Delta+9}.
\end{align}
When $\bared^{\star}(\Delta)\in[0,\bared_{-}]$, then $\bared^{\star}(\Delta)$
yields the optimized value for $\bared$, and otherwise the optimized
value is given by the minimum value for the boundary points. It is
useful to solve $\bared^{\star}(\Delta)=0$, yielding 
\begin{equation}
\Delta_{c}=\frac{1}{6}\left(3-\sqrt{3}\right)\approx0.211325.
\end{equation}
Therefore, the optimized value for $\bared$ is
\begin{equation}
\bared(\Delta)=\begin{cases}
\bared^{\star}(\Delta), & \Delta<\Delta_{c},\\
0, & \Delta\ge\Delta_{c},
\end{cases}\label{eq:deff_with_bounds}
\end{equation}
having two distinct regimes as a function of $\Delta$ (see inset
of Figure \ref{fig:visualize_d_deff}). Finally, the physical double
occupancy can be written as a function of $\Delta$ and $A$ as
\begin{equation}
d(\Delta,A)=\begin{cases}
d(\Delta,A,\bared^{\star}(\Delta)), & \Delta<\Delta_{c},\\
\frac{1}{4}-4A^{4}, & \Delta\ge\Delta_{c}.
\end{cases}\label{eq:d_with_bounds}
\end{equation}
Therefore, the total trial energy can be written purely in terms of
$\Delta$ and $A$, given as 
\begin{equation}
E(\Delta,A)=K(\Delta,A)+Ud(\Delta,A).\label{eq:qubit-trial-delta-A}
\end{equation}

In order to find the ground state for a given $D(\epsilon)$ and $U$,
it is necessary to minimize over $\Delta$ and $A$, which cannot
be performed analytically in general. However, it is important to
appreciate that the optimized value of $\Delta$ is sufficient to
determine if the system is in the Mott phase. For $\Delta=0$, the
ansatz corresponds to the Hartree-Fock wave function, while the maximum
value of $\Delta=1/4$ corresponds to a collection of isolated atoms,
and the Mott transition occurs when $\Delta=\Delta_{c}$, before the
system becomes a collection of atoms. We now verify that the system
is indeed a Mott insulator for $\Delta>\Delta_{c}$. First, the local
interaction energy is independent of $\Delta$ for $\Delta>\Delta_{c}$
(see Eq. (\ref{eq:deff_with_bounds})), dictating that $a=0$ for
the optimized $\Delta$. Second, Eq. (\ref{eq:Z_a_b}) dictates that
the quasiparticle weight is zero when $a=0$, implying a Mott insulating
state. Alternatively, when $\Delta<\Delta_{c}$, the ansatz describes
a metallic phase. 

Following the results of Sections \ref{subsec:Recovering-the-N=00003D2_gtype}
and \ref{subsec:Recovering-the-N=00003D2-bytpe}, we illustrate how
the qubit energy form for $\hat{\rho}_{G2}$ and $\hat{\rho}_{B2}$
can be recovered from $\hat{\rho}_{G3}$ when properly restricting
the variational parameters. We begin with $\hat{\rho}_{B2}$, which
can be viewed as a continuation of the Mott phase for $\hat{\rho}_{G3}$,
where $\bared=0$ and $d$ is given by $\frac{1}{4}-4A^{4}$ (see
Figure \ref{fig:visualize_d_deff}). Alternatively, the $\hat{\rho}_{G2}$
is characterized by a flat momentum density distribution determined
by $\Delta$ where $d$ and $\bared$ are identical, given as 
\begin{equation}
d_{G2}=\bared_{G2}=\frac{1}{4}\left(1-2\sqrt{\Delta}\right).\label{eq:dG=00003Dd}
\end{equation}
It is well known that for $\hat{\rho}_{G2}$ the Mott transition occurs
when the system becomes a collection of isolated atoms, meaning that
the transition happens for $\Delta=1/4$.

\subsubsection{Solution for the two-peak density of states \label{subsec:Solution-of-the-two-peak-model}}

\begin{figure}
\includegraphics[width=0.99\columnwidth]{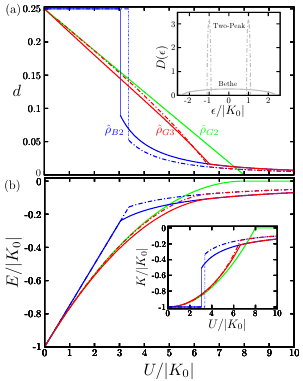}

\caption{\label{fig:compare_bethe_twopeak} A plot of the double occupancy
(panel $a$) and the total energy $E/|K_{0}|$ (panel $b$) vs. $U/|K_{0}|$
for the single band Hubbard model in $d=\infty$ at half-filling for
the Bethe lattice and the two-peak model (see Eq. (\ref{eq:DOS_two_peak})
and inset of panel ($a$)). Both models are solved using $\hat{\rho}_{G2}$,
$\hat{\rho}_{B2}$, and $\hat{\rho}_{G3}$. The inset of panel $b$
plots $K/|K_{0}|$ vs. $U/|K_{0}|$. }
\end{figure}

In Section \ref{subsec:Qubit-energy-form-delta-A}, we demonstrated
that the interaction energy can be written analytically in terms of
$\Delta$ and $A$, but the kinetic energy must be numerically determined
in terms of $\Delta$ and $A$. While the latter is a trivial numerical
problem, it still precludes a completely analytic solution for the
energy in terms of $U$. Therefore, we introduce the two-peak density
of states, where the kinetic energy can be analytically evaluated
in terms of $\Delta$ and $A$, allowing for an analytical relation
between the total energy and the Hubbard $U/|K_{0}|$ in all three
ansatz, where $K_{0}<0$ is the non-interacting kinetic energy for
the lattice. Specifically, the two-peak density of states is given
by 
\begin{equation}
D\left(\epsilon\right)=\frac{1}{2}\left(\delta\left(\epsilon-K_{0}\right)+\delta\left(\epsilon+K_{0}\right)\right).\label{eq:DOS_two_peak}
\end{equation}
While this density of states is somewhat unphysical, the results retain
all of the qualitative features of the Bethe lattice, in addition
to being quantitatively very similar assuming the same $K_{0}$ (see
Figure \ref{fig:compare_bethe_twopeak}). As expected, the $\hat{\rho}_{G2}$
solution is identical for both density of states, given that $\hat{\rho}_{G2}$
only depends on $K_{0}$, independent of the details of $D(\epsilon)$.
Interestingly, $\hat{\rho}_{G3}$ has a very similar critical value
of $U/|K_{0}|$ for both density of states, with small differences
differences in the double occupancy for a given $U/|K_{0}|$. 

\begin{figure}
\includegraphics[width=0.99\columnwidth]{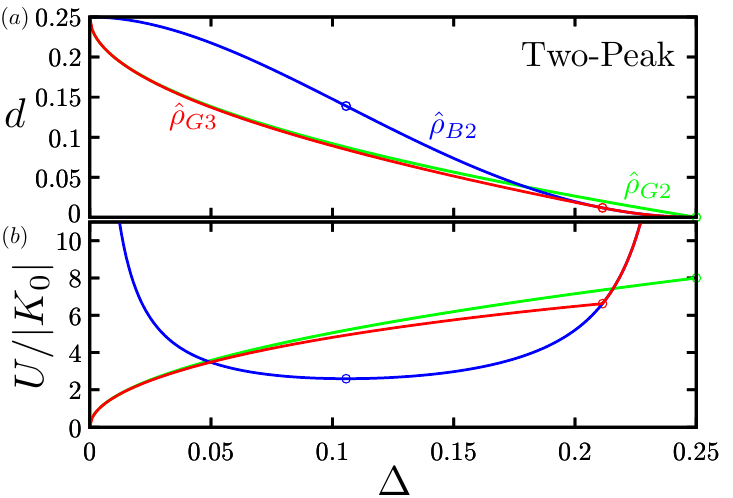}

\caption{\label{fig:d_vs_delta} Plots of the double occupancy $d$ (panel
$a$) and $U/|K_{0}|$ (panel $b$) as functions of $\Delta$ for
the two-peak model in $d=\infty$ solved using $\hat{\rho}_{G2}$,
$\hat{\rho}_{B2}$, and $\hat{\rho}_{G3}$. }
\end{figure}

We now proceed to analytically evaluate the total trial energy as
a function of $\Delta$ for the two-peak model. Using Eqns. (\ref{eq:DOS_two_peak}),
(\ref{eq:delta-half-filling}), and (\ref{eq:A-half-filling}), the
kinetic energy and $A$ are determined as $K=\left(1-4\Delta\right)K_{0}$
and $A=\sqrt{\left(2-4\Delta\right)\Delta}$, yielding the total trial
energy in terms of the single variational parameter $\Delta$, given
as
\begin{equation}
E\left(\Delta\right)=\left(1-4\Delta\right)K_{0}+Ud\left(\Delta\right),
\end{equation}
where 
\begin{align}
 & d(\Delta)=\begin{cases}
d_{1}(\Delta), & \Delta\le\Delta_{c}\\
\frac{1}{4}-16(1-2\Delta)^{2}\Delta^{2}, & \Delta>\Delta_{c}
\end{cases}\label{eq:d_delta_two_peak}
\end{align}
and $d_{1}(\Delta)$ is defined in Eq. (\ref{eq:d1_twopeak}). The
$d(\Delta)$ is plotted in Figure \ref{fig:d_vs_delta}$a$. While
for a given $U$ the energy cannot be analytically minimized over
$\Delta$, it is straightforward to find $U$ for a given value of
$\Delta$ that satisfies $dE(\Delta)/d\Delta=0$, given as 
\begin{equation}
U(\Delta)=-4\left|K_{0}\right|\left(\frac{d}{d\Delta}d(\Delta)\right)^{-1},\label{eq:UK_saddle_point}
\end{equation}
and $U/|K_{0}|$ is plotted as a function of $\Delta$ in Figure \ref{fig:d_vs_delta}$b$.
The critical value of $U$ and double occupancy at the Mott transition,
where $\Delta=\Delta_{c}$, are given as 
\begin{align}
 & U_{c}/|K_{0}|=\frac{3}{8}\left(9+5\sqrt{3}\right)\approx6.6226,\\
 & d_{c}=\frac{1}{36}\left(32\sqrt{3}-55\right)\approx0.0118229.
\end{align}
These values are the same from both the metallic and insulating sides
of the Mott transition, which can be seen in Figure \ref{fig:d_vs_delta},
confirming that the transition is continuous. 

Corresponding equations for $d$ and $U/|K_{0}|$ as functions of
$\Delta$ can be obtained for $\hat{\rho}_{G2}$ and $\hat{\rho}_{B2}$
by substituting the corresponding $d(\Delta)$ relations into Eq.
(\ref{eq:UK_saddle_point}) (see Figure \ref{fig:d_vs_delta} for
plots). For $\hat{\rho}_{G2}$, we have 
\begin{align}
 & d_{G2}(\Delta)=\frac{1}{4}\left(1-2\sqrt{\Delta}\right),\\
 & U_{G2}(\Delta)=16\left|K_{0}\right|\sqrt{\Delta},
\end{align}
which recovers the Gutzwiller approximation, and the Mott transition
occurs at $\Delta=1/4$, where $U_{c,G2}/\left|K_{0}\right|=8$ and
$d_{c,G2}=0$. For $\hat{\rho}_{B2}$, there are some subtleties to
consider. For a given $U$, there are several candidate values of
$\Delta$: the value given by the saddle point Eq. (\ref{eq:UK_saddle_point}),
which yields 
\begin{align}
 & d_{B2}(\Delta)=\frac{1}{4}-16(1-2\Delta)^{2}\Delta^{2},\\
 & U_{B2}(\Delta)=\frac{\left|K_{0}\right|}{64\Delta^{3}-48\Delta^{2}+8\Delta},\label{eq:double_peak_B-ansatz}
\end{align}
and the boundary values of $\Delta=0$ and $\Delta=1/2$. The total
energy must be used to evaluate these candidate values of $\Delta$
and select the global minimum. It should be noted that $\Delta=0$
recovers the Hartree-Fock solution. There are two critical points
to consider: the local stability of $\hat{\rho}_{B2}$ and a transition
from a saddle point solution to the Hartree-Fock solution. The local
stability of the $\hat{\rho}_{B2}$ is determined by the minimal value
of $U$ in Eq. \ref{eq:double_peak_B-ansatz}, which is given by
\begin{align}
 & U_{c,B2}/\left|K_{0}\right|=\frac{3\sqrt{3}}{2}\approx2.59808,\\
 & d_{c,B2}=\frac{5}{36}\approx0.138889,\\
 & \Delta_{c,B2}=\frac{1}{12}\left(3-\sqrt{3}\right)\approx0.105662.
\end{align}
For any $U>U_{c,B2}$, there exists a locally stable $\hat{\rho}_{B2}$
solution. However, one should compare the energy of the saddle-point
solution to the Hartree-Fock solution, which yields another critical
value of $U_{c,B2}'/|K_{0}|=3.375$. For $U>U_{c,B2}'$, the saddle
point solution is the global minimum, while for $U<U_{c,B2}'$ the
Hartree-Fock solution is the global minimum. We summarize all the
critical points for $\hat{\rho}_{G3}$, $\hat{\rho}_{G2}$, and $\hat{\rho}_{B2}$
in Table \ref{tab:The-values-of-critical-U}. 

\begin{table}
\begin{tabular}{|c|c|c|c|}
\hline 
 & $\Delta$ & $U/\left|K_{0}\right|$ & $d$\tabularnewline
\hline 
\hline 
$\hat{\rho}_{G3}$ & $\begin{matrix}\frac{1}{6}\left(3-\sqrt{3}\right)\\
\approx0.211325
\end{matrix}$ & $\begin{matrix}\frac{3}{8}\left(9+5\sqrt{3}\right)\\
\approx6.6226
\end{matrix}$ & $\begin{matrix}\frac{1}{36}\left(32\sqrt{3}-55\right)\\
\approx0.0118229
\end{matrix}$\tabularnewline
\hline 
$\hat{\rho}_{G2}$ & $\frac{1}{4}$ & $8$ & $0$\tabularnewline
\hline 
$\hat{\rho}_{B2}$ (L) & $\begin{matrix}\frac{1}{12}\left(3-\sqrt{3}\right)\\
\approx0.105662
\end{matrix}$ & $\frac{3\sqrt{3}}{2}\approx2.59808$ & $\frac{5}{36}\approx0.138889$\tabularnewline
\hline 
$\hat{\rho}_{B2}$(HF) & $\frac{1}{6}$ & $\frac{27}{8}=3.375$ & $\frac{17}{324}\approx0.0524691$\tabularnewline
\hline 
\end{tabular}

\caption{\label{tab:The-values-of-critical-U}The values of $\Delta$, $U$,
and $d$ at various critical points for the two peak Hubbard model
solved using $\hat{\rho}_{G2}$, $\hat{\rho}_{B2}$, and $\hat{\rho}_{G3}$.
For $\hat{\rho}_{B2}$, the critical point for local stability ($\hat{\rho}_{B2}$
(L)) and the transition to the Hartree-Fock solution ($\hat{\rho}_{B2}$
(HF)) are provided. For $\hat{\rho}_{G2}$ and $\hat{\rho}_{G3}$,
the critical values for the Mott transition are provided.}
\end{table}

\subsubsection{Solution for a general density of states\label{subsec:Delta-solution-for-general-DOS}}

\begin{figure}
\includegraphics[width=0.99\columnwidth]{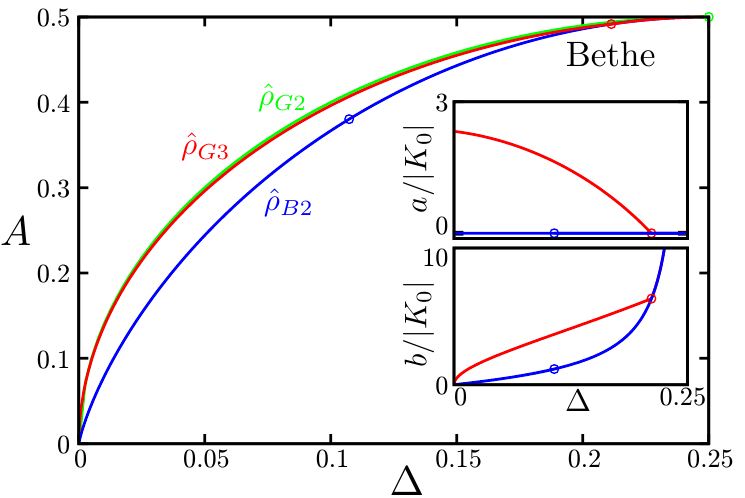}

\caption{\label{fig:Bethe_a_b_delta_d}The parameters $A$, $a$, and $b$
as a function of $\Delta$ for the single orbital Hubbard model on
the Bethe lattice in $d=\infty$ at half-filling, where $K_{0}$ is
the non-interacting kinetic energy. }
 
\end{figure}

In Section \ref{subsec:Solution-of-the-two-peak-model}, we demonstrated
that the qubit trial energy can be written solely in terms of $\Delta$
for the case of a two-peak density of states. Here we extend this
strategy for a general density of states, allowing for the optimized
value of $A$ to be evaluated as a function of $\Delta$, which is
completely independent of $U$. Therefore, this approach is particularly
useful for analyzing the Mott transition. We begin by rewriting the
qubit trial energy from Eq. (\ref{eq:qubit-trial-delta-A}) using
the result of Eq. (\ref{eq:d_delta_two_peak}), yielding 

\begin{equation}
E(\Delta,A)=K\left(\Delta,A\right)+\frac{1}{4}U+UA^{4}\tilde{O}\left(\Delta\right),\label{eq:trial-energy-delta-A-fancy-D}
\end{equation}
where $\tilde{O}\left(\Delta\right)$ is given in Eq. (\ref{eq:otilde-one-band}).
We proceed by constructing the saddle point equations of $\Delta$
and $A$ for a given $U$, yielding the following two equations
\begin{align}
 & \frac{\partial K}{\partial\Delta}=-4a=-UA^{4}\frac{\partial\tilde{O}\left(\Delta\right)}{\partial\Delta},\label{eq:saddle1_a}\\
 & \frac{\partial K}{\partial A}=2b=-4UA^{3}\tilde{O}\left(\Delta\right),\label{eq:saddle2_b}
\end{align}
where $a$ and $b$ are the Lagrange multipliers from Eq. (\ref{eq:n_epsilon_a_b}).
A practical approach for solving the two preceding equations for a
given $U$ is to express $A$ and $\Delta$ in terms of $a$ and $b$,
denoted as $A(a,b)$ and $\Delta(a,b)$, and then solve for $a$ and
$b$. However, we take an alternative approach, as our goal is to
determine the optimized value of $A$ for a given $\Delta$. Therefore,
we proceed by dividing Eq. (\ref{eq:saddle1_a}) by Eq. (\ref{eq:saddle2_b}),
which yields 

\begin{equation}
-\frac{8a}{A\left(a,b\right)b}=\frac{d\ln\left|\tilde{O}\left(\Delta\right)\right|}{d\Delta}.\label{eq:saddle_ratio}
\end{equation}
Moreover, $a$ and $b$ are required to yield the given value of $\Delta$,
such that
\begin{equation}
\Delta\left(a,b\right)=\Delta.\label{eq:delta_a_b}
\end{equation}
Simultaneously solving Eqns. (\ref{eq:saddle_ratio}) and (\ref{eq:delta_a_b})
yields $a$ and $b$ as functions of $\Delta$, and therefore all
quantities that depend on $n(\epsilon)$, including $A$ and $K$,
are now functions purely of $\Delta$. Subsequently, the double occupancy
and $U$ can be determined as a function of $\Delta$ as 
\begin{align}
 & d(\Delta)=\frac{1}{4}+UA^{4}\left(\Delta\right)\tilde{O}\left(\Delta\right),\\
 & U(\Delta)=-\frac{b(\Delta)}{2A^{3}(\Delta)\tilde{O}\left(\Delta\right)}.
\end{align}
Alternatively, the $U$ can also be expressed as 
\begin{equation}
U(\Delta)=-\frac{dK}{d\Delta}\left(\frac{d}{d\Delta}d(\Delta)\right)^{-1},
\end{equation}
which will recover Eq. (\ref{eq:UK_saddle_point}) when evaluating
the two peak model. For the case of the Bethe lattice, we plot $A$,
$a$, and $b$ as functions of $\Delta$ in Figure \ref{fig:Bethe_a_b_delta_d},
and the critical values of all quantities are listed in Table \ref{tab:Bethe_table_of_critical_values}.

\begin{table}
\begin{tabular}{|c|c|c|c|c|c|}
\hline 
 & $\Delta$ & $U/\left|K_{0}\right|$ & $U/t$ & $d$ & $A$\tabularnewline
\hline 
\hline 
$\hat{\rho}_{G3}$ & $\begin{matrix}\frac{1}{6}\left(3-\sqrt{3}\right)\\
\approx0.211325
\end{matrix}$ & $6.61836$ & $5.61784$ & $0.0162523$ & $0.491668$\tabularnewline
\hline 
$\hat{\rho}_{G2}$ & $\frac{1}{4}$ & $8$ & $6.79061$ & $0$ & $\frac{1}{2}$\tabularnewline
\hline 
$\hat{\rho}_{B2}$ (L) & $0.107273$ & $2.60206$ & $2.2087$ & $0.166367$ & $0.380259$\tabularnewline
\hline 
$\hat{\rho}_{B2}$ (HF) & $0.153687$ & $3.04071$ & $2.58103$ & $0.089871$ & $0.447304$\tabularnewline
\hline 
\end{tabular}\caption{\label{tab:Bethe_table_of_critical_values}The values of $\Delta$,
$U$, and $d$ at various critical points for the single orbital Hubbard
model on the Bethe lattice in $d=\infty$ solved using $\hat{\rho}_{G2}$,
$\hat{\rho}_{B2}$, and $\hat{\rho}_{G3}$. For $\hat{\rho}_{B2}$,
the critical point for local stability ($\hat{\rho}_{B2}$ (L)) and
the transition to the Hartree-Fock solution ($\hat{\rho}_{B2}$ (HF))
are provided. For $\hat{\rho}_{G2}$ and $\hat{\rho}_{G3}$, the critical
values for the Mott transition are provided.}
\end{table}

\subsection{Understanding the effect of Hund's coupling in the multiorbital Hubbard
model\label{subsec:Understanding-the-effect-J-two-band}}

Here we generalize the treatment from Section \ref{subsec:Understanding-the-Mott-single-band-Delta}
to the multiorbital case including the Hund coupling $J$. The key
difference is that in the multiorbital case, one cannot analytically
minimize over the local variational parameters, though these parameters
can easily be numerically minimized as a function of $\Delta$ for
a given $J/U$. The remaining procedure closely follows the single
orbital case.

Consider the multiorbital Hubbard model defined in Eq. (\ref{eq:Hubbard_multi_orbital}).
The qubit trial energy can be written as 
\begin{align}
 & E\left(\Delta,A,\rhoeff\right)=K\left(\Delta,A\right)+E_{loc}\left(\Delta,A,\rhoeff\right),\\
 & K(\Delta,A)=2N_{orb}\int_{-\infty}^{\infty}d\epsilon D\left(\epsilon\right)\epsilon n\left(\epsilon,a(\Delta,A),b(\Delta,A)\right),\\
 & E_{loc}\left(\Delta,A,\rhoeff\right)=U\mathcal{F}^{2}\left(\Delta,A,\xi(\rhoeff)\right)\langle\hat{O}\rangle_{\rhoeff},
\end{align}
where $\mathcal{F}\left(\Delta,A,\xi\right)$ is defined in Eq. (\ref{eq:F_half_filling-1}),
$\xi(\rhoeff)=\frac{1}{2}\langle\hat{\sigma}_{\ell}^{x}\rangle_{\rhoeff}$,
and 
\begin{equation}
\hat{O}=\hat{O}_{1}+\left(1-2\frac{J}{U}\right)\hat{O}_{2}+\left(1-3\frac{J}{U}\right)\hat{O}_{3},
\end{equation}
where $\hat{O}_{i}$ are defined in Eq. (\ref{eq:Hubbard_multi_orbital}).
Having defined the qubit trial energy, we proceed to obtain the solutions
for the two-peak density of states and the Bethe lattice using the
$\Delta$ parametrization. We first rewrite the local interaction
energy as 
\begin{equation}
E_{loc}\left(\Delta,A,\rhoeff\right)=UA^{4}\tilde{\mathcal{F}}^{2}\left(\Delta,\xi\left(\rhoeff\right)\right)\langle\hat{O}\rangle_{\rhoeff},\label{eq:Eloc-delta-A-rhoeff}
\end{equation}
where 
\begin{equation}
\tilde{\mathcal{F}}\left(\Delta,\xi\right)=\frac{2}{1-4\xi^{2}}\left(\sqrt{1-\frac{4\xi^{2}}{\left(1-2\Delta\right){}^{2}}}+1\right),
\end{equation}
which is obtained from $\mathcal{F}\left(\Delta,A,\xi\right)/A^{2}$.
Given the form of Eq. (\ref{eq:Eloc-delta-A-rhoeff}), the optimized
$\rhoeff$ will only depend on $\Delta$, motivating the definition
of the following function 
\begin{equation}
\tilde{O}\left(\Delta\right)=\min_{\rhoeff}\tilde{\mathcal{F}}^{2}\left(\Delta,\xi\left(\rhoeff\right)\right)\langle\hat{O}\rangle_{\rhoeff}.
\end{equation}
A convenient way to generate this function is to perform a two stage
minimization. First, we perform a constrained minimization with the
restriction $\xi\left(\rhoeff\right)=\xi$ on $\rhoeff$. Second,
we minimize the expression over $\xi$. In the first stage, given
that $\tilde{\mathcal{F}}^{2}\left(\Delta,\xi\left(\rhoeff\right)\right)$
is fixed, we only need to minimize $\langle\hat{O}\rangle_{\rhoeff}$,
which can be mathematically expressed as 
\begin{align}
 & \mathcal{O}(\xi)=\min_{\rhoeff}\{\langle\hat{O}\rangle_{\rhoeff}|\frac{1}{2}\langle\hat{\sigma}_{\ell}^{x}\rangle_{\rhoeff}=\xi,\hspace{0.2em}\langle\hat{\sigma}_{\ell}^{z}\rangle_{\rhoeff}=0\}.\label{eq:fancyO_xi}
\end{align}
To efficiently generate $\mathcal{O}\left(\xi\right)$ , we can introduce
a Lagrange multiplier $\lambda$ and determine the ground state for
$\hat{\mathcal{H}}=\hat{O}-\lambda\sum_{\alpha\sigma}\hat{\xi}_{\alpha\sigma}$,
yielding the optimal $\rhoeff$ and corresponding $\xi$ and $\langle\hat{O}\rangle_{\rhoeff}$
for a given $\lambda$. One can then perform such calculations over
a grid of $\lambda$, and then spline the relationship between $\langle\hat{O}\rangle_{\rhoeff}$
and $\xi$. A plot of $\mathcal{O}(\xi)$ is provided in Figure 2
of Ref. \cite{companion}. Finally, the partially optimized local
energy can be written as 
\begin{align}
 & E_{loc}\left(\Delta,A\right)=UA^{4}\tilde{O}\left(\Delta\right).\label{eq:Eloc-delta-A}\\
 & \tilde{O}\left(\Delta\right)=\min_{\xi\in\left[0,\frac{1}{2}-\Delta\right]}\tilde{\mathcal{F}}^{2}\left(\Delta,\xi\right)\mathcal{O}(\xi).\label{eq:tildeO_definition}
\end{align}

\begin{figure}
\includegraphics[width=0.99\columnwidth]{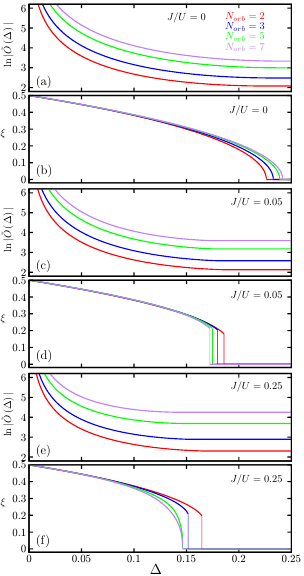}

\caption{\label{fig:Two-orbital-two-peak-O-vs-delta}Plots of $\ln|\tilde{O}\left(\Delta\right)|$
and $\xi$ as a function of $\Delta$ for the multiorbital Hubbard
model with $N_{orb}=2,3,5,7$ at half-filling for $J/U=0,0.05,0.25$,
which can be applied to all particle-hole symmetric $D(\epsilon)$. }
\end{figure}

In order to visualize $\tilde{O}\left(\Delta\right)$ and $\xi$ as
a function of $\Delta$, we plot these quantities for $J/U=0,0.05,0.25$
and $N_{orb}=2,3,5,7$ (see Figure \ref{fig:Two-orbital-two-peak-O-vs-delta}).
For $J/U=0$, one can clearly observe that $\xi$ continuously goes
to zero, while there is a discontinuity for $J/U=0.05$. For $J/U=0.25$,
$\xi$ discontinuously goes to zero for $N_{orb}=2,3$ and continuously
goes to zero for $N_{orb}=5,7$. It should be noted that $\tilde{O}\left(\Delta\right)$
can be applied to solve an arbitrary particle-hole symmetric $D(\epsilon)$,
and therefore captures the essence of the Mott transition for a given
$J/U$.

\subsubsection{Understanding the non-analytic behavior of $\tilde{O}\left(\Delta\right)$
via a Taylor series\label{subsec:Understanding-non-analytic-behavior-via-taylor}}

In Section \ref{subsec:Understanding-the-Mott-single-band-Delta},
we demonstrated that there is non-analyticity in $\tilde{O}\left(\Delta\right)$
at $\Delta=\Delta_{c}$. Here we provide a Taylor series analysis
to explain how the non-analyticity emerges. Equation (\ref{eq:tildeO_definition})
indicates that $\tilde{O}\left(\Delta\right)$ is the minimum of $\tilde{\mathcal{F}}^{2}\left(\Delta,\xi\right)\mathcal{O}(\xi)$
within the range $\xi\in\left[0,\frac{1}{2}-\Delta\right]$, and it
is convenient to study the quantity 
\begin{equation}
\mathcal{L}\left(\Delta,\xi\right)=-\frac{\tilde{\mathcal{F}}^{2}\left(\Delta,\xi\right)}{\mathcal{\tilde{F}}^{2}\left(\Delta,0\right)}\frac{\mathcal{O}\left(\xi\right)}{\mathcal{O}\left(0\right)}.
\end{equation}
Finding the minimum of $\mathcal{L}\left(\Delta,\xi\right)$ will
yield the minimum of $\tilde{\mathcal{F}}^{2}\left(\Delta,\xi\right)\mathcal{O}(\xi)$
given that $\tilde{\mathcal{F}}\left(\Delta,0\right)=4$ and $\mathcal{O}\left(0\right)=-\frac{1}{4}N_{orb}\left(1+\left(N_{orb}-1\right)J/U\right)$.

We begin by Taylor series expanding $\mathcal{L}\left(\Delta,\xi\right)$
to sixth order in $\xi$ about $\xi=0$, and the second order coefficient
in $\xi$ is expanded in $\Delta$ about $\Delta_{c,I}$ such that
it is zero for $\Delta=\Delta_{c,I}$, yielding

\begin{equation}
\mathcal{L}\left(\Delta,\xi\right)\approx-1+c_{2}\left(\Delta-\Delta_{c,I}\right)\xi^{2}+c_{4}\xi^{4}+c_{6}\xi^{6},
\end{equation}
where $c_{2}$, $c_{4}$, $c_{6}$, and $\Delta_{c,I}$ are constants
for a given $N_{orb}$ and $J/U$. Given that the optimized $\xi$
goes to zero with increasing $\Delta$ (see Figure \ref{fig:Two-orbital-two-peak-O-vs-delta}),
this requires $c_{2}>0$. Furthermore, we take $c_{6}>0$, though
there will be cases where $c_{6}$ is negative and a higher order
expansion is necessary. Therefore, we only need to understand how
the sign of $c_{4}$ influences the non-analyticity in $\tilde{O}\left(\Delta\right)$.
For $c_{4}>0$, when $\Delta>\Delta_{c;I}$, the minimum of $\mathcal{L}$
is given by $\mathcal{L}=-1$ with $\xi=0$, while for $\Delta<\Delta_{c;I}$,
the minimum of $\mathcal{L}$ is obtained with 

\begin{equation}
\xi^{2}=\frac{\sqrt{3c_{2}c_{6}\left(\Delta_{c,I}-\Delta\right)+c_{4}^{2}}-c_{4}}{3c_{6}}.\label{eq:xisaddleL}
\end{equation}
Notice that $\xi^{2}$ continuously increases from $0$ when $\Delta$
decreases from $\Delta_{c;I}$. Therefore the critical value of $\Delta$
is given by $\Delta_{c}=\Delta_{c;I}$. Moreover, using 
\begin{equation}
\frac{d\tilde{O}\left(\Delta\right)}{d\Delta}=\left(\frac{\partial}{\partial\xi}\tilde{\mathcal{F}}^{2}\left(\Delta,\xi\right)\right)\mathcal{O}\left(\xi\right),
\end{equation}
we find that $\frac{d\tilde{O}\left(\Delta\right)}{d\Delta}$ is continuous
and only has a kink at $\Delta=\Delta_{c}$. For $c_{4}<0$, when
$\Delta>\Delta_{c;I}$ there is a local minimum at $\xi=0$, while
for $\Delta<\Delta_{c,I}+\frac{c_{4}^{2}}{3c_{2}c_{6}}$ there is
a local minimum given by Eq. (\ref{eq:xisaddleL}). The two saddle
points need to be compared to obtain the global minimum, which yields
$\Delta_{c}=\Delta_{c,I}+\frac{c_{4}^{2}}{4c_{2}c_{6}}.$ Therefore,
$\xi$ jumps from zero to a finite value when $\Delta$ decreases
from $\Delta_{c}$, implying that $\frac{d\tilde{O}\left(\Delta\right)}{d\Delta}$
is discontinuous at $\Delta=\Delta_{c}$. It should be noted that
for $c_{4}>0$, $\Delta_{c}=\Delta_{c;I}$ is exact, while for $c_{4}<0$,
the expression $\Delta_{c}=\Delta_{c,I}+\frac{c_{4}^{2}}{4c_{2}c_{6}}$
is an approximation, and in this case one should use the exact form
of $\mathcal{L}\left(\Delta,\xi\right)$ to determine $\Delta_{c}$
if precision is needed.

We now proceed to analytically compute the expansion for two cases:
$J/U=0$ with $N_{orb}\ge1$ , and $J/U>0$ with $N_{orb}=2$ . We
begin by expanding
\begin{equation}
\frac{\mathcal{\tilde{F}}^{2}\left(\Delta,\xi\right)}{\mathcal{\tilde{F}}^{2}\left(\Delta,0\right)}=1+a_{2}\xi^{2}+a_{4}\xi^{4}+a_{6}\xi^{6}+\dots,
\end{equation}
where

\begin{align}
 & a_{2}=8-\frac{2}{(1-2\Delta)^{2}},\\
 & a_{4}=-\frac{16}{(1-2\Delta)^{2}}-\frac{1}{(1-2\Delta)^{4}}+48,\\
 & a_{6}=-\frac{96}{(1-2\Delta)^{2}}-\frac{8}{(1-2\Delta)^{4}}-\frac{2}{(1-2\Delta)^{6}}+256.
\end{align}
It should be noted that $a_{2}$, $a_{4}$, and $a_{6}$ all monotonically
decrease with increasing $\Delta$ for $\Delta\in\left[0,1/4\right]$.
It is straightforward to show that $a_{2}=0$ when $\Delta=1/4$ and
$a_{4}=0$ when $\Delta=\frac{1}{4}\left(2-\sqrt{\frac{1}{3}\left(2+\sqrt{7}\right)}\right)\approx0.188895$
and $a_{6}=0$ when $\Delta=0.155281$. The remaining task is to compute
$-\frac{\mathcal{O}\left(\xi\right)}{\mathcal{O}\left(0\right)}$,
which we separately consider for the two aforementioned cases. 

For $J/U=0$ and $N_{orb}\ge1$, symmetry can be used to parametrize
$\rhoeff=|\Psi\rangle\langle\Psi|$ with $|\Psi\rangle=\sum_{\Gamma}\sqrt{p_{\left|N_{\Gamma}-N_{orb}\right|}}|\Gamma\rangle$,
where $p_{i}$ is a variational parameter with $i=0,\dots,N_{orb}$
and $N_{\Gamma}=\langle\Gamma|\sum_{\ell=0}^{2N_{orb}}\hat{n}_{\ell}|\Gamma\rangle$
counts the number of electrons in state $\Gamma$. It is convenient
to reparametrize $p_{i}$ as
\begin{equation}
x_{i}=\sqrt{\begin{pmatrix}2N_{orb}\\
N_{orb}
\end{pmatrix}}\left(2-\delta_{i,0}\right)\frac{\left(-i+N_{orb}+1\right)_{i}}{\left(N_{orb}+1\right)_{i}}\sqrt{p_{i}},
\end{equation}
such that $\text{Tr}\left(\rhoeff\right)=\sum_{i=0}^{N_{orb}}x_{i}^{2}$,
where $(m)_{n}=\frac{\Gamma(m+n)}{\Gamma(m)}=m(m+1)...(m+n-1)$ is
the Pochhammer symbol. When taking an expectation value of an operator
$\hat{A}$ in the qubit space, it is convenient to use a matrix representation
$\left[\hat{A}\right]_{ij}$, where $i$ and $j$ take values from
$0,\dots,N_{orb}$, such that 
\begin{equation}
\text{Tr}\left(\rhoeff\hat{A}\right)=\sum_{i,j=0}^{N_{orb}}x_{i}\left[\hat{A}\right]_{ij}x_{j}.
\end{equation}
The non-zero entries of the effective matrices for $\hat{O}$ and
$\hat{\sigma}_{\ell}^{x}$ are given as 

\begin{equation}
\left[\hat{O}\right]_{ii}=\frac{1}{4}\left(2i^{2}-N_{orb}\right),
\end{equation}

\begin{equation}
\left[\hat{\sigma}_{\ell}^{x}\right]_{i,i+1}=\left[\hat{\sigma}_{\ell}^{x}\right]_{i+1,i}=\frac{\sqrt{\left(N_{orb}-i\right)\left(i+N_{orb}+1\right)}}{\sqrt{2}N_{orb}\sqrt{2-\delta_{0,i}}}.
\end{equation}
Perturbation theory can then be used to obtain 
\begin{equation}
-\frac{\mathcal{O}\left(\xi\right)}{\mathcal{O}\left(0\right)}=-1+\frac{4\xi^{2}}{N_{orb}+1}+\frac{4b_{4}\xi^{4}}{N_{orb}}+\frac{4b_{6}\xi^{6}}{N_{orb}}+O\left(\xi^{7}\right),
\end{equation}
where 

\begin{align}
 & b_{4}=\frac{N_{orb}\left(7N_{orb}\left(N_{orb}+1\right)+2\right)}{4\left(N_{orb}+1\right){}^{3}},\\
 & b_{6}=\frac{N_{orb}\left(N_{orb}\left(N_{orb}+1\right)h+24\right)}{36\left(N_{orb}+1\right){}^{5}},\\
 & h=209N_{orb}\left(N_{orb}+1\right)+146.
\end{align}
The expansion coefficients of $\mathcal{L}$ are obtained as 
\begin{alignat}{1}
 & \Delta_{c;I}=\frac{4N_{orb}-\sqrt{4N_{orb}^{2}+6N_{orb}+2}+2}{8N_{orb}+4},\\
 & c_{2}=\frac{16\sqrt{2}\left(2N_{orb}+1\right){}^{3}}{\left(\left(N_{orb}+1\right)\left(2N_{orb}+1\right)\right){}^{3/2}},\\
 & c_{4}=\frac{N_{orb}\left(N_{orb}\left(32N_{orb}+55\right)+27\right)+6}{\left(N_{orb}+1\right){}^{3}},\\
 & c_{6}=\frac{N_{orb}\left((h+9211)N_{orb}+2774\right)+384}{9\left(N_{orb}+1\right){}^{5}},\\
 & h=N_{orb}\left(N_{orb}\left(3456N_{orb}+11729\right)+15070\right).
\end{alignat}
Given that $c_{4}>0$, we have $\Delta_{c}=\Delta_{c;I}$ and $\frac{d\tilde{O}\left(\Delta\right)}{d\Delta}$
has a kink at $\Delta=\Delta_{c}$.

We now discuss the case of $J/U>0$ with $N_{orb}=2$. Using perturbation
theory, we find
\begin{equation}
-\frac{\mathcal{O}\left(\xi\right)}{\mathcal{O}\left(0\right)}=-1+4\xi^{2}+b_{4}\xi^{4}+b_{6}\xi^{6},
\end{equation}
where 
\begin{align}
 & b_{4}=-\frac{4}{3}\left(\frac{8}{r}-\frac{6}{r-1}-7\right),\\
 & b_{6}=\frac{32(r(r(7r(2r-3)+38)-42)+29)}{9(r-1)^{2}r^{2}},
\end{align}
where $r=J/U$. Therefore, we can compute the coefficients in the
expansion of $\mathcal{L}$ as 
\begin{align}
 & \Delta_{c;I}=\frac{1}{4}\left(2-\sqrt{2}\right)\approx0.146447,\\
 & c_{2}=16\sqrt{2},\\
 & c_{4}=\frac{8}{3}\left(-\frac{4}{r}+\frac{3}{r-1}+5\right),\\
 & c_{6}=\frac{16}{9}\left(\frac{8r+58}{r^{2}}+\frac{36}{(r-1)^{2}}+67\right).
\end{align}
We can see that $c_{4}<0$, and therefore $\Delta_{c}>\Delta_{c,I}$,
indicating that there is a discontinuity in $\frac{d\tilde{O}\left(\Delta\right)}{d\Delta}$
for $\Delta=\Delta_{c}$.

Finally, we numerically explore the cases of $N_{orb}>2$, where $\Delta_{c;I}$
and $c_{2}$ are identical to the case of $N_{orb}=2$. To be concrete,
we consider $J/U=0.25$, and we numerically compute the expansion
coefficients by fitting $\mathcal{O}(\xi)$ to a sixth order polynomial.
For $N_{orb}=3$, we have $c_{4}\approx-2.9$ and $c_{6}\approx-147$.
Notice that in this case $c_{6}<0$, and thus an expansion beyond
sixth order is necessary, though plotting $\mathcal{L}\left(\Delta,\xi\right)$
indicates that $\Delta_{c}>\Delta_{c,I}$, yielding a discontinuity
in $\frac{d\tilde{O}\left(\Delta\right)}{d\Delta}$ for $\Delta=\Delta_{c}$
(not shown). For $N_{orb}=4$, $c_{4}\approx-0.63$ and $c_{6}\approx39$.
For $N_{orb}=5$, $c_{4}\approx0.44$ and $c_{6}\approx35$. For $N_{orb}=6$,
$c_{4}\approx1.1$ and $c_{6}\approx33$. For $N_{orb}=7$, $c_{4}\approx1.5$
and $c_{6}\approx34$. It should be noted that for $N_{orb}\geq5$,
we have $\Delta_{c}=\Delta_{c;I}$ and $\frac{d\tilde{O}\left(\Delta\right)}{d\Delta}$
has a kink at $\Delta=\Delta_{c}$. 

\subsubsection{Solution for the two-peak density of states\label{subsec:Two-orbital-two-peak}}

\begin{figure}
\includegraphics[width=0.99\columnwidth]{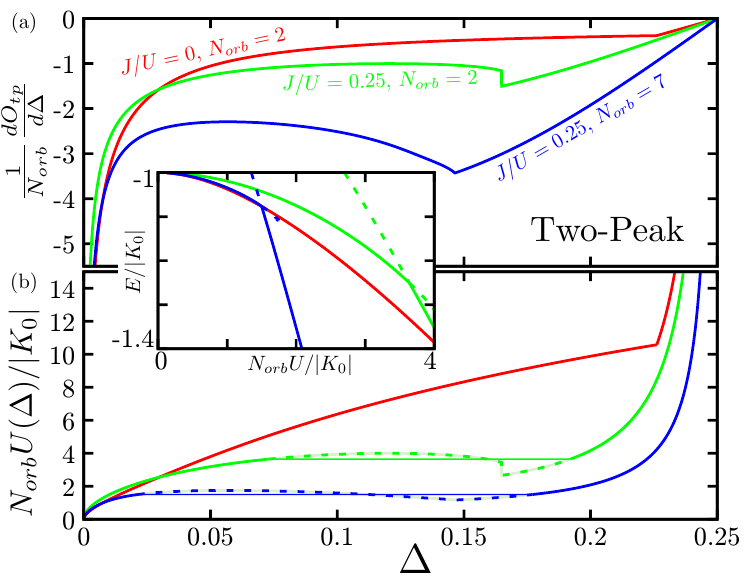}

\caption{\label{fig:dOtildedDelta_vs_Delta} Plots of $\frac{1}{N_{orb}}\frac{dO_{tp}}{d\Delta}$
(panel $a$) and $N_{orb}U(\Delta)/|K_{0}|$ (panel $b$) as functions
of $\Delta$ for various $N_{orb}$ and $J/U$ for the two-peak model
in $d=\infty$ solved using $\hat{\rho}_{G3}$. The inset plots $E/|K_{0}|$
vs. $N_{orb}U/|K_{0}|$, where $E$ is the total energy per site.
The dotted lines represent unstable or metastable solutions.}
\end{figure}

\global\long\def\odelta{O_{tp}(\Delta)}%
We now consider the two peak density of states
\begin{equation}
D\left(\epsilon\right)=\frac{1}{2}\left(\delta\left(\epsilon-\frac{1}{N_{orb}}K_{0}\right)+\delta\left(\epsilon+\frac{1}{N_{orb}}K_{0}\right)\right),\label{eq:DOS_two_peak-1}
\end{equation}
where $K_{0}$ is the total non-interacting kinetic energy per site.
For the two-peak density of states, $A=\sqrt{\Delta(2-4\Delta)}$,
and the qubit trial energy can be written purely in terms of $\Delta$
as 

\begin{align}
 & E\left(\Delta\right)=\left(1-4\Delta\right)K_{0}+U\odelta,\\
 & \odelta=\left(2-4\Delta\right)^{2}\Delta^{2}\widetilde{O}\left(\Delta\right).
\end{align}
Following the single orbital case, $\Delta$ can be used to determine
$U$ from $dE\left(\Delta\right)/d\Delta=0,$ which yields
\begin{equation}
U(\Delta)=-4\left|K_{0}\right|\left(\frac{d\odelta}{d\Delta}\right)^{-1},\label{eq:U_delta_tp}
\end{equation}
thus providing a succinct solution parametrized by $\Delta$. The
relation $U(\Delta)$ can be used to determine the nature of the Mott
transition from $\widetilde{O}\left(\Delta\right)$, given that this
quantity will allow any observable to be expressed in terms of $U$.
We plot $\frac{dO_{tp}}{d\Delta}$ versus $\Delta$ for various $N_{orb}$
and $J/U$, which demonstrates three types of non-analytical scenarios
for $O_{tp}\left(\Delta\right)$ (see Fig. \ref{fig:dOtildedDelta_vs_Delta},
panel $a$). First, for all $J/U=0$, the $\frac{dO_{tp}}{d\Delta}$
is continuous with a positive slope and has a kink at $\Delta=\Delta_{c}$.
Second, for $J/U>0$ and small $N_{orb}$, the $\frac{dO_{tp}}{d\Delta}$
is discontinuous at $\Delta=\Delta_{c}$ with a negative slope for
$\Delta_{c}^{-}$. Third, for $J/U=0.25$ and $N_{orb}=7$, the $\frac{dO_{tp}}{d\Delta}$
is continuous and has a kink at $\Delta=\Delta_{c}$, and the slope
is negative for $\Delta_{c}^{-}$. The $\frac{dO_{tp}}{d\Delta}$
can now be used to determine $U(\Delta)$, which yields the order
of the Mott transition (see Fig. \ref{fig:dOtildedDelta_vs_Delta},
panel $b$). First, for all $J/U=0$, $U$ increases monotonically
and continuously with $\Delta$, with a kink at $\Delta=\Delta_{c}$,
and therefore there are no metastable regions and the Mott transition
is continuous. Second, for $J/U>0$, there is an unstable region in
the metal phase where $\frac{dU}{d\Delta}<0$, and the total energy
can be used to determine the transition between the metal and insulating
phase, corresponding to a horizontal line, and the transition is first-order. 

In summary, we have demonstrated that the nature of the Mott transition
for the two peak density of states is determined purely by $\widetilde{O}\left(\Delta\right)$,
and below we demonstrate that the Bethe lattice has the same behavior.
Therefore, it appears that $\widetilde{O}\left(\Delta\right)$ is
the essence of what determines the nature of the Mott transition in
$d=\infty$.

\subsubsection{Solution for a general density of states}

We now execute a similar strategy for a general density of states.
Using Eq. (\ref{eq:Eloc-delta-A}), the qubit trial energy can be
written as 
\begin{equation}
E\left(\Delta,A\right)=K\left(\Delta,A\right)+UA^{4}\widetilde{O}\left(\Delta\right).
\end{equation}
The saddle point equations are given as 
\begin{align}
 & \frac{\partial K}{\partial\Delta}=-4N_{orb}a=-UA^{4}\frac{d\widetilde{O}\left(\Delta\right)}{d\Delta},\label{eq:partialK-partial-delta}\\
 & \frac{\partial K}{\partial A}=2N_{orb}b=-4UA^{3}\widetilde{O}\left(\Delta\right).\label{eq:partialK-partial-A}
\end{align}
Given that $\Delta$ and $A$ are functions of $a$ and $b$, one
can solve $a$ and $b$ from Eqns. (\ref{eq:partialK-partial-delta})
and (\ref{eq:partialK-partial-A}) for a given $U$, and then determine
all physical quantities. We now demonstrate that $\frac{d\widetilde{O}\left(\Delta\right)}{d\Delta}=0$
indicates that the system is in the Mott phase. Given that the quasiparticle
weight is given as $Z=a/\sqrt{a^{2}+b^{2}}$ (see Eq. \ref{eq:Z_a_b}),
and that $A>0$ for finite $U$ (see Eq. \ref{eq:A-a-b}), the only
scenario where $Z=0$ is when $\frac{d\widetilde{O}\left(\Delta\right)}{d\Delta}=0$.
Therefore, when $\Delta<\Delta_{c}$, the system is metallic, while
$\Delta>\Delta_{c}$ the system is insulating. For a given $U$, one
must minimize over $\Delta$ in order to determine nature of the ground
state.

An alternate approach is to parametrize the solution in terms of $\Delta$.
Equations (\ref{eq:partialK-partial-delta}) and (\ref{eq:partialK-partial-A}),
in addition to the constraint on $a$ and $b$ for a given $\Delta$,
yield 

\begin{align}
 & -\frac{8a}{A\left(a,b\right)b}=\frac{d\ln\left|\widetilde{O}\left(\Delta\right)\right|}{d\Delta},\label{eq:abdelta1}\\
 & \Delta\left(a,b\right)=\Delta.\label{eq:abdelta2}
\end{align}
For a given $\Delta$, $a$ and $b$ can be determined from Eqns.
(\ref{eq:abdelta1}) and (\ref{eq:abdelta2}), $A$ can be determined
from $A\left(a,b\right)$, and $U$ can be determined using Eq. (\ref{eq:partialK-partial-A})
as 
\begin{equation}
U(\Delta)=-\frac{N_{orb}b}{2A^{3}(a,b)\widetilde{O}\left(\Delta\right)},
\end{equation}
allowing for the evaluation of the total energy. 

\begin{figure}
\includegraphics[width=0.99\columnwidth]{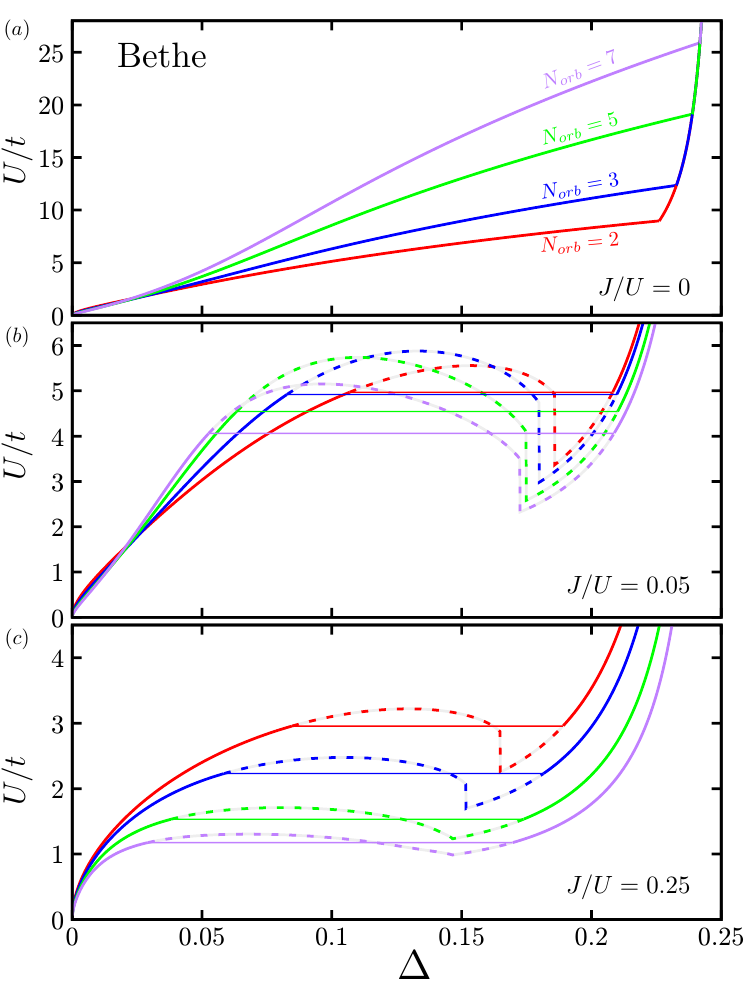}

\caption{\label{fig:multi-orbital-U-vs-Delta}Plots of $U/t$ as a function
of $\Delta$ for the multiorbital Hubbard model on the Bethe lattice
in $d=\infty$ using $\hat{\rho}_{G3}$ for $N_{orb}=2,3,5,7$ with
$J/U=0$ (panel $a$), $J/U=0.05$ (panel $b$), and $J/U=0.25$ (panel
$c$). The dotted lines indicate metastable or unstable solutions.}
\end{figure}

\begin{figure}
\includegraphics[width=0.99\columnwidth]{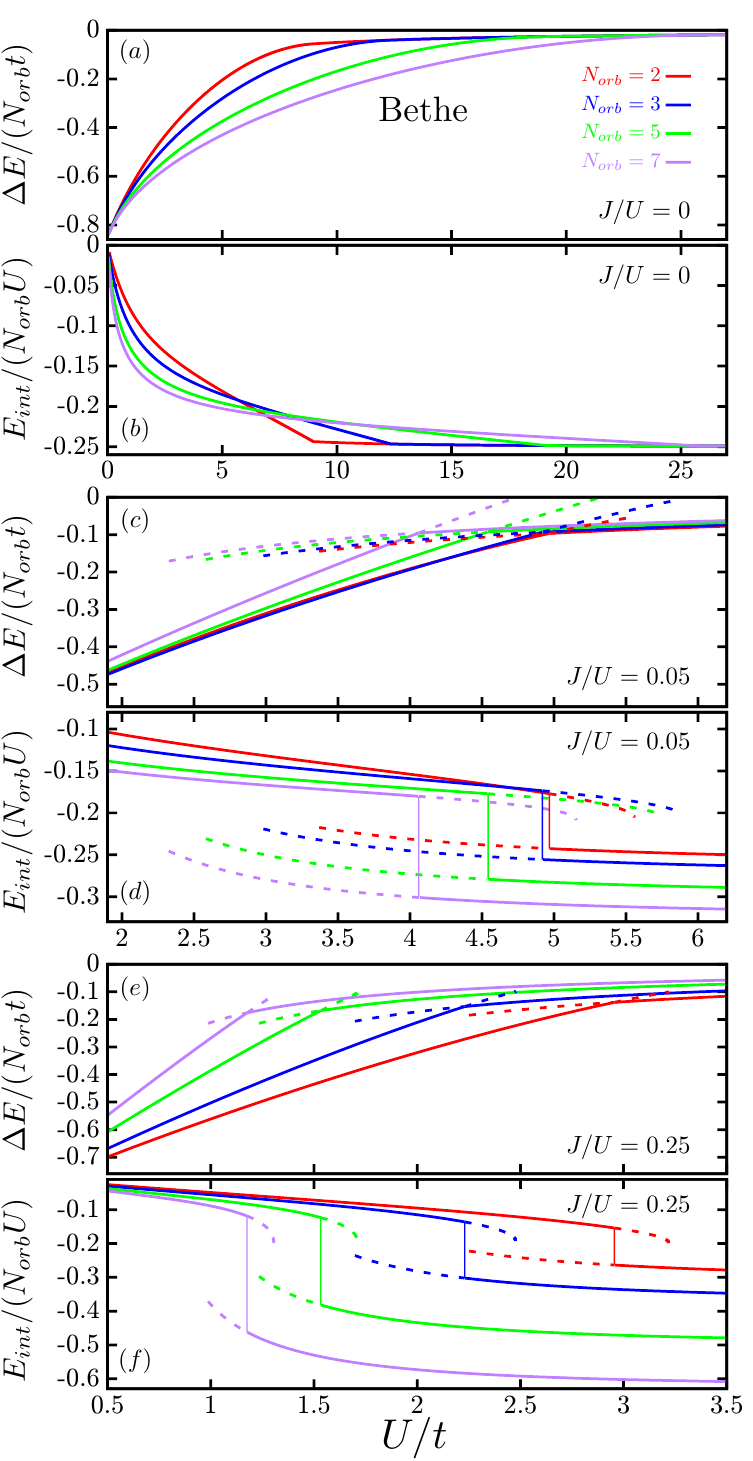}

\caption{\label{fig:multiorbital-energy-vs-U} Plots of the total energy and
the interaction energy as functions of $U/t$ for the multiorbital
Hubbard model on the Bethe lattice in $d=\infty$ using $\hat{\rho}_{G3}$
for $N_{orb}=2,3,5,7$ with $J/U=0$ (panel $a$, $b$), $J/U=0.05$
(panel $c$, $d$), and $J/U=0.25$ (panels $e$, $f$). The dotted
lines indicate metastable or unstable solutions.}
\end{figure}

We now consider the Bethe lattice in $d=\infty$ for $N_{orb}=2,3,5,7$.
Recall that for a given $J/U$, the $\tilde{O}\left(\Delta\right)$
yields a $\Delta_{c}$ which divides the metallic and insulating states,
where $\Delta>\Delta_{c}$ indicates an insulating phase. For the
case of a continuous transition, the $U_{c}$ will be determined by
$\Delta_{c}$, while for a first-order transition, one must explicitly
determine the $U_{c}$ where the insulating and metallic states cross
in energy. The algorithm is executed by evaluating $A$, $U$, the
interaction energy, and the total energy as functions of $\Delta$.
We begin by plotting the $U/t$ as a function of $\Delta$ (see Figure
\ref{fig:multi-orbital-U-vs-Delta}, panel $a$). For $J/U=0$, $U$
is a monotonic and continuous function of $\Delta$, implying a continuous
phase transition at $\Delta_{c}$, which can be identified as a kink.
Alternatively, for $J/U=0.05$, the $U$ is not a monotonic nor a
continuous function of $\Delta$, implying that the there are regions
of phase coexistence and unstable regions. The metallic curve exists
for $\Delta<\Delta_{c}$, and the solution is only stable for $\Delta<\Delta_{c;1}$,
where $\Delta_{c;1}$ is determined from $\frac{dU}{d\Delta}=0$,
and therefore the metallic solution is only stable for $U<U(\Delta_{c;1})$.
The insulating curve exists for $\Delta>\Delta_{c}$ without any unstable
regions, and therefore the insulating phase exists for $U>U(\Delta_{c}+0^{+})$.
Given that $U(\Delta_{c}+0^{+})<U(\Delta_{c;1})$, there exists a
region of coexistence for the metallic and insulating solutions, and
the total energy dictates the lowest energy solution. We now proceed
to present the total energy and the interaction energy as functions
of $U/t$ for various $J/U$ and $N_{orb}$ (see Figure \ref{fig:multiorbital-energy-vs-U}),
and the results are very similar to the two-peak case. Consistent
with previous Gutzwiller \cite{Bunemann19974011}, slave boson \cite{Hasegawa19971391},
and DMFT \cite{Ono2003035119} studies, the Mott transition is continuous
for $J/U=0$ and first-order for $J/U>0$. 

Finally, we evaluate $U_{c}$ for $J/U=0$, which is explicitly given
by $U_{c}=\frac{b_{c}}{8A_{c}^{3}}$, where $b_{c}$ is determined
from $\Delta(0,b_{c})=\Delta_{c}$ and $A_{c}=A(0,b_{c})$, where
$\Delta(a,b)$ and $A(a,b)$ are defined in Eqns. (\ref{eq:delta-a-b})
and (\ref{eq:A-a-b}), respectively. The critical values for the Bethe
lattice are listed in Table \ref{tab:Critical-values-for-SUN}. For
the case of large $N_{orb}$, we have
\begin{align}
U_{c} & /t\approx\frac{32N_{orb}}{3\pi}+\frac{20}{3\pi}\\
 & \approx3.39531N_{orb}+2.12207,
\end{align}
consistent with our numerical results in Ref. \cite{Cheng2023035127}.
For a systematic exploration of how $U_{c}$ depends on $J/U$ and
$N_{orb}$, see Fig. 5 of Ref. \cite{companion}.

\begin{table}
\begin{tabular}{|c|c|c|c|c|}
\hline 
$N_{orb}$ & $\Delta_{c}$ & $b_{c}/t$ & $A_{c}$ & $U_{c}/t$\tabularnewline
\hline 
\hline 
$1$ & $\frac{1}{6}\left(3-\sqrt{3}\right)\approx0.211325$ & $5.34164$ & $0.491668$ & $5.61784$\tabularnewline
\hline 
$2$ & $\frac{1}{20}\left(10-\sqrt{30}\right)\approx0.226139$ & $8.80352$ & $0.496835$ & $8.97283$\tabularnewline
\hline 
$3$ & $\frac{1}{2}-\frac{1}{\sqrt{14}}\approx0.232739$ & $12.2288$ & $0.498345$ & $12.3511$\tabularnewline
\hline 
$4$ & $\frac{1}{12}\left(6-\sqrt{10}\right)\approx0.236477$ & $15.6412$ & $0.498984$ & $15.7369$\tabularnewline
\hline 
$5$ & $\frac{1}{22}\left(11-\sqrt{33}\right)\approx0.238884$ & $19.0475$ & $0.499314$ & $19.1261$\tabularnewline
\hline 
$6$ & $\frac{1}{2}-\frac{\sqrt{\frac{7}{26}}}{2}\approx0.240563$ & $22.4505$ & $0.499505$ & $22.5172$\tabularnewline
\hline 
$7$ & $\frac{1}{2}-\frac{1}{\sqrt{15}}\approx0.241801$ & $25.8515$ & $0.499627$ & $25.9094$\tabularnewline
\hline 
\end{tabular}

\caption{\label{tab:Critical-values-for-SUN}Critical values for the SU(2$N_{orb}$)
Hubbard model ($J/U=0$) on the Bethe lattice in $d=\infty$ solved
using $\hat{\rho}_{G3}$.}
\end{table}

\section{\label{subsec:Conclusion}Summary and Conclusions}

We begin by providing a high level overview of the variational discrete
action theory (VDAT), such that the developments of the present work
can be properly understood. VDAT is a variational approach to the
many-body body problem that consists of two main components: a variational
ansatz for the many-body wave function or density-matrix, known as
the sequential product density matrix (SPD), and a formalism for evaluating
expectation values under the SPD, known as the discrete action theory
\cite{Cheng2021195138,Cheng2021206402}. The SPD has a natural mechanism
to trade off between efficiency and accuracy, where the integer $\mathcal{N}$
monotonically increases the variational power of the SPD and guarantees
the ability to recover the ground state solution for $\mathcal{N}\rightarrow\infty$.
Moreover, there are two distinct types of SPD which satisfy the properties
of a many-body density matrix, denoted as G-type and B-type. The G-type
$\mathcal{N}=1$, $\mathcal{N}=2$, and $\mathcal{N}=3$ SPD encapsulate
the Hartree-Fock wave function, the Gutzwiller wave function, and
the Gutzwiller-Baeriswyl wave function, respectively. The key breakthrough
using VDAT was the demonstration that the SPD can be \emph{exactly}
evaluated for multiorbital Hubbard models in $d=\infty$. We demonstrated
that the G-type $\mathcal{N}=3$ SPD accurately solves the Anderson
impurity model on a ring \cite{Cheng2021206402}, the single band
Hubbard model over all parameter space \cite{Cheng2021206402}, the
two orbital Hubbard model including a crystal field and the full rotationally
invariant Hund's coupling \cite{Cheng2022205129}, and the $SU(2N_{orb})$
Hubbard model for $N_{orb}\le8$ \cite{Cheng2023035127}. Moreover,
we demonstrated that the computational cost of solving a G-type $\mathcal{N}=3$
SPD is comparable to a G-type $\mathcal{N}=2$ SPD, meaning that VDAT
can provide a sufficiently accurate solution at a cost not far beyond
the Gutzwiller approximation. The success of VDAT at $\mathcal{N}=3$
motivated a search for the best possible algorithm for executing calculations
using a G-type $\mathcal{N}=3$ SPD \cite{Cheng2023035127}, which
is essential for detailed exploration of the multiorbital Hubbard
model and merging VDAT with realistic electronic structure methods. 

The VDAT algorithm in $d=\infty$ consists of two steps: the exact
evaluation of the SPD via the self-consistent canonical discrete action
theory (SCDA) and the optimization of the energy with respect to the
variational parameters. The SCDA requires the numerical solution of
a set of self-consistency conditions, and therefore can be inconvenient
when minimizing over the variational parameters. For the case of a
G-type $\mathcal{N}=3$ SPD with certain restrictions (see Sections
\ref{sec:Introduction}, \ref{subsec:Comparing-original-gauge}, and
\ref{subsec:Review-of-the-gauge-const-SCDA} for further details),
the SCDA self-consistency condition can be automatically satisfied,
which we refer to as the gauge constrained SCDA algorithm \cite{Cheng2023035127}.
In the present work, we introduce the so-called qubit parametrization
of the gauge constrained SCDA algorithm, which is mathematically equivalent
to the original gauge constrained SCDA algorithm. The qubit parametrization
offers several key improvements. The qubit parametrization analytically
resolves some constraints over the variational parameters, thus reducing
the number of variational parameters by one per spin orbital. Additionally,
the variational parameters are physically intuitive and facilitate
a deeper understanding of how the SPD captures Mott and Hund physics.
Therefore, the qubit parametrization achieves the long sought goal
of resolving the shortcomings of the Gutzwiller approximation while
maintaining the computational simplicity and physical appeal. 

The variational parameters of the qubit parametrization consist of
the momentum density distribution, the non-interacting reference momentum
density distribution, and the pure state of a qubit system with a
dimension of the local Hilbert space. The qubit system naturally arises
from reparametrizing the variational parameters of the interacting
projector, and the renormalized correlations within the qubit space
yield the physical local correlations. The variational parameters
are restricted by two constraints per spin orbital, requiring that
the local density computed from the momentum density distribution
is the same as that computed from the non-interacting reference momentum
density distribution, and the same as the density computed from the
qubit system. The qubit trial energy has a very intuitive form: the
kinetic energy is determined by the momentum density distributions,
while the local interaction energy is the expectation value of an
effective Hamiltonian within the qubit system. Interestingly, the
effective Hamiltonian has the same form as the local interacting Hamiltonian
where the local density operator is substituted by an effective density
operator of the qubit system. The effective density operator for a
given spin-orbital $\ell$ depends on five parameters: the density
$n_{\ell}$, the magnetization in the $x$ direction for the $\ell$-th
qubit, denoted $\xi_{\ell}$, and three quantities determined from
the momentum density distribution, denoted as $\Delta_{\ell}$, $\mathcal{A}_{<\ell}$,
and $\mathcal{A}_{>\ell}$. The quantity $\Delta_{\ell}$ characterizes
the number of electrons promoted across the reference Fermi surface,
while $\mathcal{A}_{<\ell}$ and $\mathcal{A}_{>\ell}$ characterize
the momentum density distribution below and above the reference Fermi
surface, respectively. The quantity $\xi_{\ell}$ is an important
variable which differentiates between a zero and non-zero quasiparticle
weight when all variables are fully optimized, where $\xi_{\ell}=0$
indicates zero quasiparticle weight for spin-orbital $\ell$. The
main computational cost for evaluating the ansatz is dictated by computing
expectation values within the qubit system.

While evaluating the trial qubit energy ansatz is a straightforward
task, optimizing over all the variational parameters remains nontrivial.
In general, the qubit trial energy can be partially optimized over
the momentum density distribution by introducing four Lagrange multipliers
per spin-orbital, replacing the continuous momentum density distribution
with four variables. For a system with $2N_{orb}$ spin-orbitals per
site, there remain $2^{2N_{orb}}+3\times2N_{orb}-1$ variational parameters
which must be optimized in general. However, this number may be greatly
reduced by symmetry, and it is likely possible to compress these variables
into a smaller number of parameters without a serious loss of fidelity.
For the special case of half-filled orbitals with particle hole symmetry,
we demonstrate that one can efficiently minimize over all variational
parameters, with a computational cost proportional to computing the
ground state of a Hamiltonian defined within the qubit system.

In order to demonstrate the power of the qubit parametrization, we
studied the ground state properties of the multiorbital Hubbard model
at half-filling with particle-hole symmetry for various $J/U$ and
$N_{orb}=2-7$. For a given $J/U$, the majority of the energy minimization
can be encapsulated into the computation of a single variable function
$\tilde{O}(\Delta)$, which can then be used to obtain the solution
at a negligible cost for an arbitrary $U$ and density-of-states.
The entire function $\tilde{O}(\Delta)$ is evaluated by solving a
collection of qubit systems, which has a relatively small computational
cost. For example, for $N_{orb}=7$, the ground state for a given
qubit system can be solved in several seconds on a typical single
desktop computer core, and taking on the order of 100 samples, the
entire function $\tilde{O}(\Delta)$ can be accurately obtained on
the order of hundreds of seconds. The extreme computational efficiency
of the qubit parametrization in this case allows one to easily map
out all of parameter space, which is not possible with DMFT given
the lack of efficient impurity solvers for the zero temperature multiorbital
problem. We find that for $J/U=0$, the Mott transition is continuous,
while it is first-order for $J/U>0$, consistent with previous Gutzwiller
\cite{Bunemann19974011}, slave boson \cite{Hasegawa19971391}, and
DMFT \cite{Ono2003035119} studies. 

While the key result of this paper is formulating the qubit parametrization
of the gauge constrained SCDA algorithm for a G-type $\mathcal{N}=3$
SPD, we also demonstrate that the qubit parametrization can be applied
to the G-type and B-type $\mathcal{N}=2$ SPD. Moreover, we demonstrate
that properly restricting the variational parameters of the qubit
trial energy for the G-type $\mathcal{N}=3$ SPD can recover the corresponding
qubit trial energy for the G-type and B-type $\mathcal{N}=2$ SPD.
Interestingly, the qubit trial energy for the G-type $\mathcal{N}=2$
SPD has an identical form to the slave spin mean-field theory (see
Appendix \ref{app:SSMF}), and thus the $\mathcal{N}=3$ qubit trial
energy may provide insights for proceeding beyond mean-field theory
in the slave spin formalism.

The qubit parametrization of the gauge constrained SCDA algorithm
at $\mathcal{N}=3$ is likely the optimal form when evaluating an
SPD with a kinetic projector that is diagonal in both the momentum
and spin-orbital indices and an interacting projector that consists
of diagonal Hubbard operators. For the Hamiltonians treated in the
present study, which have density-density interactions and hopping
parameters that are diagonal in the spin-orbital index, the aforementioned
restrictions on the SPD do not limit the variational power. When solving
a general Hamiltonian which includes the full rotationally invariant
form of the Hund exchange or non-diagonal hopping terms, the qubit
parametrization can still be applied and it will still yield an upper
bound on the energy in $d=\infty$, but it will not contain the full
variational power of $\mathcal{N}=3$. Ongoing research is addressing
how to generalize the qubit parametrization to handle an arbitrary
$\mathcal{N}=3$ G-type SPD, with aspirations of completely superseding
our general decoupled minimization algorithm for $\mathcal{N}=3$
\cite{Cheng2022205129}.

\section{Acknowledgments}

This work was a supported by a RISE-LDRD grant from Columbia University
and Brookhaven National Laboratory. This research used resources of
the National Energy Research Scientific Computing Center, a DOE Office
of Science User Facility supported by the Office of Science of the
U.S. Department of Energy under Contract No. DE-AC02-05CH11231.

\appendix

\section{One-body reduced density matrix functional for the Hubbard model}

An important application of the qubit energy form for $\hat{\rho}_{G3}$
is the construction of a one body reduced density matrix functional
(1RDMF) for the multi-orbital Hubbard model, which is the focus of
our companion manuscript \cite{companion}. Here we derive the corresponding
results for $\hat{\rho}_{G2}$ and $\hat{\rho}_{B2}$. Additionally,
we evaluate existing 1RDMF's from the literature for the single band
Hubbard model at half-filling in $d=\infty$. 

\subsection{The 1RDMF's for the multi-orbital Hubbard model from $\hat{\rho}_{G2}$
and $\hat{\rho}_{B2}$}

We begin by presenting the 1RDMF from $\hat{\rho}_{B2}$ using the
qubit parametrization. The interaction energy is given as 
\begin{align}
 & E_{int}\left(\left\{ n_{k\ell}\right\} \right)=\min_{\rhoeff}\left\langle H_{loc}\left(\left\{ \hat{n}_{eff,\ell}\right\} \right)\right\rangle _{\rhoeff}\\
 & \textrm{subject to\ }\left\langle \hat{n}_{\ell}\right\rangle _{\rhoeff}=n_{\ell},
\end{align}
where $n_{\ell}=\int dkn_{k\ell}$, the effective density operator
is $\hat{n}_{eff,\ell}=n_{\ell}+\frac{A_{\ell}^{2}}{\xi_{\ell,0}^{2}}\left(\hat{n}_{\ell}-n_{\ell}\right)$
with $A_{\ell}=\int dk\sqrt{n_{k\ell}\left(1-n_{k\ell}\right)}$ and
$\xi_{\ell,0}=\sqrt{n_{\ell}(1-n_{\ell})}$, and both $\rhoeff$ and
$H_{loc}\left(\left\{ \hat{n}_{eff,\ell}\right\} \right)$ are diagonal
in the Pauli-Z basis of the qubit system. For the case of half-filling,
the minimization yields 
\begin{equation}
E_{int}\left(\left\{ n_{k\ell}\right\} \right)=16UA^{4}\mathcal{O}\left(0\right),
\end{equation}
where $\mathcal{O}\left(0\right)=-\frac{1}{4}N_{orb}\left(1+\left(N_{orb}-1\right)J/U\right)$. 

We now present the 1RDMF for $\hat{\rho}_{G2}$. Unlike $\hat{\rho}_{G3}$
or $\hat{\rho}_{B2}$, the variational parameters do not explicitly
contain $n_{k\ell},$ but instead $n_{k\ell}-n_{\ell}=\xi_{\ell}^{2}/\xi_{\ell,0}^{2}\left(n_{k\ell,0}-n_{\ell}\right)$,
where $n_{k\ell,0}\in\left[0,1\right]$. Therefore, for a given $\left\{ n_{k\ell}\right\} $,
we have $\left|\xi_{\ell}/\xi_{\ell,0}\right|\geq\sqrt{Z_{\ell,min}}$,
where $Z_{\ell,min}=\max\left(\frac{n_{\ell,max}-n_{\ell}}{1-n_{\ell}},\frac{n_{\ell}-n_{\ell,min}}{n_{\ell}}\right)$,
where $n_{\ell,max}$ and $n_{\ell,min}$ are the maximum and minimum
values of $\left\{ n_{k\ell}\right\} $ for a given $\ell$. It should
be noted that the gauge symmetry can be used to restrict $\xi_{\ell}\ge0$.
The interaction energy can then be written as 
\begin{align}
 & E_{int}\left(\left\{ n_{k\ell}\right\} \right)=\min_{\rho}\left\langle H_{loc}\left(\left\{ \hat{n}_{\ell}\right\} \right)\right\rangle _{\rhoeff}\\
 & \textrm{subject to}\hspace{1em}\frac{1}{2}\langle\hat{\sigma}_{\ell}^{x}\rangle_{\rhoeff}\geq\xi_{\ell;min},\hspace{1em}\left\langle \hat{n}_{\ell}\right\rangle _{\rhoeff}=n_{\ell},\label{eq:restriction}
\end{align}
where $\xi_{\ell;min}=\xi_{\ell;0}\sqrt{Z_{min}}$. Notice that the
interaction energy will decrease with decreasing $\xi_{\ell}$, and
therefore the restriction in Eq. (\ref{eq:restriction}) can be replaced
with $\frac{1}{2}\langle\hat{\sigma}_{\ell}^{x}\rangle_{\rhoeff}=\xi_{\ell;min}$
. For the case of half-filling, we have 
\begin{equation}
E_{int}\left(\left\{ n_{k\ell}\right\} \right)=U\mathcal{O}\left(\frac{1}{2}\sqrt{1-2n_{\ell;min}}\right),
\end{equation}
where $\mathcal{O}(\xi)$ is defined in Eq. (\ref{eq:fancyO_xi}). 

Using the above interaction energy functionals will yield results
identical to the corresponding VDAT results, such as the ones provided
in Fig. \ref{fig:compare_bethe_twopeak}.

\subsection{Results using published 1RDMF's for the single-orbital Hubbard model
in $d=\infty$}

We are not aware of any applications of existing 1RDMF's to the Hubbard
model in $d=\infty$, though there are various studies of Hubbard
clusters \cite{Kamil2016085141,Mitxelena2017425602} and $d=1$ \cite{Lopezsandoval20001764,Mitxelena20201701}
and $d=2$ Hubbard models \cite{Saubanere2016045102,Mitxelena2020064108}.
It is important to emphasize that $d=\infty$ is the most relevant
test of local electronic correlations, and is emblematic of typical
three dimensional strongly correlated materials in that $d=\infty$
hosts a standard Fermi liquid in the metal phase and exhibits a Mott
transition at a finite value of $U$. While $d=1$ is exactly solvable
via the Bethe ansatz, the Mott transition occurs at an infinitesimal
$U$ \cite{Lieb19681445}. Alternatively, $d=2$ is a testbed for
the most advanced and expensive computational approaches, and the
properties at half-filling and low temperatures are still actively
studied \cite{Tanaka2019205133,Schafer2021011058,Chatzieleftheriou2024236504,Geng2025115143}.
While it is still very interesting to compare total energies from
1RDMF's in $d=1$ and $d=2$, if the goal is to determine whether
or not a 1RDMF can describe the Mott transition, then it is most critical
to first benchmark in $d=\infty$. 

In this section, we discuss how to use various 1RDMF's to solve the
one band Hubbard model at half-filling, including the MBB \cite{Muller1984446},
CA \cite{Csanyi20007348}, CGA \cite{Csanyi2002032510}, power \cite{Sharma2008201103},
PNOF5 \cite{Piris2011164102}, PNOF7 \cite{Piris2017063002}, and
dimer \cite{Saubanere2016045102} functionals. The interaction energy
is given by $Ud$, where $d$ is the double occupancy and is a functional
of $\left\{ n_{k\sigma}\right\} $. If the interaction Hamiltonian
is rewritten as $\frac{U}{2}\sum_{i\sigma\sigma'}\hat{a}_{i\sigma}^{\dagger}\hat{a}_{i\sigma'}^{\dagger}\hat{a}_{i\sigma'}\hat{a}_{i\sigma}$,
the interaction energy for the MBB, CA, CGA, and power functionals
can be viewed as a modification of the Hartree-Fock (HF) energy where
the Fock term is altered \cite{Mitxelena2017425602}. We begin by
writing the local interaction as $E_{int}=\frac{U}{2}\sum_{\sigma\sigma'}\left\langle \hat{a}_{i\sigma}^{\dagger}\hat{a}_{i\sigma'}^{\dagger}\hat{a}_{i\sigma'}\hat{a}_{i\sigma}\right\rangle $,
which can be written in momentum space as 
\begin{equation}
E_{int}=\frac{U}{2L^{2}}\sum_{k_{1}k_{2}k'_{1}k'_{2}\sigma\sigma'}\hspace{-1em}\delta_{k_{1}+k_{2},k'_{1}+k'_{2}}\left\langle \hat{a}_{k_{1}\sigma}^{\dagger}\hat{a}_{k_{2}\sigma'}^{\dagger}\hat{a}_{k'_{2}\sigma'}\hat{a}_{k'_{1}\sigma}\right\rangle ,\label{eq:1RDMF_interaction_energy}
\end{equation}
where $L$ is the number of $k$-points. The MBB, CA, CGA, and power
functionals are all defined using the following approximation 
\begin{align}
 & \left\langle \hat{a}_{k_{1}\sigma}^{\dagger}\hat{a}_{k_{2}\sigma'}^{\dagger}\hat{a}_{k'_{2}\sigma'}\hat{a}_{k'_{1}\sigma}\right\rangle \approx n_{k_{1}}n_{k_{2}}\delta_{k_{1}k'_{1}}\delta_{k_{2}k'_{2}}\nonumber \\
 & -\mathcal{F}\left(n_{k_{1}},n_{k_{2}}\right)\delta_{k_{1}k'_{2}}\delta_{k_{2}k'_{1}}\delta_{\sigma\sigma'},
\end{align}
where $n_{k}$ is defined through $\left\langle \hat{a}_{k\sigma}^{\dagger}\hat{a}_{k'\sigma'}\right\rangle =n_{k}\delta_{kk'}\delta_{\sigma\sigma'}$.
The $\mathcal{F}\left(n_{i},n_{j}\right)$ and $E_{int}$ are given
for each functional in Table \ref{tab:1RDMF_functionals}, and our
values of $\mathcal{F}\left(n_{i},n_{j}\right)$ are identical to
those provided in Ref. \cite{Mitxelena2017425602}. It is worth noting
that the MBB, CA, and CGA recover the atomic limit where $E_{int}=0$
when $n_{k}=1/2$, while the power functional does not if $\alpha\neq1/2$. 

\begin{table}
\begin{tabular}{|c|c|}
\hline 
Functional & $\mathcal{F}\left(n_{i},n_{j}\right)$\tabularnewline
\hline 
\hline 
HF & $n_{i}n_{j}$\tabularnewline
\hline 
MBB & $\sqrt{n_{i}n_{j}}$\tabularnewline
\hline 
power & $\left(n_{i}n_{j}\right)^{\alpha}$\tabularnewline
\hline 
CA & $n_{i}n_{j}+\sqrt{n_{i}\left(1-n_{i}\right)n_{j}\left(1-n_{j}\right)}$\tabularnewline
\hline 
CGA & $\frac{1}{2}\left(n_{i}n_{j}+\sqrt{n_{i}\left(2-n_{i}\right)n_{j}\left(2-n_{j}\right)}\right)$\tabularnewline
\hline 
\hline 
 & $E_{int}$\tabularnewline
\hline 
\hline 
HF & $\frac{U}{4}$\tabularnewline
\hline 
MBB & $\frac{U}{2}\left(1-2\left(\int dk\sqrt{n_{k}}\right)^{2}\right)$\tabularnewline
\hline 
power & $\frac{U}{2}\left(1-2\left(\int dkn_{k}^{\alpha}\right)^{2}\right)$\tabularnewline
\hline 
CA & $U\left(\frac{1}{4}-\left(\int dk\sqrt{n_{k}\left(1-n_{k}\right)}\right)^{2}\right)$\tabularnewline
\hline 
CGA & $U\left(\frac{3}{8}-\frac{1}{2}\left(\int dk\sqrt{n_{k}\left(2-n_{k}\right)}\right)^{2}\right)$\tabularnewline
\hline 
\end{tabular}

\caption{$\mathcal{F}\left(n_{i},n_{j}\right)$ and $E_{int}$ for the HF,
MBB, Power, CA, and CGA functionals for the single orbital Hubbard
model at half-filling in the paramagnetic phase.\label{tab:1RDMF_functionals}}
 
\end{table}

Now we discuss how to minimize the total energy over $\{n_{k}\}$.
Given that MBB is a special case of the power functional, we first
consider the power functional. Given that the interaction energy for
the power functional is fixed for a given $\int dkn_{k}^{\alpha}$
and $\int dkn_{k}$, the kinetic energy can be minimized by introducing
two Lagrange multipliers $a$ and $b$, yielding a target functional
$K=\int dk\left(\epsilon_{k}-a\right)n_{k}-b\int dkn_{k}^{\alpha}$
to be minimized, and the partially optimized $n_{k}$ are obtained
as 
\begin{equation}
n_{k}=\begin{cases}
1 & \epsilon_{k}\leq a+\alpha b\\
\left(\frac{\epsilon_{k}-a}{\alpha b}\right)^{\frac{1}{\alpha-1}} & \epsilon_{k}>a+\alpha b
\end{cases}.
\end{equation}
The ground state energy for a given $U$ is then obtained by optimizing
the total energy over $a$ and $b$. 

We now consider the CA and CGA, which can be viewed in a unified way
by writing the interaction energy functional as
\begin{equation}
E_{int}=U\left(\left(\frac{1}{2}-\frac{1}{4\theta}\right)-\frac{1}{\theta}\left(\int dk\sqrt{n_{k}\left(\theta-n_{k}\right)}\right)^{2}\right),
\end{equation}
where $\theta=1$ corresponds to CA and $\theta=2$ corresponds to
CGA. We can similarly introduce two Lagrange multipliers, yielding
a target functional 
\begin{equation}
K=\int dk\left(\epsilon_{k}-a\right)n_{k}-b\int dk\sqrt{n_{k}\left(\theta-n_{k}\right)},
\end{equation}
where $\theta\geq1$ and the partially optimized value of $n_{k}$
is given as
\begin{equation}
n_{k}=\begin{cases}
1 & \epsilon\leq a+\frac{b(\theta-2)}{2\sqrt{\theta-1}}\\
\frac{\theta}{2}\left(1-\frac{(\epsilon_{k}-a)}{\sqrt{(\epsilon_{k}-a)^{2}+b^{2}}}\right) & \epsilon>a+\frac{b(\theta-2)}{2\sqrt{\theta-1}}
\end{cases}.
\end{equation}
For a given $U$, one can then optimize the Lagrange multipliers to
obtain the ground-state energy and other observables. Plots of the
double occupancy as a function of $U/t$ are provided in Ref. \cite{companion}.

An important drawback of the MBB, power, CA, and CGA functionals is
that the anti-symmetry of the two-body reduced density matrix (2RDM)
is violated when $\mathcal{F}\left(n_{i},n_{j}\right)$ deviates from
the HF value \cite{Mitxelena2017425602}. The Piris natural orbital
functionals (PNOF) \cite{Piris2013620} resolve this issue by approximating
the form of the 2RDM $D_{ij,kl}^{\alpha\beta}=\frac{1}{2}\langle\hat{a}_{i\alpha}^{\dagger}\hat{a}_{j\beta}^{\dagger}\hat{a}_{l\beta}\hat{a}_{k\alpha}\rangle$
as \cite{Mitxelena2017425602}
\begin{align}
 & D_{ij,kl}^{\sigma\sigma}=\frac{n_{i}n_{j}-\Delta_{ij}}{2}\left(\delta_{ik}\delta_{jl}-\delta_{il}\delta_{jk}\right),\\
 & D_{ij,kl}^{\sigma\bar{\sigma}}=\frac{n_{i}n_{j}-\Delta_{ij}}{2}\delta_{ik}\delta_{jl}+\frac{\Pi_{ik}}{2}\delta_{ij}\delta_{kl},
\end{align}
where $\Delta_{ij}$ and $\Pi_{ik}$ are explicitly defined in PNOF5
and PNOF7. In the previous calculations \cite{Mitxelena2017425602,Piris20131298},
it was found that the optimized natural orbital often breaks symmetry
in order to further minimize the energy, and translational symmetry
breaking in the Hubbard model allows PNOF5 and PNOF7 to recover the
correct atomic limit \cite{Mitxelena2017425602}. In our context,
we require translational symmetry to be respected, so we will examine
PNOF5 and PNOF7 in this case. PNOF5 only includes intra-pair correlation
(i.e. $\Pi_{ik}=0$ when $i,k$ belong to different pairs), and therefore
the deviation of the summation in Eq. (\ref{eq:1RDMF_interaction_energy})
from the corresponding Hartree-Fock value scales like $L$ in the
thermodynamic limit while there is a prefactor of $1/L^{2}$, and
thus PNOF5 will yield the same energy form as HF in the thermodynamic
limit. PNOF7 accounts for interpair correlation, which will provide
a thermodynamic contribution, and for half-filling we find $E_{int}=\frac{1}{4}-\frac{1}{2}A^{2}$,
which is similar to the CA but with different prefactor for $A^{2}$. 

Finally, we discuss the interaction energy of the dimer functional
at half-filling, given as 
\begin{equation}
E_{int}=\frac{U}{4}\left(1-\sqrt{1-\left(K/K_{0}\right)^{2}}\right),
\end{equation}
where $K=2\int dk\epsilon_{k}n_{k}$ and $K_{0}$ is the non-interacting
kinetic energy. Minimizing the total energy over $\{n_{k}\}$ yields
the double occupancy as a function of $U$ as 
\begin{equation}
d=\frac{1}{4}\left(1-\sqrt{1-\frac{1}{1+\left(U/\left(4K_{0}\right)\right)^{2}}}\right).
\end{equation}

\section{Equivalence between the Qubit energy form for $\hat{\rho}_{G2}$
and the Slave Spin Mean Field Theory\label{app:SSMF}}

Here we prove that the slave spin mean-field theory (SSMF) is identical
to the qubit energy form for $\rho_{G2}$. We begin by describing
the SSMF using the conventions in our work, as there are trivial differences.
We associate the spin state $|\uparrow\rangle$ with the fermionic
state $|0\rangle$, while the standard SSMF associates $|\downarrow\rangle$
with $|0\rangle$. To restore the enlarged Hilbert space of the SSMF
back to the physical space, we require that $\hat{n}_{i\ell}=\frac{1-\hat{\sigma}_{i\ell}^{z}}{2}$
holds for every local site $i$ and spin-orbital index $\ell$. Then,
the local Hamiltonian at site $i$ can be written as $H_{loc}\left(\left\{ \hat{n}_{i\ell}\right\} \right)=H_{loc}\left(\left\{ \frac{1-\hat{\sigma}_{i\ell}^{z}}{2}\right\} \right)$,
while the hopping term can be written as $\hat{a}_{i\ell}^{\dagger}\hat{a}_{j\ell}=\hat{f}_{i\ell}^{\dagger}\hat{O}_{i\ell}^{\dagger}\hat{O}_{j\ell}\hat{f}_{j\ell}$,
where $\hat{O}_{i\ell}=\hat{S}_{i\ell}^{+}+c_{i\ell}\hat{S}_{i\ell}^{-}$
, $\hat{S}_{i\ell}^{\pm}=\frac{\hat{\sigma}_{i\ell}^{x}\pm\hat{\sigma}_{i\ell}^{y}}{2}$,
$c_{i\ell}$ is an arbitrary complex number, and $\hat{f}_{j\ell}$
is a fermionic annihilation operator. While the preceding transformations
are exact for any $c_{i\ell}$, an appropriate choice of $c_{i\ell}$
is indeed critical when making mean-field approximations. In Ref.
\cite{De'medici2005205124}, which only addressed half-filling, the
$c_{i\ell}=1$ and $\hat{O}_{i\ell}=\hat{\sigma}_{i\ell}^{x}$, while
later work generalized this result \cite{Hassan2010035106}, choosing
$c_{i\ell}$ such that the non-interacting limit can always be correctly
recovered within the mean field approximation. We follow the latter
choice. Finally, the Hamiltonian in the slave spin representation
is given as 

\begin{equation}
\hat{H}=\sum_{ij\ell}t_{i\ell,j\ell}\hat{f}_{i\ell}^{\dagger}\hat{O}_{i\ell}^{\dagger}\hat{O}_{j\ell}\hat{f}_{j\ell}+\sum_{i}H_{loc}\left(\left\{ \frac{1-\hat{\sigma}_{i\ell}^{z}}{2}\right\} \right),\label{eq:H_slave_spin}
\end{equation}
with the operator constraint $\hat{f}_{i\ell}^{\dagger}\hat{f}_{i\ell}=\frac{1-\hat{\sigma}_{i\ell}^{z}}{2}$
for every site and spin-orbital. There are two ways to derive the
SSMF. The first approach introduces a mean-field decoupling for the
hopping term and treats the operator constraint in the mean field
level, yielding decoupled Hamiltonians for electrons and spins that
must be solved self-consistently. The second approach \cite{Georgescu2017165135,Maurya2021425603,Maurya2022055602,Crispino2023155149}
uses the variational principle, assuming a trial wave-function $|\Psi\rangle=|\Psi_{f}\rangle\otimes|\Psi_{s}\rangle$
which is a direct product of a fermionic part $|\Psi_{f}\rangle$
and a spin part $|\Psi_{s}\rangle$. Furthermore, the $|\Psi_{f}\rangle$
is assumed to be Slater determinant and the $|\Psi_{s}\rangle$ is
assumed to be a direct product in real space, such that $|\Psi_{s}\rangle=\otimes_{i}|\Psi_{s;i}\rangle$.
Therefore, the trial energy of Eq. (\ref{eq:H_slave_spin}) is given
as 
\begin{align}
E= & \sum_{ij\ell}t_{i\ell,j\ell}\langle\Psi_{f}|\hat{f}_{i\ell}^{\dagger}\hat{f}_{j\ell}|\Psi_{f}\rangle\langle\Psi_{s;i}|\hat{O}_{i\ell}^{\dagger}|\Psi_{s;i}\rangle\langle\Psi_{s;j}|\hat{O}_{j\ell}|\Psi_{s;j}\rangle\nonumber \\
 & +\sum_{i}\langle\Psi_{s;i}|H_{loc}\left(\left\{ \frac{1-\hat{\sigma}_{i\ell}^{z}}{2}\right\} \right)|\Psi_{s;i}\rangle,\label{eq:SSMF_trial_energy}
\end{align}
subject to the constraint 
\begin{equation}
\left\langle \Psi_{f}|\hat{f}_{i\ell}^{\dagger}\hat{f}_{i\ell}|\Psi_{f}\right\rangle =\langle\Psi_{s;i}|\frac{1-\hat{\sigma}_{i\ell}^{z}}{2}|\Psi_{s;i}\rangle=n_{\ell}.\label{eq:SSMF_constraint}
\end{equation}
Given translational symmetry and the correspondence $\rhoeff\leftrightarrow|\Psi_{s;j}\rangle\langle\Psi_{s;j}|$
and $n_{k\ell;0}\leftrightarrow\langle\Psi_{f}|f_{k\ell}^{\dagger}f_{k\ell}|\Psi_{f}\rangle$,
we can prove that trial energy in Eq. (\ref{eq:SSMF_trial_energy})
is identical to the qubit trial energy for $\hat{\rho}_{G2}$ given
in Eq. (\ref{eq:energygansatz}) and the constraint in Eq. (\ref{eq:SSMF_constraint})
is identical to Eq. (\ref{eq:ga constraint}). The key step is to
evaluate 
\begin{equation}
\langle\Psi_{s;j}|\hat{O}_{j\ell}|\Psi_{s;j}\rangle=\left(1+c_{\ell}\right)\langle\frac{1}{2}\hat{\sigma}_{\ell}^{x}\rangle_{\rhoeff},\label{eq:SSMF_factor}
\end{equation}
where gauge symmetry allows $\langle\hat{\sigma}_{\ell}^{y}\rangle_{\rhoeff}=0$.
To ensure $\langle\Psi_{s;j}|\hat{O}_{j\ell}|\Psi_{s;j}\rangle=1$
in the non-interacting limit, we must choose $c_{\ell}$ such that
$\left(1+c_{\ell}\right)\xi_{\ell;0}=1$, where $\xi_{\ell;0}=\sqrt{n_{\ell}\left(1-n_{\ell}\right)}$
is the value for $\xi_{\ell}=\langle\frac{1}{2}\hat{\sigma}_{\ell}^{x}\rangle_{\rhoeff}$
in the non-interacting limit. Plugging $c_{\ell}$ into Eq. (\ref{eq:SSMF_factor}),
we have $\langle\Psi_{s;j}|\hat{O}_{j\ell}|\Psi_{s;j}\rangle=\xi_{\ell}/\xi_{\ell;0}$,
completing the proof.

The preceding proof demonstrates the equivalence of the qubit energy
form for $\hat{\rho}_{G2}$ and the trial energy for the SSMF. Now
we demonstrate how to obtain a Hamiltonian form for the SSMF using
the saddle point equations for the trial energy. The total energy
under a fixed $\{n_{\ell}\}$ can be minimized by introducing Lagrange
multipliers for the electron and spin systems, yielding 
\begin{align}
E & =\frac{1}{L}\sum_{k\ell}\epsilon_{k\ell}\left(n_{\ell}+\frac{\xi_{\ell}^{2}\left(n_{k\ell,0}-n_{\ell}\right)}{n_{\ell}\left(1-n_{\ell}\right)}\right)+\left\langle H_{loc}\left(\left\{ \hat{n}_{\ell}\right\} \right)\right\rangle _{\rhoeff}\nonumber \\
 & -\sum_{\ell}\left(\lambda_{\ell}\left(\frac{1}{L}\sum_{k}n_{k\ell;0}-n_{\ell}\right)+\lambda'_{\ell}\left(\left\langle \hat{n}_{\ell}\right\rangle _{\rhoeff}-n_{\ell}\right)\right).
\end{align}
Taking the derivative with respect to $\rhoeff$ yields the mean-field
Hamiltonian for the spin system as

\begin{equation}
\hat{H}_{eff}^{s}=\frac{\partial E}{\partial\rhoeff}=H_{loc}\left(\left\{ \hat{n}_{\ell}\right\} \right)+2\sum_{\ell}h_{\ell}\hat{\xi}_{\ell}-\lambda'_{\ell}\hat{n}_{\ell}+C,\label{eq:SSMF-spin_hamiltonian}
\end{equation}
where $\hat{n}_{\ell}=\frac{1}{2}\left(1-\hat{\sigma}_{\ell}^{z}\right)$
, $\hat{\xi}_{\ell}=\frac{1}{2}\hat{\sigma}_{\ell}^{x}$, $h_{\ell}=\frac{1}{L}\frac{\xi_{\ell}}{n_{\ell}\left(1-n_{\ell}\right)}\sum_{k}\epsilon_{k\ell}n_{k\ell,0}$,
$C$ is a constant that does not influence the results, and we assume
$\sum_{k}\epsilon_{k\ell}=0$. Now consider the derivative of the
electron components as 

\begin{equation}
\frac{\partial E}{\partial n_{k\ell;0}}=Z_{\ell}\epsilon_{k\ell}-\lambda_{\ell},
\end{equation}
where $Z_{\ell}=\frac{\xi_{\ell}^{2}}{n_{\ell}\left(1-n_{\ell}\right)}$,
which can be connected to a non-interacting fermionic Hamiltonian
as 
\begin{equation}
\hat{H}_{eff}^{f}=\sum_{k\ell}\left(Z_{\ell}\epsilon_{k\ell}-\lambda_{\ell}\right)\hat{n}_{k\ell}.\label{eq:SSMF-electron_hamiltonian}
\end{equation}
The ground states for Eqns. (\ref{eq:SSMF-spin_hamiltonian}) and
(\ref{eq:SSMF-electron_hamiltonian}) will yield an updated $\rhoeff$
and $n_{k\ell}$, yielding new Hamiltonians for Eqns. (\ref{eq:SSMF-spin_hamiltonian})
and (\ref{eq:SSMF-electron_hamiltonian}), and this procedure is iterated
until self-consistency is achieved. 

\section{The Central Point Expansion (CPE) \label{app:CPE}}

The central point expansion (CPE) can be viewed as an approach to
evaluate both a G-type and B-type SPD at $\mathcal{N}=2$ by expanding
about a reference SPD referred to as the central point. The CPE can
be applied in arbitrary dimensions, and it is formally exact if all
orders are summed, though it has only ever been applied at first-order.
Interestingly, for the G-type $\mathcal{N}=2$ SPD, the first-order
CPE yields the same result as the Gutzwiller approximation (GA) and
the SCDA in any dimension, and the derivation offers an alternative
perspective from the GA and the SCDA. For the B-type $\mathcal{N}=2$
SPD, the first-order CPE yields the same result as the SCDA in $d=\infty$,
while providing a different approximation in finite dimensions. The
CPE was originally developed in the context of the off-shell effective
energy theory (OET) \cite{Cheng2020081105}, where the CPE was renormalized
using both weak coupling and strong coupling perturbation theory to
ensure the correct limiting behavior, yielding excellent results for
the Hubbard model in $d=1,2,\infty$.

\subsection{The CPE for $\hat{\rho}_{G2}$ \label{appendix:The-CPE-for-G}}

In this section, we use the first-order CPE to evaluate $\hat{\rho}_{G2}$,
which will be shown to be equivalent to the GA and the SCDA. The first-order
CPE can be motivated by the fact that the GA relation between $n_{k\ell}-n_{\ell}$
and $n_{k\ell;0}-n_{\ell}$ is a linear form (see Eq. (\ref{eq:nk}))
given by
\begin{equation}
n_{k\ell}-n_{\ell}=\mathcal{Z}_{\ell}\left(n_{k\ell;0}-n_{\ell}\right),\label{eq:nk 1}
\end{equation}
which is valid for an arbitrary $n_{k\ell;0}\in\left[0,1\right]$
with a constraint $\left(1/L\right)\sum_{k}n_{k\ell;0}=n_{\ell}$,
where $L$ is the number of sites in the lattice. This remarkably
simple relation motivates the use of $n_{k\ell;0}-n_{\ell}$ as the
expansion parameters, where $\mathcal{Z}_{\ell}$ can be determined
using a $\hat{\rho}_{G2}$ with $n_{k\ell;0}$ slightly deviating
from the uniform distribution $n_{\ell}$ at first order. Alternatively,
the relation between $n_{k\ell;0}$ and $\gamma_{k\ell}$ is highly
nonlinear, suggesting that a first-order approximation in $\gamma_{k\ell}$
cannot be applied to the Gutzwiller wavefunction.

The CPE begins by choosing an expansion point for $\hat{\rho}_{G2}=\hat{P}_{1}\hat{K}_{2}\hat{P}_{1}$
as $\hat{\rho}_{G2}^{\star}=\hat{P}_{1}\hat{K}_{2}^{\star}\hat{P}_{1}$,
where $\hat{K}_{2}^{\star}=\exp\left(\sum_{k\ell}\gamma_{\ell}^{\star}\hat{n}_{k\ell}\right)$
and $\gamma_{\ell}^{\star}$ is chosen to reproduce the non-interacting
local density $n_{\ell;0}\equiv\left\langle \hat{n}_{i\ell}\right\rangle _{\hat{K}_{2}}=\left\langle \hat{n}_{i\ell}\right\rangle _{\hat{K}_{2}^{\star}}$
through $n_{\ell;0}=1/\left(1+\exp\left(-\gamma_{\ell}^{\star}\right)\right)$.
Considering a kinetic projector that deviates slightly from $\hat{K}_{2}^{\star}$
as $\hat{K}_{2}=\exp\left(\sum_{k\ell}\gamma_{k\ell}\hat{n}_{k\ell}\right)$,
where $\gamma_{k\ell}=\gamma_{\ell}^{\star}+\delta\gamma_{k\ell}$
, we compute the response of the expectation value $\langle\hat{O}\rangle_{\hat{\rho}_{G2}}$
to $\gamma_{k\ell}$ to the first order about the central point, given
as 
\begin{align}
 & \frac{\partial\langle\hat{O}\rangle_{\hat{\rho}_{G2}}}{\partial\gamma_{k\ell}}\bigg|_{\hat{\rho}_{G2}=\hat{\rho}_{G2}^{*}}=\frac{\partial}{\partial\gamma_{k\ell}}\frac{\text{Tr}\left(\hat{P}_{1}\hat{K}_{2}\hat{P_{1}}\hat{O}\right)}{\text{Tr}\left(\hat{P}_{1}\hat{K}_{2}\hat{P_{1}}\right)}\bigg|_{\hat{\rho}_{G2}=\hat{\rho}_{G2}^{*}}\label{eq:dOdgamma}\\
 & =\frac{\text{Tr}\left(\hat{P}_{1}\sqrt{\hat{K}_{2}^{\star}}\hat{n}_{k\ell}\sqrt{\hat{K}_{2}^{\star}}\hat{P_{1}}\hat{O}\right)}{\text{Tr}\left(\hat{\rho}_{G2}^{\star}\right)}\nonumber \\
 & -\frac{\text{Tr}\left(\hat{P}_{1}\sqrt{\hat{K}_{2}^{\star}}\hat{n}_{k\ell}\sqrt{\hat{K}_{2}^{\star}}\hat{P_{1}}\right)\text{Tr}\left(\hat{\rho}_{G2}^{\star}\hat{O}\right)}{\text{Tr}\left(\hat{\rho}_{G2}^{\star}\right)^{2}}\\
 & =\langle\hat{n}_{k\ell};\hat{O}\rangle_{\hat{\rho}_{G2}^{\star}}-\left\langle \hat{n}_{k\ell}\right\rangle _{\hat{\rho}_{G2}^{\star}}\langle\hat{O}\rangle_{\hat{\rho}_{G2}^{\star}},\label{eq:dOdgamma-response}
\end{align}
where we have utilized that fact that $\hat{K}_{2}^{\star}$ commutes
with $\hat{P}_{1}$ \cite{Cheng2020081105}, $\sqrt{\hat{\rho}_{G2}^{\star}}=\hat{P}_{1}\sqrt{\hat{K}_{2}^{\star}}=\sqrt{\hat{K}_{2}^{\star}}\hat{P_{1}}$,
and the notation $\langle\hat{A};\hat{B}\rangle_{\hat{\rho}}$ is
defined as 
\begin{equation}
\langle\hat{A};\hat{B}\rangle_{\hat{\rho}}=\langle\hat{B};\hat{A}\rangle_{\hat{\rho}}=\text{Tr}\left(\sqrt{\hat{\rho}}\hat{A}\sqrt{\hat{\rho}}\hat{B}\right)/\text{Tr}\left(\hat{\rho}\right).\label{eq:correlator}
\end{equation}
The response coefficient of $\langle\hat{O}\rangle_{\hat{\rho}_{G2}}$
to $\gamma_{k\ell}$ at a general $\hat{\rho}_{G2}$ can be conveniently
expressed via the integer time correlation function in the compound
space as
\begin{equation}
\frac{\partial\langle\hat{O}\rangle_{\hat{\rho}_{G2}}}{\partial\gamma_{k\ell}}=\left\langle \barhat[n]_{k\ell}^{\left(1\right)}\barhat[O]^{\left(2\right)}\right\rangle _{\barhat[\rho]_{G2}}-\left\langle \barhat[n]_{k\ell}^{\left(1\right)}\right\rangle _{\barhat[\rho]_{G2}}\left\langle \barhat[O]^{\left(2\right)}\right\rangle _{\barhat[\rho]_{G2}}.\label{eq:dodgammdat}
\end{equation}
While Eq. (\ref{eq:dodgammdat}) is difficult to evaluate in general,
the CPE circumvents this problem by observing that Eq. (\ref{eq:dodgammdat})
becomes trivial to evaluate at the central point given that $\hat{\rho}_{G2}^{\star}$
and $\sqrt{\hat{\rho}_{G2}^{\star}}$ are direct product states in
real space. Though it is not yet clear, it will prove critical to
exploit the gauge symmetry of the SPD to restrict $\hat{P}_{1}$ such
that $\left\langle \hat{n}_{i\ell}\right\rangle _{\hat{\rho}_{G2}^{\star}}=\left\langle \hat{n}_{i\ell}\right\rangle _{\hat{K}_{2}^{\star}}=n_{\ell;0}$.
The response for the momentum density distribution $n_{k\ell}$ to
$\gamma_{k'\ell'}$ can be obtained as 

\begin{equation}
\frac{\partial\left\langle \hat{n}_{k\ell}\right\rangle _{\hat{\rho}_{G2}}}{\partial\gamma_{k'\ell'}}\bigg|_{\hat{\rho}_{G2}=\hat{\rho}_{G2}^{*}}=\langle\hat{n}_{k\ell};\hat{n}_{k'\ell'}\rangle_{\hat{\rho}_{G2}^{\star}}-n_{\ell;0}n_{\ell';0},\label{eq:dndg def}
\end{equation}
where the correlation function $\langle\hat{n}_{k\ell};\hat{n}_{k'\ell'}\rangle_{\hat{\rho}_{G2}^{\star}}$
can be computed by transforming to real space as 
\begin{align}
 & \langle\hat{n}_{k\ell};\hat{n}_{k'\ell'}\rangle_{\hat{\rho}_{G2}^{\star}}=\frac{1}{L^{2}}\sum_{j_{1}j_{2}j_{3}j_{4}}e^{ik\left(j_{1}-j_{2}\right)+ik'\left(j_{3}-j_{4}\right)}\nonumber \\
 & \times\left\langle \hat{a}_{j_{1}\ell}^{\dagger}\hat{a}_{j_{2}\ell};\hat{a}_{j_{3}\ell'}^{\dagger}\hat{a}_{j_{4}\ell'}\right\rangle _{\hat{\rho}_{G2}^{\star}}.\label{eq:nklnkplpfourier}
\end{align}
Utilizing the fact that $\hat{\rho}_{G2}^{\star}=\otimes_{i}\hat{\rho}_{G2;i}^{\star}$,
where $\hat{\rho}_{G2;i}^{\star}$ is the local reduced density matrix
for $\hat{\rho}_{G2}^{\star}$, then $\langle\hat{a}_{j_{1}\ell}^{\dagger}\hat{a}_{j_{2}\ell};\hat{a}_{j_{3}\ell'}^{\dagger}\hat{a}_{j_{4}\ell'}\rangle_{\hat{\rho}_{G2}^{\star}}$
is only non-zero in the following cases: (1) when $j_{1}=j_{2}$,
$j_{3}=j_{4},$ and $j_{1}\neq j_{4}$, the value is $n_{\ell;0}n_{\ell';0}$;
(2) when $j_{1}=j_{4}$, $j_{2}=j_{3}$, $j_{1}\neq j_{2}$, and $\ell=\ell'$,
the value is $A_{\ell}^{2}$, where $A_{\ell}\equiv\langle\hat{a}_{j\ell}^{\dagger};\hat{a}_{j\ell'}\rangle_{\hat{\rho}_{G2}^{\star}}=\langle\hat{a}_{j\ell};\hat{a}_{j\ell'}^{\dagger}\rangle_{\hat{\rho}_{G2}^{\star}}$;
(3) when $j_{1}=j_{2}=j_{3}=j_{4},$ and $\ell=\ell'$, the value
is $n_{\ell;0}=A_{\ell;0}^{2}-n_{\ell;0}^{2}$, where $A_{\ell;0}\equiv\langle\hat{a}_{j\ell}^{\dagger};\hat{a}_{j\ell'}\rangle_{\hat{K}_{2}^{\star}}=\sqrt{\left(1-n_{\ell;0}\right)n_{\ell;0}}$;
(4) when $j_{1}=j_{2}=j_{3}=j_{4}$, and $\ell\neq\ell'$, the value
is $\left\langle \hat{n}_{i\ell}\hat{n}_{i\ell'}\right\rangle _{\hat{\rho}_{G2}^{\star}}$.
Combining these four cases, we have 
\begin{align}
 & \langle\hat{a}_{j_{1}\ell}^{\dagger}\hat{a}_{j_{2}\ell};\hat{a}_{j_{3}\ell'}^{\dagger}\hat{a}_{j_{4}\ell'}\rangle_{\hat{\rho}_{G2}^{\star}}=\delta_{j_{1}j_{2}}\delta_{j_{3}j_{4}}n_{\ell;0}n_{\ell';0}\nonumber \\
 & +\delta_{j_{1}j_{4}}\delta_{j_{2}j_{3}}\delta_{\ell\ell'}A_{\ell}^{2}+\delta_{j_{1}j_{4}}\delta_{j_{2}j_{3}}\delta_{j_{1}j_{3}}C_{\ell\ell'},\label{eq:real space j1j2j3j4rhog2}
\end{align}
where $C_{\ell\ell'}$ is defined as 
\begin{equation}
C_{\ell\ell'}=\begin{cases}
A_{\ell;0}^{2}-A_{\ell}^{2} & \ell=\ell'\\
\left\langle \hat{n}_{i\ell}\hat{n}_{i\ell'}\right\rangle _{\hat{\rho}_{G2}^{\star}}-n_{\ell;0}n_{\ell';0} & \ell\neq\ell'
\end{cases}.
\end{equation}
Therefore, the response coefficient of $n_{k\ell}$ to $\gamma_{k'\ell'}$
is given as 

\begin{align}
 & \frac{\partial\left\langle \hat{n}_{k\ell}\right\rangle _{\hat{\rho}_{G2}}}{\partial\gamma_{k'\ell'}}\bigg|_{\hat{\rho}_{G2}=\hat{\rho}_{G2}^{*}}=\delta_{\ell\ell'}\delta_{kk'}A_{\ell}^{2}+C_{\ell\ell'}/L,\label{eq:dnkdgkplp}
\end{align}
which consists of two contributions: (1) the coherent contribution
$\delta_{\ell\ell'}\delta_{kk'}A_{\ell}^{2}$, which reflects the
hopping renormalization captured in the GA, and (2) the incoherent
contribution $C_{\ell\ell'}/L$, which has no contribution to the
momentum density distribution given that the constraint $\left\langle \hat{n}_{i\ell}\right\rangle _{\hat{K}_{2}}=\left\langle \hat{n}_{i\ell}\right\rangle _{\hat{K}_{2}^{\star}}$
requires 
\begin{equation}
\sum_{k}\delta\gamma_{k\ell}=0.\label{eq:constraint on g}
\end{equation}
When applying Eq. (\ref{eq:dnkdgkplp}) with $\hat{P}_{1}\rightarrow1$
such that $\hat{\rho}_{G2}^{\star}\rightarrow\hat{K}_{2}^{\star}$,
the response for the bare momentum density distribution $n_{k\ell;0}$
to $\gamma_{k'\ell'}$ is given by 
\begin{equation}
\frac{\partial\left\langle \hat{n}_{k\ell}\right\rangle _{\hat{K}_{2}}}{\partial\gamma_{k'\ell'}}\bigg|_{\hat{K}_{2}=\hat{K}_{2}^{*}}=\delta_{\ell\ell'}\delta_{kk'}A_{\ell;0}^{2}.\label{eq:dnkdgkplpInK2}
\end{equation}

The next stage is to find the relation between $\delta n_{k\ell}=\left\langle \hat{n}_{k\ell}\right\rangle _{\hat{\rho}_{G2}}-\left\langle \hat{n}_{k\ell}\right\rangle _{\hat{\rho}_{G2}^{\star}}$
and $\delta n_{k\ell;0}=\left\langle \hat{n}_{k\ell}\right\rangle _{\hat{K}_{2}}-\left\langle \hat{n}_{k\ell}\right\rangle _{\hat{K}_{2}^{\star}}$.
We first compute the response of $\delta n_{k\ell}$ and $\delta n_{k\ell;0}$
for a given $\delta\gamma_{k\ell}$ with the constraint given by Eq.
(\ref{eq:constraint on g}), and then solve $\delta\gamma_{k\ell}$
from $\delta n_{k\ell;0}$ and express $\delta n_{k\ell}$ in terms
of $\delta n_{k\ell;0}$, given as 
\begin{equation}
n_{k\ell}-n_{\ell;0}=\frac{A_{\ell}^{2}}{A_{\ell;0}^{2}}\left(n_{k\ell;0}-n_{\ell;0}\right).\label{eq:dnk in A A0}
\end{equation}
The constraint given by Eq. (\ref{eq:constraint on g}) for $\delta\gamma_{k\ell}$
naturally yields the following relation
\begin{equation}
n_{\ell}\equiv\frac{1}{L}\sum_{k}n_{k\ell}=n_{\ell;0}=\frac{1}{L}\sum_{k}n_{k\ell;0},\label{eq:constraint on nk}
\end{equation}
which is the constraint imposed in the GA. 

Finally, to connect Eq. (\ref{eq:dnk in A A0}) with Eq. (\ref{eq:r matrix}),
we need to show $\mathcal{R}_{\ell}=A_{\ell}/A_{\ell;0}$, which can
be accomplished by proving that $\hat{\rho}_{G2}$ and $\hat{\rho}_{G2}^{\star}$
have the same local reduced density matrix. To prove this, we need
to consider the expectation of a given Hubbard operator under $\hat{\rho}_{G2}$,
which can be accomplished by considering the linear response of a
general diagonal Hubbard operator $\hat{X}_{i\Gamma}$ at site $i$
to $\gamma_{k\ell}$ as 
\begin{align}
 & \frac{\partial\langle\hat{X}_{i\Gamma}\rangle_{\hat{\rho}_{G2}}}{\partial\gamma_{k\ell}}\bigg|_{\hat{\rho}_{G2}=\hat{\rho}_{G2}^{*}}=\langle\hat{X}_{i\Gamma};\hat{n}_{k\ell}\rangle_{\hat{\rho}_{G2}^{\star}}-\langle\hat{X}_{i\Gamma}\rangle_{\hat{\rho}_{G2}^{\star}}n_{\ell},\label{eq:dkdgkldef}\\
 & \langle\hat{X}_{i\Gamma};\hat{n}_{k\ell}\rangle_{\hat{\rho}_{G2}^{\star}}=\frac{1}{L}\sum_{jj'}e^{ik\left(j-j'\right)}\langle\hat{X}_{i\Gamma};\hat{a}_{j\ell}^{\dagger}\hat{a}_{j'\ell}\rangle_{\hat{\rho}_{G2}^{\star}},\label{eq:xnklcorrelation}\\
 & \langle\hat{X}_{i\Gamma};\hat{a}_{j\ell}^{\dagger}\hat{a}_{j'\ell}\rangle_{\hat{\rho}_{G2}^{\star}}=\delta_{jj'}\langle\hat{X}_{i\Gamma}\rangle_{\hat{\rho}_{G2}^{\star}}n_{\ell}+\delta_{jj'}\delta_{ij}C_{\Gamma\ell},\label{eq:xajajpcorrelation}\\
 & C_{\Gamma\ell}=\left\langle \hat{X}_{i\Gamma}\hat{n}_{i\ell}\right\rangle _{\hat{\rho}_{G2}^{\star}}-\langle\hat{X}_{i\Gamma}\rangle_{\hat{\rho}_{G2}^{\star}}n_{\ell},\label{eq:cxl}\\
 & \frac{\partial\left\langle \hat{X}_{i\Gamma}\right\rangle _{\hat{\rho}_{G2}}}{\partial\gamma_{k\ell}}\bigg|_{\hat{\rho}_{G2}=\hat{\rho}_{G2}^{*}}=\frac{1}{L}C_{\Gamma\ell}.\label{eq:dxdgkl}
\end{align}
Given the constraint on $\gamma_{k\ell}$ (see Eq. \ref{eq:constraint on g}),
the first order change in $\langle\hat{X}_{i\Gamma}\rangle_{\hat{\rho}_{G2}}$
is zero and therefore we have 
\begin{equation}
\left\langle \hat{X}_{i\Gamma}\right\rangle _{\hat{\rho}_{G2}}=\left\langle \hat{X}_{i\Gamma}\right\rangle _{\hat{\rho}_{G2}^{\star}}.\label{eq:x=00003Dx}
\end{equation}
Similarly, with $\hat{P}_{1}\rightarrow1$ and $\hat{\rho}_{G2}\rightarrow\hat{K_{2}},$we
have 
\begin{equation}
\left\langle \hat{X}_{i\Gamma}\right\rangle _{\hat{K}_{2}}=\left\langle \hat{X}_{i\Gamma}\right\rangle _{\hat{K}_{2}^{\star}}.\label{eq:x=00003Dx K2}
\end{equation}
Therefore, we have recovered Eq. (\ref{eq:pgamma}). Moreover, we
have 
\begin{equation}
\frac{A_{\ell}}{A_{\ell;0}}=\frac{\text{Tr}\left(\sqrt{\hat{\rho}_{G2}^{\star}}\hat{a}_{i\ell}^{\dagger}\sqrt{\hat{\rho}_{G2}^{\star}}\hat{a}_{i\ell}\right)/\text{Tr}\left(\hat{\rho}_{G2}^{\star}\right)}{\text{Tr}\left(\sqrt{\hat{K}_{2}^{\star}}\hat{a}_{i\ell}^{\dagger}\sqrt{\hat{K}_{2}^{\star}}\hat{a}_{i\ell}\right)/\text{Tr}\left(\hat{K}_{2}^{\star}\right)}=\mathcal{R}_{\ell},\label{eq:a/a0}
\end{equation}
where we have used the fact that the local reduced density matrix
of $\hat{\rho}_{G2}^{\star}$ is same as the local reduced density
matrix of $\hat{\rho}_{G2}$, as indicated in Eq. (\ref{eq:x=00003Dx}),
and represented as $\rho_{loc}$ in Eq. (\ref{eq:rholoc}). Similarly,
the local reduced density matrix of $\hat{K}_{2}^{\star}$ is the
same as the local reduced density matrix of $\hat{K}_{2}$, as indicated
in Eq. (\ref{eq:x=00003Dx K2}), and represented as $\rho_{loc;0}$
in Eq. (\ref{eq:rholoc0}). 

In summary, we have demonstrated that the first order CPE is equivalent
to the GA and the SCDA. The key steps in the proof include: (1) the
local reduced density matrix for $\hat{\rho}_{G2}$ is the same as
for $\hat{\rho}_{G2}^{\star}$, with no dependency on the details
of $n_{k\ell;0}$ except its average value and (2) the momentum density
distribution $n_{k\ell}$ is uniformly shrunk towards $n_{\ell}$
through $\mathcal{Z}_{\ell}$, which is uniquely determined by $\hat{\rho}_{G2}^{\star}$.
Finally, it would be interesting to explore the behavior of the CPE
beyond first order in finite dimensions, which provides insight beyond
the GA for evaluating $\hat{\rho}_{G2}$.

\subsection{The CPE for $\hat{\rho}_{B2}$ \label{subsec:Appendix-the-CPE-for-B}}

In this section, we use the first-order CPE to evaluate $\hat{\rho}_{B2}=\hat{K}_{1}\hat{P}_{1}\hat{K}_{1}$,
where $\hat{P}_{1}=\exp\left(\sum_{i\Gamma}\upsilon_{\Gamma}\hat{X}_{i\Gamma}\right)$
and $\hat{K}_{1}=\exp\left(\sum_{k\ell}\gamma_{k\ell}\hat{n}_{k\ell}\right)$,
demonstrating the equivalence to the SCDA in $d=\infty$. The CPE
for $\hat{\rho}_{B2}$ can be viewed as a dual version of the CPE
for the $\hat{\rho}_{G2}$, which has various correspondences \cite{Cheng2020081105}.
First, in the CPE for the $\hat{\rho}_{G2}$, the central projector
$\hat{K}_{2}$ determines the local density, and is invariant after
applying the interacting projector $\hat{P}_{1}$. Correspondingly,
in the CPE for $\hat{\rho}_{B2}$, the central projector $\hat{P}_{1}$
determines the local density, and is invariant after applying $\hat{K}_{1}$.
Second, in the CPE for the $\hat{\rho}_{G2}$, the physical momentum
density distribution is linearly related to the bare momentum density
distribution, and the local reduced density matrix is independent
of the details of the momentum density distribution. Correspondingly,
in the CPE for the $\hat{\rho}_{B2}$, the local reduced density matrix
is linearly related to the reference local reduced density matrix,
and the momentum density distribution is independent of the details
of the local reduced density matrix. 

We proceed in evaluating $\hat{\rho}_{B2}$ via the CPE by choosing
the central point for $\hat{\rho}_{B2}$ as $\hat{\rho}_{B2}^{\star}=\hat{K}_{1}\hat{P}_{1}^{\star}\hat{K}_{1}$,
where $\hat{P}_{1}^{\star}=\exp\left(\sum_{i\ell}\gamma_{\ell}^{\star}\hat{n}_{i\ell}\right)$
and $\gamma_{\ell}^{\star}$ is chosen such that $n_{\ell}^{\star}\equiv\left\langle \hat{n}_{i\ell}\right\rangle _{\hat{P}_{1}}=\left\langle \hat{n}_{i\ell}\right\rangle _{\hat{P}_{1}^{\star}}$.
We can also rewrite $\hat{P}_{1}^{\star}=\exp\left(\sum_{i\Gamma}\upsilon_{\Gamma}^{\star}\hat{X}_{i\Gamma}\right)$
by expressing $\hat{n}_{i\ell}$ as a linear combination of Hubbard
operators, where $\upsilon_{\Gamma}^{\star}=\sum_{\ell}\gamma_{\ell}^{\star}\Gamma(\ell)$.
Considering the linear response of $\upsilon_{\Gamma}$ about $\upsilon_{\Gamma}^{\star}$,
similar to Eqns. (\ref{eq:dOdgamma})-(\ref{eq:dOdgamma-response}),
we have 
\begin{align}
 & \frac{\partial\langle\hat{O}\rangle_{\hat{\rho}_{B2}}}{\partial\upsilon_{i\Gamma}}\bigg|_{\hat{\rho}_{B2}=\hat{\rho}_{B2}^{*}}=\frac{\partial}{\partial\upsilon_{i\Gamma}}\frac{\text{Tr}\left(\hat{K}_{1}\hat{P}_{1}\hat{K_{1}}\hat{O}\right)}{\text{Tr}\left(\hat{K}_{1}\hat{P}_{1}\hat{K_{1}}\right)}\label{eq:dOdupsilon}\\
 & =\frac{\text{Tr}\left(\hat{K}_{1}\sqrt{\hat{P}_{1}^{\star}}\hat{X}_{i\Gamma}\sqrt{\hat{P}_{1}^{\star}}\hat{K}_{1}\hat{O}\right)}{\text{Tr}\left(\hat{\rho}_{B2}^{\star}\right)}\nonumber \\
 & -\frac{\text{Tr}\left(\hat{K}_{1}\sqrt{\hat{P}_{1}^{\star}}\hat{X}_{i\Gamma}\sqrt{\hat{P}_{1}^{\star}}\hat{K}_{1}\right)\text{Tr}\left(\hat{\rho}_{B2}^{\star}\hat{O}\right)}{\text{Tr}\left(\hat{\rho}_{B2}^{\star}\right)^{2}}\\
 & =\langle\hat{X}_{i\Gamma};\hat{O}\rangle_{\hat{\rho}_{B2}^{\star}}-\langle\hat{X}_{i\Gamma}\rangle_{\hat{\rho}_{B2}^{\star}}\langle\hat{O}\rangle_{\hat{\rho}_{B2}^{\star}},\label{eq:XOB2}
\end{align}
where we similarly utilize the fact that $\hat{P}_{1}^{\star}$ commutes
with $\hat{K}_{1}$ and $\sqrt{\hat{\rho}_{B2}^{\star}}=\hat{K}_{1}\sqrt{\hat{P}_{1}^{\star}}=\sqrt{\hat{P}_{1}^{\star}}\hat{K}_{1}$.
We proceed by computing the response 
\begin{align}
\frac{\partial\langle\hat{X}_{i'\Gamma'}\rangle_{\hat{\rho}_{B2}}}{\partial\upsilon_{i\Gamma}}\bigg|_{\hat{\rho}_{B2}=\hat{\rho}_{B2}^{*}} & =\langle\hat{X}_{i\Gamma};\hat{X}_{i'\Gamma'}\rangle_{\hat{\rho}_{B2}^{\star}}\nonumber \\
 & -\langle\hat{X}_{i\Gamma}\rangle_{\hat{\rho}_{B2}^{\star}}\langle\hat{X}_{i'\Gamma'}\rangle_{\hat{\rho}_{B2}^{\star}}.\label{eq:X X correlation}
\end{align}
Directly computing Eq. (\ref{eq:X X correlation}) is cumbersome,
and this can be avoided by decomposing $\hat{X}_{i\Gamma}$ into an
alternate form using Eq. (\ref{eq:X decomposition}), resulting in
\begin{equation}
\hat{X}_{i\Gamma}=\prod_{\ell}\hat{X}_{i\Gamma;\ell}=\prod_{\ell}\left(X_{i\Gamma;\ell}^{\star}+\delta\hat{X}_{i\Gamma;\ell}\right),\label{eq:X decomposition fluctuation}
\end{equation}
where 
\begin{align}
 & X_{i\Gamma;\ell}^{\star}\equiv\langle\hat{X}_{i\Gamma;\ell}\rangle_{\hat{\rho}_{B2}^{\star}}=\left(1-n_{\ell}^{\star}\right)\delta_{0,\Gamma\left(\ell\right)}+n_{\ell}^{\star}\delta_{1,\Gamma\left(\ell\right)},\label{eq:Xigstar}\\
 & \delta\hat{X}_{i\Gamma;\ell}\equiv\hat{X}_{i\Gamma;\ell}-X_{i\Gamma;\ell}^{\star}=\left(\hat{n}_{i\ell}-n_{\ell}^{\star}\right)\left(\delta_{1,\Gamma\left(\ell\right)}-\delta_{0,\Gamma\left(\ell\right)}\right).\label{eq:dXig}
\end{align}
We introduce the density fluctuation operator $\delta\hat{D}_{iI}=\prod_{\ell\in I}\delta\hat{n}_{i\ell}$,
where $\delta\hat{n}_{i\ell}=\hat{n}_{i\ell}-n_{\ell}^{\star}$, and
$\delta\hat{D}_{I}=\hat{1}$ when $I=\left\{ \right\} $, and we use
$\delta\hat{D}_{iI}$ as a new basis for the CPE expansion. The diagonal
Hubbard operator $\hat{X}_{i\Gamma}$ can be written as 
\begin{equation}
\hat{X}_{i\Gamma}=\sum_{I}\left(\prod_{\ell\in\bar{I}}X_{i\Gamma;\ell}^{\star}\right)\left(\prod_{\ell\in I}\left(\delta_{1,\Gamma\left(\ell\right)}-\delta_{0,\Gamma\left(\ell\right)}\right)\right)\delta\hat{D}_{iI},\label{eq:X in deltaD}
\end{equation}
where $I$ enumerates over all subsets of $\left\{ 1,2,\dots,2N_{orb}\right\} $
and $\bar{I}=\left\{ 1,2,\dots,N_{orb}\right\} -I$ . The interacting
projector can be written as $\hat{P}_{1}=\exp\left(\sum_{iI}\eta_{iI}\delta\hat{D}_{iI}\right)$,
and at the central point $\eta_{i\left\{ \ell\right\} }^{\star}=\gamma_{\ell}^{\star}$,
with $\eta_{i\left\{ \right\} }^{\star}$ absorbing the constant contribution,
while $\eta_{iI}^{\star}=0$ if $\left|I\right|>1$. Similar to Eq.
(\ref{eq:XOB2}), for $\left|I\right|>0$ we have 
\begin{equation}
\frac{\partial\langle\hat{O}\rangle_{\hat{\rho}_{B2}}}{\partial\eta_{iI}}\bigg|_{\hat{\rho}_{B2}=\hat{\rho}_{B2}^{*}}=\langle\delta\hat{D}_{iI};\hat{O}\rangle_{\hat{\rho}_{B2}^{\star}},
\end{equation}
given that $\langle\delta\hat{D}_{iI}\rangle_{\hat{\rho}_{B2}^{\star}}=0$
in this case. It should be noted that $\eta_{i\{\}}$ has no effect
on the expectation values so we implicitly only consider cases with
$\left|I\right|>0$. The response of $\langle\delta\hat{D}_{i'I'}\rangle_{\hat{\rho}_{B2}}$
to $\eta_{iI}$ is given as 
\begin{equation}
\frac{\partial\langle\delta\hat{D}_{i'I'}\rangle_{\hat{\rho}_{B2}}}{\partial\eta_{iI}}\bigg|_{\hat{\rho}_{B2}=\hat{\rho}_{B2}^{\star}}=\langle\delta\hat{D}_{iI};\delta\hat{D}_{i'I'}\rangle_{\hat{\rho}_{B2}^{\star}}.\label{eq:ddDdeta}
\end{equation}
Since $\hat{\rho}_{B2}^{\star}$ is diagonal in $\ell$, the correlation
function on the right side of Eq. (\ref{eq:ddDdeta}) is only non-zero
when $I=I'$, and it can be written as the product of contributions
from each relevant spin-orbital as 
\begin{equation}
\langle\delta\hat{D}_{iI};\delta\hat{D}_{i'I'}\rangle_{\hat{\rho}_{B2}^{\star}}=\delta_{II'}\prod_{\ell\in I}\langle\delta\hat{n}_{i\ell};\delta\hat{n}_{i'\ell}\rangle_{\hat{\rho}_{B2}^{\star}},\label{eq:dDdD}
\end{equation}
requiring the evaluation of 
\begin{equation}
\langle\delta\hat{n}_{i\ell};\delta\hat{n}_{i'\ell}\rangle_{\hat{\rho}_{B2}^{\star}}=\langle\hat{n}_{i\ell};\hat{n}_{i'\ell}\rangle_{\hat{\rho}_{B2}^{\star}}-(n_{\ell}^{\star})^{2},\label{eq:dndn}
\end{equation}
and $\langle\hat{n}_{i\ell};\hat{n}_{i'\ell}\rangle_{\hat{\rho}_{B2}^{\star}}$
can be evaluated similarly to Eq. (\ref{eq:nklnkplpfourier}) as
\begin{align}
 & \langle\hat{n}_{j\ell};\hat{n}_{j'\ell}\rangle_{\hat{\rho}_{B2}^{\star}}=\frac{1}{L^{2}}\sum_{k_{1}k_{2}k_{3}k_{4}}e^{i\left(k_{2}-k_{1}\right)j+i\left(k_{4}-k_{3}\right)j'}\nonumber \\
 & \times\langle\hat{a}_{k_{1}\ell}^{\dagger}\hat{a}_{k_{2}\ell};\hat{a}_{k_{3}\ell}^{\dagger}\hat{a}_{k_{4}\ell}\rangle_{\hat{\rho}_{B2}^{\star}}.\label{eq:njlnjpl}
\end{align}
Given that $\hat{\rho}_{B2}^{\star}=\otimes_{k\ell}\hat{\rho}_{B2;k\ell}^{\star}$
is a direct product state in momentum space, $\langle\hat{a}_{k_{1}\ell}^{\dagger}\hat{a}_{k_{2}\ell};\hat{a}_{k_{3}\ell}^{\dagger}\hat{a}_{k_{4}\ell}\rangle_{\hat{\rho}_{B2}^{\star}}$
is only non-zero in the following cases. (1) $k_{1}=k_{2}$, $k_{3}=k_{4}$,
and $k_{1}\neq k_{4}$ results in $n_{k_{1}\ell}^{\star}n_{k_{3}\ell}^{\star}$,
where $n_{k\ell}^{\star}=\langle\hat{n}_{k\ell}\rangle_{\hat{\rho}_{B2}^{\star}}$,
(2) $k_{1}=k_{4}$, $k_{2}=k_{3}$, and $k_{1}\neq k_{2}$ results
in $A_{k_{1}\ell}A_{k_{2}\ell}$, where $A_{k\ell}\equiv\langle\hat{a}_{k\ell}^{\dagger};\hat{a}_{k\ell}\rangle_{\hat{\rho}_{B2}^{\star}}=\langle\hat{a}_{k\ell};\hat{a}_{k\ell}^{\dagger}\rangle_{\hat{\rho}_{B2}^{\star}}$
and $A_{k\ell}=\sqrt{n_{k\ell}^{\star}\left(1-n_{k\ell}^{\star}\right)}$,
(3) $k_{1}=k_{2}=k_{3}=k_{4}$ results in $n_{k\ell}^{\star}=\left(n_{k\ell}^{\star}\right)^{2}+A_{k\ell}^{2}$.
In summary, we have 
\begin{align}
 & \langle\hat{a}_{k_{1}\ell}^{\dagger}\hat{a}_{k_{2}\ell};\hat{a}_{k_{3}\ell}^{\dagger}\hat{a}_{k_{4}\ell}\rangle_{\hat{\rho}_{B2}^{\star}}=\delta_{k_{1}k_{2}}\delta_{k_{3}k_{4}}n_{k_{1}\ell}^{\star}n_{k_{3}\ell}^{\star}\nonumber \\
 & +\delta_{k_{1}k_{4}}\delta_{k_{2}k_{3}}A_{k_{1}\ell}A_{k_{2}\ell},\label{eq:momentum space k1 k2 k3 k4 rhob2}
\end{align}
which can be used to obtain 
\[
\langle\hat{n}_{j\ell};\hat{n}_{j'\ell}\rangle_{\hat{\rho}_{B2}^{\star}}=\left(\frac{1}{L}\sum_{k}n_{k\ell}^{\star}\right)^{2}+\left|\frac{1}{L}\sum_{k}e^{ik\left(j-j'\right)}A_{k\ell}\right|^{2},
\]
where $L$ is the number of sites in the lattice. A constraint is
imposed on $\hat{K}_{1}$ such that 
\begin{equation}
\langle\hat{n}_{i\ell}\rangle_{\hat{\rho}_{B2}^{\star}}=n_{\ell}^{\star},\label{eq:constraint on K}
\end{equation}
resulting in $\left(1/L\right)\sum_{k}n_{k\ell}^{\star}=n_{\ell}^{\star}$
and 
\begin{equation}
\langle\delta\hat{n}_{j\ell};\delta\hat{n}_{j'\ell}\rangle_{\hat{\rho}_{B2}^{\star}}=\left|\frac{1}{L}\sum_{k}e^{ik\left(j-j'\right)}A_{k\ell}\right|^{2}.\label{eq:dnjdnjp}
\end{equation}
In infinite dimensions, only local contributions need to be accounted
for, which is used in the following steps. Considering the response
$\delta D_{iI}=\langle\delta\hat{D}_{iI}\rangle_{\hat{\rho}_{B2}}$
and $\delta D_{iI;0}=\langle\delta\hat{D}_{iI}\rangle_{\hat{P}_{1}}$
to first order in $\eta_{iI}$ and solving for $\eta_{iI}$ as a function
of $\delta D_{iI;0}$, the response in infinite dimensions is 
\begin{equation}
\delta D_{iI}=\left(\prod_{\ell\in I}\mathcal{F}_{\ell}\right)\delta D_{iI;0},\label{eq:deltaD}
\end{equation}
where the renormalization factor is given by 
\begin{equation}
\mathcal{F}_{\ell}\equiv\left(\sum_{k}A_{k\ell}\right)^{2}\bigg/\left(\sum_{k}A_{k\ell;0}\right)^{2},\label{eq:Fl}
\end{equation}
where $A_{k\ell}=\sqrt{n_{k\ell}^{\star}\left(1-n_{k\ell}^{\star}\right)}$
and $A_{k\ell;0}=\sqrt{n_{\ell}^{\star}\left(1-n_{\ell}^{\star}\right)}$. 

Finally, we need to consider how the momentum density distribution
is influenced by $\eta_{iI}$ using the response 
\begin{align}
 & \frac{\partial\langle\hat{n}_{k\ell}\rangle_{\hat{\rho}_{B2}}}{\partial\eta_{iI}}\bigg|_{\hat{\rho}_{B2}=\hat{\rho}_{B2}^{*}}=\langle\delta\hat{D}_{iI};\hat{n}_{k\ell}\rangle_{\hat{\rho}_{B2}^{\star}}\\
 & =\prod_{\ell'\in I}\langle\delta\hat{n}_{i\ell'};\hat{n}_{k\ell}\rangle_{\hat{\rho}_{B2}^{\star}},
\end{align}
which is only non-zero when $I=\left\{ \ell\right\} $. To first order,
$\eta_{i\{\ell\}}=\eta_{i\{\ell\}}^{\star}$ and the contribution
from $\eta_{iI}$ where $\left|I\right|>0$ yields
\begin{equation}
n_{k\ell}\equiv\langle\hat{n}_{k\ell}\rangle_{\hat{\rho}_{B2}}=\langle\hat{n}_{k\ell}\rangle_{\hat{\rho}_{B2}^{\star}}\equiv n_{k\ell}^{\star},\label{eq:nkl nklstar}
\end{equation}
and
\begin{equation}
n_{\ell}\equiv\langle\hat{n}_{i\ell}\rangle_{\hat{\rho}_{B2}}=\langle\hat{n}_{i\ell}\rangle_{\hat{\rho}_{B2}^{\star}}=n_{\ell}^{\star}.\label{eq:nkl p1}
\end{equation}

In conclusion, if we write the local Hamiltonian as $\hat{H}_{loc}=\sum_{iI}E_{loc;I}\delta\hat{D}_{iI}$,
the total energy per site for $\hat{\rho}_{B2}$ is given as 
\[
E=\sum_{\ell}\int dk\epsilon_{k\ell}n_{k\ell}+\sum_{I}E_{loc;I}\left(\prod_{\ell\in I}\mathcal{F}_{\ell}\right)\delta D_{iI;0},
\]
where $\delta D_{iI;0}=\langle\delta\hat{D}_{iI}\rangle_{\hat{P}_{1}}$
with $\delta D_{iI;0}=0$ for $\left|I\right|=1$, and $n_{k\ell}\equiv\langle\hat{n}_{k\ell}\rangle_{\hat{\rho}_{B2}}\in\left[0,1\right]$
is the physical momentum density distribution and is constrained by
$\left(1/L\right)\sum_{k}n_{k\ell}=n_{\ell}$. The $n_{k\ell}$ can
be viewed as variational parameters determined from the $\gamma_{k\ell}$
within $\hat{K}_{1}$. Combining Eq. (\ref{eq:Fl}) and Eq. (\ref{eq:nkl nklstar}),
we can explicitly express $\mathcal{F}_{\ell}$ as 
\begin{equation}
\mathcal{F}_{\ell}=\left|\frac{\int dk\sqrt{n_{k\ell}\left(1-n_{k\ell}\right)}}{\sqrt{n_{\ell}\left(1-n_{\ell}\right)}}\right|^{2}.\label{eq:Fl explicite}
\end{equation}
To evaluate the Hamiltonian in the usual Hubbard operator representation
for a given $X_{i\Gamma;0}=\langle\hat{X}_{i\Gamma}\rangle_{\hat{P}_{1}}$
while respecting the density constraint, we can compute $\delta D_{iI;0}$
from $X_{i\Gamma;0}$, and then use Eq. (\ref{eq:deltaD}) to evaluate
$\delta D_{iI}$ and then use Eq. (\ref{eq:X in deltaD}) to compute
$X_{i\Gamma}=\langle\hat{X}_{i\Gamma}\rangle_{\hat{\rho}_{B2}}.$
Finally, we remark that Eq. (\ref{eq:deltaD}) is derived by ignoring
non-local contributions, which is equivalent to taking the limit of
infinite dimensions. One could straightforwardly account for non-local
contributions using Eq. (\ref{eq:dnjdnjp}), which will be explored
in the future work.


\end{document}